\documentclass[journal]{rmaa}
\usepackage{amsmath}
\usepackage{xspace}
\usepackage[caption=false]{subfig}
\usepackage{paralist}

\usepackage{psfrag,color}

\usepackage[latin1]{inputenc}

\usepackage{graphicx}
\usepackage{lscape}
\usepackage{txfonts}
\usepackage{natbib}
\usepackage{url}
\usepackage[T1]{fontenc}

\usepackage{url}
\usepackage{xcolor}

\DeclareRobustCommand{\ion}[2]{%
\relax\ifmmode
\ifx\testbx\f@series
{\mathbf{#1\,\sc{#2}}}\else
{\mathrm{#1\,\sc{#2}}}\fi
\else\textup{#1\,{\mdseries\textsc{#2}}}%
\fi}

\DeclareRobustCommand{\ION}[2]{%
\relax\ifmmode
\ifx\testbx\f@series
{\mathbf{#1\,\mathsc{#2}}}\else
{\mathrm{#1\,\mathsc{#2}}}\fi
\else\textup{#1\,{\mdseries\textsc{#2}}}%
\fi}

\newcommand{\hii}{\ION{H}{ii}}

\newcommand{\nii}{[\ION{N}{ii}]}
\newcommand{\oi}{[\ION{O}{i}]}
\newcommand{\oii}{[\ION{O}{ii}]}
\newcommand{\oiii}{[\ION{O}{iii}]}
\newcommand{\sii}{[\ION{S}{ii}]}

\newcommand{\Ha}{$\rm{H}\alpha$}
\newcommand{\Hb}{$\rm{H}\beta$}
\newcommand{\Hd}{$\rm{H}\delta$}
\newcommand{\Hg}{$\rm{H}\gamma$}

\newcommand{\EWa}{EW$(\rm{H}\alpha)$}

\newcommand{\HII}{\ion{H}{ii}}

\newcommand{\ageLW}{\ifmmode \mathcal{A}_{\star,L} \else $\mathcal{A}_{\star,L}$\fi\xspace}
\newcommand{\age}{\ifmmode \mathcal{A}_\star \else $\mathcal{A}_\star$\fi\xspace}
\newcommand{\met}{\ifmmode \mathcal{Z}_\star \else $\mathcal{Z}_\star$\fi\xspace}
\newcommand{\metLW}{\ifmmode \mathcal{Z}_{\star,L} \else $\mathcal{Z}_{\star,L}$\fi\xspace}
\newcommand{\ageMW}{\ifmmode \mathcal{A}_{\star,M} \else $\mathcal{A}_{\star,M}$\fi\xspace}
\newcommand{\metMW}{\ifmmode \mathcal{Z}_{\star,M} \else $\mathcal{Z}_{\star,M}$\fi\xspace}

\newcommand{\funitsI}{10$^{-16}$\,erg/s/\AA/cm$^2$}

\newcommand{\kms}{km\,s$^{-1}$} 

\newcommand{\flux}{erg\,s$^{-1}$\,cm$^{-2}$}

\def\pyf{\texttt{pyFIT3D}\xspace}
\def\pyp{\texttt{pyPipe3D}\xspace}

\newcommand{\edr}{\citet{edr}}
\newcommand{\edrp}{\citep{edr}}
\newcommand{\edrnp}{\citep{edr}}

\title{The Calar Alto Legacy Integral Field Area Survey: spatial resolved properties.}

\author{S. F. S\'anchez\altaffilmark{1}, J. K. Barrera-Ballesteros\altaffilmark{1},L. Galbany\altaffilmark{2}\altaffilmark{3}   %C.J. Walcher\altaffilmark{4},  
R.Garc\'\i a-Benito\altaffilmark{4}, E. Lacerda\altaffilmark{1}, A. Camps-Fari\~na\altaffilmark{5}}%, A. Mej\'ia-Narv\'aez\altaffilmark{1}}
\altaffiltext{1}{Instituto de Astronom\'ia, Universidad Nacional Aut\'onoma de  M\'exico}
\altaffiltext{2}{Institute of Space Sciences (ICE, CSIC), Campus UAB}
\altaffiltext{3}{Institut d'Estudis Espacials de Catalunya (IEEC).}
\altaffiltext{4}{Instituto de Astrof\'isica de Andaluc\'ia (IAA/CSIC)}
\altaffiltext{5}{Universidad Complutense de Madridd (UCM)}
\shortauthor{S\'anchez et al.}
\shorttitle{eCALIFA spatial resolved properties}

\fulladdresses{
\item Instituto de Astronom\'\i a,
  Universidad Nacional Auton\'oma de M\'exico,
  A.P. 70-264, 04510 M\'exico, D.F.,  Mexico (sfsanchez@astro.unam.mx).

 \item Institute of Space Sciences (ICE, CSIC),
  Campus UAB, Carrer de Can Magrans, s/n, E-08193 Barcelona, Spain.

\item Institut d'Estudis Espacials de Catalunya (IEEC),
  E-08034 Barcelona, Spain.
  
\item Leibniz-Institut f\"ur Astrophysik Potsdam (AIP),
  An der Sternwarte 16,
  D-14482 Potsdam, Germany.

\item Instituto de Astrof\'isica de Andaluc\'ia (IAA/CSIC),
  Glorieta de la Astronom\'{\i}a s/n Aptdo. 3004,
  E-18080 Granada, Spain.

\item Departamento de FÃ­sica de la Tierra y Astrof\'isica, Universidad Complutense de Madrid, Pl. Ciencias, 1, Madrid, 28040, Madrid.
  
}

\listofauthors{S. F. S\'anchez}
\indexauthor{S. F. S\'anchez}

\abstract{
  We present the analysis using the
  \pyp pipeline for the 895 galaxies that comprises the eCALIFA data release
  \edrp, data with a significantly improved spatial resolution (1.0-1.5$\arcsec$/FWHM). We include a description of (i) the analysis performed by the pipeline, (ii)
  the adopted datamodel for the derived spatially resolved properties
  and (iii) the catalog of integrated, characteristics and slope of
  the radial gradients for a set of observational and physical
  parameters derived for each galaxy. We illustrate the results of the
  analysis (i) using the NGC\,2906 galaxy, showing the
  spatial distribution of the different derived parameters, and (ii) showing the 
  distribution of the spatial resolved ionized gas across the \oiii/\Hb\ vs. \nii/\Ha\ diagram for the whole galaxy sample. A general agreement is found with published results, with a clear improvement in the tracing of radial patterns and the segregation of individual ionized structures. Access to all the discussed dataproducts: \url{http://ifs.astroscu.unam.mx/CALIFA_WEB/public_html/}}

\resumen{
Presentamos el an\'alisis utilizando \pyp en las 985 galaxies de la dstribuci\'on de datos eCALIFA \edrp, los cuales presentan una mejora significativa en la resoluci\'on espacial  (1.0-1.5$\arcsec$/FWHM), incluyendo una descripci\'on de (i) el an\'alisis realizado por el dataducto, (ii) el modelo de datos adoptado para las propiedades espacialmente resueltas obtenidas, y (iii) el cat\'alogo de
propiedades f\'\i sicas y observacionales, integradas, caracter\'\i siticas y
las pendientes de sus gradientes radiales. Ilustramos los resultados del
an\'alisis (i) utilizando la galaxia
NGC\,2906, mostrando la distribuci\'on espacial de los diferentes
par\'ametros obtenidas, y (ii) explorando la distribuci\'on del
gas ionizado espacialmente resuelto a lo largo del diagrama  \oiii/\Hb\ vs. \nii/\Ha\ para la muestra completa de
galaxias. Se reporta un acuerdo general con resultados anteriores, mejorando la determinación de patrones radiales y la segregación de estructuras ionizadas. Acceso a los productos del analisis: \url{http://ifs.astroscu.unam.mx/CALIFA_WEB/public_html/}.
}

\addkeyword{galaxies: evolution}
\addkeyword{galaxies: star formation}
\addkeyword{galaxies: stellar content}
\addkeyword{galaxies: ISM}
\addkeyword{techniques: imaging spectroscopy}

\begin{document}

% Page header
%\markboth{S.F.S\'anchez}{Properties of SFGs}

% Title
%\title{Spatially Resolved Properties of Low-Redshift Star-Forming Galaxies in the Local Universe}

%\affil{$^2$Department/Institute, University, City, Country, Postal code}
%\affil{$^3$Department/Institute, University, City, Country, Postal code}}

%Abstract

%Keywords, etc.

\maketitle

%Table of Contents
%\tableofcontents

\section{Introduction}
\label{sec:intro}

The Calar Alto Legacy Integral Field Area (CALIFA) project is an
integral field spectroscopic (IFS) galaxy survey focused on the
exploration of the spatial resolved spectroscopic properties of a
statistical well defined and significant sample of galaxies in the
nearby survey \citep{califa}. It selected galaxies using their
optical extension (diameter) as the primary criterion
(D$_{iso}\sim$60$\arcsec$), in order to fit them within the field-of-view
of the adopted integral-field unit \citep[PPAK;][]{kelz06}. Besides
that, the sample was restricted to a redshift range around
z$\sim$0.015, and a lower-limit in the apparent magnitude was imposed
to limit the contribution of dwarf galaxies, that would in other case
dominate the statistics \citep{walcher14}. This survey gathered data
for a time period between 2010 and 2016, presenting its last formal
data release in \citet{sanchez16}. Afterwards, a set of programs
continue using similar selection criteria and the same instrumental
and observational setup to explore galaxies under-represented in the
original sample, like companion galaxies in interacting systems,
cluster members, and in particular supernovae hosts
\citep[e.g.][]{pisco}. All those additional sub-samples of galaxies can
be considered as extensions of the original sample, comprising the
so-called extended CALIFA
compilation \citep[e.g.][]{sanchez16,lacerda20,espi20}.

In a recent companion article \edr\ it was presented the final
compilation of the extended CALIFA dataset (eCALIFA), releasing the
datacubes using an improved version of the data-reduction that
significantly improves the spatial resolution and image quality. This
data release (DR) comprises the datacubes corresponding to the V500
setup \citep[e.g., low-resolution][]{califa} of 895 individual
galaxies, together with a set of integrated and central aperture
observational and physical parameters included in a single table
\footnote{\url{http://ifs.astroscu.unam.mx/CALIFA_WEB/public_html/}}.

Following a similar approach adopted in previous DRs of IFS galaxy
surveys \citep[e.g.][]{sanchez22}, we distribute in this paper the
dataproducts of the analysis for this dataset using the {\tt pyPipe3D}
pipeline \citep[][]{pypipe3d}. {\tt pyPipe3D} is a recently
updated version of {\tt Pipe3D} fully coded in {\tt python}. The {\tt
  Pipe3D} pipeline makes use of the routines and algorithms included
in the {\tt FIT3D} package \citep{pipe3d}, which main goal is to extract
the properties of the ionized gas and the stellar component from IFS
data in the optical range. {\tt Pipe3D} has been frequently used to
explore the data from different surveys: e.g., CALIFA
\citep{laura16,espi20}, SAMI \citep{sanchez19}, AMUSING++
\citep{laura18,carlos20} and MaNGA \citep{sanchez18,sanchez22}.

The structure of this articles is as follows: (i) Sec. \ref{sec:data}
includes a description of the explored data, with a brief summary of
the data reduction; (ii) the analysis performed on the data is
included in Sec. \ref{sec:ana}, comprising (iii) a summary of the main
properties of \pyp (Sec. \ref{sec:pyPipe3D}), that includes (iii.1) a
description of the adopted spatial binning (Sec. \ref{sec:sp_bin}),
(iii.2) details on the stellar population fitting
(Sec. \ref{sec:st_fit}, (iii.3) how the emission line properties are
derived (Sec. \ref{sec:el_fit}, (iii.4) a description of the
derivation of stellar indices (Sec. \ref{sec:st_ind}), and finally
(iii.5) how the errors of the different parameters are derived
(Sec. \ref{sec:ana_error}) and (iii.6) the final masks generate
excluding field stars and low signal-to-noise (S/N) regions
(Sec. \ref{sec:ana_mask}); (iv) a description of the physical
quantities derived from \pyp dataproducts (such as the stellar mass,
M$_\star$ and the star-formation rate, SFR) is included in
Sec. \ref{sec:ana_phy}, with a particular description on the
derivation of some kinematics parameters as the specific angular
momentum ($\lambda_R$) included in Sec. \ref{sec:cat_kin}; (v)
Finally, we describe how we derive the integrated, characteristics and
slope of radial gradients of different properties in
Sec. \ref{sec:ana_int}; (vi) The results of all these analyses is
included in Sec. \ref{sec:res}, comprising a description of the
adopted data format (Sec. \ref{sec:cubes}), and each of its extensions
(Sec. \ref{sec:ssp_cube} to Sec. \ref{sec:masks}); (vii)
Sec. \ref{sec:NGC2906} illustrates the content of the derived
dataproducts using NGC\,2906 as an archetype galaxy, with (vii.1) a
detailed exploration of the properties of the ionized gas
(Sec. \ref{sec:res_NGC2906}) and (vii.2) a comparison of the detectability of
\HII\ regions between the current data, the former reduced data set, and data
of a much better spatial resolution (Sec. \ref{sec:HII_NGC2906}; (viii) an exploration of the different ionizing sources across the extension of galaxies is included in Sec. \ref{sec:ion_nat}; (ix) the conclusions of the current study are included in Sec. \ref{sec:dis}.

When required, along the current study, we adopted an standard $\Lambda$ Cold 
Dark Matter cosmology with the parameters: $H_0$=71 km/s/Mpc, $\Omega_M$=0.27, $\Omega_\Lambda$=0.73.

%\citep{ARAA}

\section{Data}
\label{sec:data}

As described previously the data analyzed along this article comprises the
IFS datacubes included in the eCALIFA data release \edr. All these
data was acquired using the 3.5m telescope at the Calar Alto
observatory using the PPAK integral field unit \citep{kelz06} of the
PMAS spectrograph \citep{roth05}. All observations were performed with
the V500 grating, setup at the same goniometer angle, using a similar
exposure time, and the same dithering scheme (to cover completely the
FoV) as the one designed and adopted by CALIFA survey \citep[][for
more details]{califa,sanchez16}.  This setup allows to cover a
wavelength range between 3745-7500\AA, with a spectral resolution of
R$\sim$850 (FWHM $\sim$6.5\AA), covering an hexagonal area of
$74\arcsec \times 64\arcsec$.

The data were reduced using version 2.3 of the CALIFA pipeline,
described in detail in \edr, and previous studies \citep[][]{sanchez06a,dr2,
  sanchez16}. We present here a brief summary of the most relevant
steps: (i) pre-processing the raw data to join in a single frame the
data read by different amplifiers, removing the bias,
normalizing by the GAIN, and cleaning the cosmic-rays, (ii)
identifying and tracing the spectra corresponding to different fibers
in the CCD, obtaining the width of the light projected along the
dispersion and cross-dispersion axis, (iii) extraction of the spectra
using the trace and widths estimated during the previous setup,
removing the effects of the stray-light, (iv) performing a wavelength
calibration and resampling of the data to follow a linear wavelength
solution (2\AA/spectral pixel), (v) homogenization of the spectral
resolution along the wavelength range fixing the final resolution to
FWHM = 6.5\AA, (vi) correct for the differential fiber-to-fiber
transmission, (vii) flux-calibrate the spectra, (viii) separate the
science fibers, covering the central hexagonal region, from those
fibers that sample the sky (and those coupled to the calibration), generating and subtracting a night-sky spectrum for each dither
pointing, (ix) glue the three dithering pointings into a single
spectral frame with its corresponding position table, (x) derivation
of the astrometric solution for each dithering pointing using the data
from the PanStarrs survey as astrometry reference (PS;
\citealt{PS1,PS1_data}), correcting the nominal position table based
on this registration, (xi) correction of the broken fibers and
vignetted regions of the spectra using the information of adjacent
fibers, and finally (xii) implementing a new image reconstruction
procedure that modifies the previously adopted interpolation kernel
\citep{sanchez16}, introduces a low-order deconvolution procedure, and
a final monochromatic flat-fielding.

The major differences in this new reduction procedure with respect to
previous versions are introduced in the last three steps. As a result
it is obtained a considerable improvement in the image quality and the
final spatial resolution, that approaches to that provided by the
natural seeing (FWHM$_{psf}\sim$1.5$\arcsec$, in average). This
improved spatial resolution requires that the new datacube have a
smaller spaxel size (0.5$\arcsec$/spaxel) for a proper sampling of the
point-spread-function (PSF). This is half of the size of the CALIFA
datacubes delivered in previous DR \citep{dr1,dr2,sanchez16}. Finally,
the absolute photometry is anchored to that of the PS survey, in
contrast to previous reductions in which the Sloan Digital Sky Survey
\citep[SDSS][]{york+2000} photometry was adopted.

The final dataset, once estabilished a set of quality control
criteria, comprises 895 cubes each one corresponding to an individual
galaxy. \edr~demonstrate that this compilation, covering a wide range of
galaxy morphologies and masses ($\sim$10$^7$-10$^{12}$ M$_\odot$),
behaves as a diameter selected sample, being roughly representative of
the population of galaxies in the nearby universe ($z\sim 0.015$).
For more details on the data reduction, its quality and the properties of the
sample we refer the reader to \edr.

\section{Analysis}
\label{sec:ana}

As indicated before, the analysis performed on the described dataset
follows \citet{sanchez22}. We estimate the main spatial resolved,
characteristics and integrated properties of each galaxy by applying
\pyp to each datacube. Then, the dataproducts provided by this
pipeline were processed to obtain a set of physical quantities.
We integrated them galaxy wide, and exploring their radial distributions
to derive their value at the effective radius (Re), i.e., the
characteristic value, and the radial gradient. We present here a brief
summary of the analysis and the derived quantities, referring to
previous publications for more details to avoid repetition \citep[in
particular to][]{sanchez16,sanchez21,sanchez22}.

\subsection{{\tt pyPipe3D} analysis}
\label{sec:pyPipe3D}

We use \pyp \citep{pypipe3d}, the new version of the {\tt Pipe3D}
pipeline \citep{pipe3d_ii}. This version adapts to python the
algorithms and analysis sequence of the precedent one, improving its
performance thanks to the computational capabilities of this coding
language, and correcting some bugs when necessary. {\tt Pipe3D} is a
well proved pipeline, that has been used to study IFS data from
different datasets, including the massive exploration of CALIFA
\citep[e.g.][]{mariana16,espi20}, MaNGA
\citep[e.g.][]{ibarra16,jkbb18,sanchez18b,bluck19,laura19}, SAMI
\citep[e.g.][]{sanchez19}, and AMUSING++ datasets
\citep{laura18,carlos20}. Besides that, it has been tested using
mock datasets and ad-hoc simulations based on
hydrodynamical simulations \citep{guidi18,ibarra19,sarm23}. We present
here just a brief description of the code to avoid unnecessary repetitions.

\subsubsection{Spatial binning/tessellation}
\label{sec:sp_bin}

\pyp pipeline uses the algorithms included in the \pyf package to
analyze automatically each data cube. However, to provide with a
reliable result in the analysis of the stellar population (that we will describe
below) it is required to perform a spatial binning to increase the
siganl-to-noise (S/N) above a minimum value \citep[see ][for
details]{pipe3d,pypipe3d}. The binning procedure adopted by \pyp
 preserves the shape of the original light
distribution to a larger extend than other methods frequently
implemented in the literature \citep[see][ for detailed discussion on
the topic]{pipe3d}. Then the fitting procedure is applied to each
binned spectrum, recovering the parameters of the stellar population
and emission lines in each tessella/spatial-bin.

{ We need to recall that the new reduced data \edr, have a
  spaxel scale 4-times smaller than the original one. Beside that,
  the whole image reconstruction algorithm has change considerable. This means
  that the covariance pattern between adjacent spaxels has changed too.
Thus, a new covariance correction has to be introduced during the binning/tessallation processed, as discussed in Sec. 5.4 of \edr (shown in  Fig. 26). The code has been modified to take into account this modification. }

\subsubsection{Stellar population analysis}
\label{sec:st_fit}

As already mentioned previously each binned spectrum is analyzed using \pyf. This
code separates the stellar and ionized-gas emission in each spectrum
by creating a model of the former one using a linear combination of a
set of pre-defined single stellar populations (SSP) spectra
($S_{ssp}$). Following \citet{sanchez22} and \edr, we use the {\sc
  MaStar\_sLOG} SSP library for the current analysis. This library
comprises 273 SSP templates with 39 ages, $\mathcal{A}_\star$/Gyr~=~(0.001, 0.0023, 0.0038, 0.0057, 0.008, 0.0115, 0.015, 0.02, 0.026, 0.033, 0.0425, 0.0535, 0.07, 0.09, 0.11, 0.14, 0.18, 0.225, 0.275, 0.35, 0.45, 0.55, 0.65, 0.85, 1.1, 1.3, 1.6, 2, 2.5, 3, 3.75, 4.5, 5.25, 6.25, 7.5, 8.5, 10.25, 12, 13.5), and 7 metallicities, $Z_\star$~=~(0.0001, 0.0005, 0.002, 0.008, 0.017, 0.03, 0.04) \citep[Details on how the library and its ingredients are 
included in Appendix A of][]{sanchez22}. Before doing this modelling,
the SSP templates are shifted and broadened to take into account the
systemic velocity ($v_\star$) and the velocity dispersion
($\sigma_\star$), assuming that the line-of-sight velocity
distribution follows a Gaussian distribution
$G(v_\star,\sigma_\star)$. In addition, the dust attenuation
affecting the stellar populations (A$_{V,\star}$) is modelled using
the \citet{cardelli89} extinction law ($E(\lambda)$). The derivation
of these three parameters is performed by selecting a limited sub-set
of the SSP spectra, that limits the range of ages and metallicities
avoiding possible degenerancies.

Then, the shifted, broadened and dust attenuated SSP templates are fitted to
the data by an iterative set of linear decompositions that includes or
removes templates into the library forcing the procedure to provide
with only positive coefficients ($w_{ssp,\star,L}$) in the linear
decomposition as explained in \citet{pipe3d} and \citet{pypipe3d}.
Finally,  the observed spectrum ($S_{obs}$) is fitted by a stellar
model ($S_{mod}$) that follows the
expression \begin{equation}\label{eq:dec}
  \begin{aligned}
S_{obs}(\lambda) \approx\ & S_{mod}(\lambda) = \\ &\left[ \sum_{ssp} w_{ssp,\star,L} S_{ssp}(\lambda) \right]
10^{-0.4\ A_{\rm V,\star}\ E(\lambda)} \ast G(v_\star,\sigma_\star),
  \end{aligned}
\end{equation}
\noindent In the particular implementation for the current analysis
the light-fractions or weights of the stellar population decomposition(i.e., $w_{\star,L}$),
are normalized to the 5450-5550\AA\ spectral range (approximately the $V-$band central wavelength).

%As discussed in detail in previous studies
%\citep[e.g.][]{sanchez21,sanchez22},
%The coefficients of the decomposition are used to

\pyf derives the luminosity- and mass-weighted parameters of the
stellar population ($P_{LW}$ and $P_{MW}$) using the coefficients of the stellar decomposition
described before, using the equation:
\begin{equation}
 \begin{array}{l}
   {\rm log} P_{LW}  = \sum_{ssp} w_{ssp,\star,L} {\rm log} P_{ssp}\\
   \\
{\rm log} P_{MW}  = \frac{\sum_{ssp} w_{ssp,\star,L} \Upsilon_{ssp,\star} {\rm log}P_{ssp}}{\sum_{ssp} w_{ssp,\star,L} \Upsilon_{ssp,\star}}\\
\end{array}
\label{eq:par}    
\end{equation}
\noindent where $P_{ssp}$ is the value of a particular parameter for
each SSP (for instance, age, \age, or metallicity, \met), and  $\Upsilon_{ssp,\star}$ is the stellar mass-to-light ratio. It also estimates stellar mass across the considered aperture, using the equation:
\begin{equation}
\begin{array}{l}
  M_\star= L_V \sum_{ssp} w_{ssp,\star,L} \Upsilon_{ssp,\star}\\
\end{array}
  \label{eq:M}  
\end{equation}
where $L_V$ is the dust-corrected $V$-band luminosity.

%, dust corrected
%($L_V = 4\pi DL(z)^2 f_{V} 10^{0.4 A_{\rm V,*}}$), (ii) 
%$f_{V}$ is the observed flux intensity in the $V$-band, $A_{\rm V,*}$ is the dust attenuation affecting the stellar populations (derived by the fitting procedure as described before) and $DL(z)$ is the luminosity distance already described in the previous section; and (iii) to estimate mass- and metallicity-assembly histories of galaxies or regions within galaxies, by applying Equations \ref{eq:M} and \ref{eq:par} to a restricted age range, that corresponds to a particular look-back time \citep[for instance as described in][]{sanchez21,camps22}. From them it is possible to derive the
%star-formation and chemical enrichment history, as the derivatives of the previous distributions \citep[e.g.][]{ibarra19}.

\subsubsection{Emission line analysis}
\label{sec:el_fit}

To derive the properties of the ionized gas emission the code
subtracts the best stellar population model from the observed
one. Then, it fits each emission line ($el$) extracted from a
pre-defined list \footnote{in this analysis, \oii $\lambda$3727, \Hd, \Hg, \Hb, \oiii
$\lambda$$\lambda$4959,5007, \Ha, \nii $\lambda$$\lambda$6548,84 and \sii $\lambda$$\lambda$6717,31} with a Gaussian
function ($G(vel_{el},\sigma_{el}$), where $v_{el}$ and $\sigma_{el}$
are the $el$ velocity and velocity dispersion, respectively, scaled by
the integrated flux intensity ($F_{el}$). The bulk procedure (analysis
of stellar population and emission lines) is repeated iteratively,
using the best emission line model to decontaminate the original
spectrum than then is modelled by the combination of SSPs. In each
iteration the $\chi^2$ considering both the stellar population and
emission line model is adopted as the figure of merit of the quality
of the fitting.

The described procedure provides with the properties of the emission
lines for each tessella/spatial-bin. In order to recover the
properties of the emission lines spaxel-wise, an additional procedure
is implemented. First, the best fitted stellar population model
derived for each tessella is scaled to each spaxel within the bin
using the so-called deszonification parameter
\citep[DZ,][]{cid-fernandes13}.  Then this spaxel-wise stellar
population model is subtracted to the original datacube to create the
so-called pure-GAS cube, which comprises the emission from the ionized
gas, plus noise and residuals of the stellar population
fitting. Finally, two different procedures are implemented to extract
the parameters of the ionized gas emission lines from this cube: (i) a
Gaussian model similar to the one described before is applied to the
strongest emission lines in the wavelength range (listed before); and
(ii) a weighted-moment analysis is performed for a large set of
emission lines, including both the strong and weak emission
lines. Both methods provides with the same three parameters ($F_{el}$,
$v_{el}$ and $\sigma_{el}$). However, the second one determines also
the equivalent width of each analyzed emission line ($EW_{el}$). The
complete list of emission lines analyzed using this latter method will
be described later on (Sec.~\ref{sec:flux_elines_cube}).

%
% indices
%

\subsubsection{Stellar Indices}
\label{sec:st_ind}

In addition, \pyp estimates a set of the stellar indices for each
voxel/tessella.  For doing so, it is first generated a stellar
spectrum decontaminated by the ionized gas contribution by removing
the best emission line model derived along the fitting procedure
described before. Afterwards, the equivalent width is derived 
for each stellar index defining three wavelength ranges: (i) a
central one at which the stellar index is defined, and (ii) two adjacent ones
at which it is estimated the density flux of the continuum.
The particular list of stellar indices included in the current analysis is
described in Sec. \ref{sec:ind_cube}.

\subsubsection{Error estimation}
\label{sec:ana_error}

The errors of all the derived parameters are estimated by performing
a Monte Carlo (MC) procedure. Each original spectrum (spaxel-wise or
binned) is perturbed by co-adding a random value using the errors
provided by the data reduction. Then, the full procedure is repeated
for each perturbed spectrum, defining the errors as the standard
deviation of the individual values derived for each set of modified
spectra.  We should note that when an ionized-gas or stellar
population model is subtracted to the original spectrum the errors are
updated, considering the uncertainties of the model. For more details
on the estimation of the errors we refer the reader to
\citet{pypipe3d}.

{ Once more we remind the reader that the new data reduction
  introduce a new image reconstruction scheme that improves the
  spatial resolution, and therefore the datacubes have a spaxel size
  two times smaller than the one adopted in previous reductions:
  0.5$\arcsec$/spaxel \edr, rather than 1.0$\arcsec$/spaxel
  \citep{sanchez16}. Thus, the flux per spaxel at any wavelength is
  now four times smaller than in the previous versions of the data
  reduction. On the other hand, the S/N per arcsec$^2$ is roughly the
  same in the two versions of the data reduction, as seeing when
  comparing Fig. 5 of \edr\ with Fig. 14 of
  \citet{sanchez16}. Therefore, the S/N per spaxel is $\sim$4 times
  lower at any wavelength in the new dataset, what affects equally the
  continuum and emission lines.  This has to be taken into account in
  any spatial resolved analysis performed with the new dataproducts.}

%We
%adopt the definition of the different stellar indices and the
%procedures described in \citet{Cardiel:2003p3435}. We note that we
%depart from the classical definition of D4000, derived using the flux
%intensity in units of frequency \citep[i.e., $F_\nu$][]{bruzual83},
%and adopt the more convenient functional form proposed by
%\citet{gorgas99} (their Eq. 2).
%Like in the previous cases, the
%pipeline provides with maps of the spectral index and its estimated
%error.
%Finally, the pipeline obtains the spatial distribution (map) of a set
%of stellar spectral indices (see 
%). To do so, we
%subtract the best model for the emission lines from the spectrum in
%each tessella, in order to generate a stellar spectrum without the
%contamination by the ionized gas contribution.
%\subsubsection{Masking}
%\label{sec:ana_error}

\subsection{Data masks}
\label{sec:ana_mask}

The errors in each parameter, and in particular the error in the flux
intensity, are used to define different masks. In the case of the
continuum we select those regions above a signal-to-noise of one prior
to the spatial binning described in Sec. \ref{sec:sp_bin}.  This
defines a selection mask that is applied prior to any analysis of the
stellar population properties. In addition to this mask, regions strongly
affected by foreground field stars should be masked too. As we
describe in \edr~we follow the procedure included in  \citet{sanchez22},
searching for  possible foreground
field stars in the FoV of the analyzed datacubes, using the Gaia DR3
catalog\footnote{https://www.cosmos.esa.int/web/gaia/dr3}
\citep{gaia1,gaia3}. This catalog comprises the most accurate and complete
list of stars with good quality astrometry covering the full sky. Only those sources
with an accurate parallax (five times larger than the uncertainty) are selected, generating
a circular mask of 2.5$\arcsec$ around each of them. This mask is then used ignoring
the values reported by the analysis in those regions and replacing them by an interpolated
version of the values in the adjacent spaxels.

\subsection{Physical quantities}
\label{sec:ana_phy}

The observational properties estimated by \pyp for each spaxel or
tessella within the datacubes for both the stellar populations an
ionized gas emission lines are used to estimate further physical
parameters. Details of the derivation of all them have been
extensively discussed in previous articles, in particular in
\citet{sanchez21} and \citet{sanchez22}. We provide here with a brief
summary to avoid repetition:

{ Stellar Masses:} The stellar mass is derived directly by \pyf as one of the
quantities directly estimated from the stellar decomposition analysis, as indicated
in Sec. \ref{sec:st_fit}. From this estimation it is derived the stellar mass density
($\Sigma_\star$), and the corresponding values in different apertures. Based on this
value it is possible to derive the integrated $\Upsilon_\star$ (i.e., M$_\star$/L$_\star$ for
each spaxel), that it is different than the LW or MW average value that would be provided
by equation \ref{eq:par}.

{ Star-formation and chemical enrichment histories:} Equations
\ref{eq:par} and \ref{eq:M} can be integrated up to a certain look-back
time, correcting for the mass loss when needed
\citep[e.g.][]{courteau96} obtaining the mass-assembly
\citep[MAH,][]{eperez13,ibarra16,ibarra19}, star-formation
{\citep[SFH,][]{panter07,rosa17,lopfer18,sanchez18b} and
  chemical-enrichment histories
  \citep[ChEH,][]{vale09,camps20,camps22}. From those distributions we
  estimate (i) the current SFR (${\rm SFR}_{\rm ssp,t}$), by
  restricting the SFH to the most recent (short) time scales (e.g.,
  ${\rm SFR}_{\rm ssp, 10Myr}$, average SFR in the last 10 Myr), (ii)
  a set of time-scales that characterize the SFHs, defined as the time
  in which a fraction of the stellar mass is formed ({\tt T\%}, e.g.,
  T80, time in which 80\% of the stellar mass was formed); and (iii)
  the metallicity at those time scales ({\tt a\_ZH\_T\%}).

{ Emission line ratios:} Line ratios are frequently used to estimate
  the nature of the ionizing source and the physical properties of the ionized
  gas itself. From the emission line fluxes derived using the moment analysis
  we estimate a set of line ratios including the most popular ones, including
  \oii/\Hb, \oiii/\Hb, \oi/\Ha, \nii/\Ha, \sii/\Ha~and \Ha/\Hb.

  { Ionized gas Dust extinction:} The \Ha/\Hb~line ratio is used to estimate
  the dust extinction  (A$_{\rm V,gas}$). For doing so we adopt a nominal ratio for
 \Ha/\Hb~of 2.86, that corresponds to case-B recombination case with  an electron density of n$_e$=100
 cm$^{-3}$ and temperature of T$_e$=10$^4$ K
 \citep{osterbrock89}. The \citet{cardelli89} dust attenuation law was applied, with a
 total-to-selective extinction value of  R$_V=$3.1.

 { Oxygen and nitrogen abundances and ionization parameter:} The
 line ratios described before are frequently used to estimate the
 oxygen and nitrogen abundances and the ionization parameter using
 strong-line calibrators. Those calibrators are only valid for those
 ionization sources for which they are derived (i.e., regions ionized
 by OB stars related with recent SF events). We classify the ionizing
 source following \citet{sanchez21}, selecting as SF regions those
 located below the demarcation line defined by \citet{kewley01} in the
 \oiii/\Hb~vs \nii/\Ha~diagnostic diagram with an
 EW(H$\alpha$)$>3$\AA. Then, for those regions we estimate the oxygen
 ($O/H$) and nitrogen ($N/H$ and/or $N/O$) abundance and the
 ionization parameter ($U$) using a set of strong-line calibrators. In
 total we use 28 calibrators for oxygen abundances, 3 calibrators for
 the nitrogen abundance and 4 for the ionization parameter. The
 complete list of calibrators, the line ratios that are used to derive
 them, and a detailed description of their nature is included in
 \citet[][in particular in Appendix D and Table 15]{sanchez22}. We
 refer the reader to that reference to avoid repetition.

 { H$\alpha$ based Star-formation rate:} The SFR is derived
 spaxel-wise using the \citet{kennicutt98} relation between this
 quantity and the dust-corrected H$\alpha$ luminosity:
 \begin{equation}\label{eq:sfr}
  {\rm SFR}\ (M_\odot\ yr^{-1})\ =\ 0.79\ 10^{-41}\ {\rm L}_{\rm H\alpha}\ (erg/s)
\end{equation}
This relation corresponds to a \citet{salpeter55} IMF. Like in the
case of other extensive quantities (e.g. M$_\star$), from the SFR
estimated at each spaxel it is possible to derive the SFR surface
density ($\Sigma_{\rm SFR}$), and the corresponding quantities at
different apertures. Following \citet{sanchez21} we derive the SFR for
all spaxels, irrespectively of its ionizing source. However, it is
possible to select only those regions compatible with being ionized
by young massive OB-stars as indicated in the previous section in the
derivation of the integrated or aperture limited SFR. A correction to
the possible contribution of other ionizing sources can be also
applied, either applying a mask that exclude the contaminating regions
or estimating its contribution.  Following \citet{sanchez22} we also
estimate the SFR decontaminated by the contribution of the diffuse
ionization due to old/evolved post-AGB stars \citep[e.g.][]{singh13}.

{ Molecular gas estimation:} We use the recent calibrators proposed by
\citet{jkbb20} and \citet{jkbb21b} to estimate the molecular gas mass density from
the dust attenuation:
\begin{equation}\label{eq:mol}
   \begin{aligned} 
    \Sigma_{\rm mol}\ (M_\odot\ pc^{-2})\ =\ 23\ A_{\rm V,gas}\ (mag)\\
     \Sigma_{\rm mol}\ (M_\odot\ pc^{-2})\ =\ 1.06\ A_{\rm V,gas}^{2.58}\ (mag)
    \end{aligned}
\end{equation}
Different variations of this calibrator have been applied, including possible corrections
taking into account the \EWa, the oxygen abundance, and using as tracer of the gas
the dust attenuation derived for the stellar populations (A$_{\rm V,\star}$) and the ionized gas
(A$_{\rm V,gas}$).

{ Electron density:} The electron density is derived spaxel-wise using the [SII]$\lambda$6717,31 line ratio solving the equation:
\begin{equation}\label{eq:ne}
  \frac{[{\rm SII}]\lambda 6717}{[{\rm SII}]\lambda 6731} = 1.49 \frac{1+3.77x}{1+12.8x},
\end{equation}
where $x=10^{-4}n_{e}t^{-1/2}$, with $t$ being the electron temperature in units of $10^{4}$ K \citep{McCa85}, and $n_e$ being the electron density in units of cm$^{-3}$. We solve this equation assuming  a temperature of T$_e=10^{4}$ K (i.e., $t=$1). As the dependence of the electron density with T$_e$  is rather weak this assumption does not affect significantly the derivation of this parameter.

\subsubsection{Kinematics parameters}
\label{sec:cat_kin}

\pyp provides with the spatial distribution of the stellar and gas velocity and velocity dispersion.
From this distributions we estimate (i) the velocity to velocity dispersion ratio $\left(\frac{v}{\sigma}_R\right)$ for both the stellar and ionized gas components, and (ii) the apparent stellar angular momentum parameter ($\lambda_R$) at different deprojected galactocentric distances ($R$). For $\frac{v}{\sigma}_R$ we adopt the formulae:
\begin{equation}\label{eq:v_s}
\frac{v}{\sigma} = \sqrt{\frac{\sum_{r<1Re} f_\star v_\star^2}{\sum_{r<1Re} f_\star \sigma_\star^2}},
\end{equation}
where $f_\star$ is the stellar flux intensity in the $V$-band, and $v_\star$ and $\sigma_\star$ are the the velocity and velocity dispersions. A similar for the ionized gas changing each parameter in eq. \ref{eq:v_s} by the corresponding values derived for the \Ha~emission lines. Finally, for $\lambda_R$ we followed \citet{emsellem07}, using the formula:
\begin{equation}\label{eq:lambda}
\lambda_{\rm R} = \frac{\sum_{R<1.15Re} f_\star\ r\ |v_\star|}{\sum_{r<1.15Re} f_\star\ R\ \sqrt{v_\star^2+\sigma_\star^2}},
\end{equation}
where $f_\star$, $v_\star$ and $\sigma_\star$ are the same parameters
adopted in Eq. \ref{eq:v_s}. An inclination correction was applied, following the prescriptions included in the Appendix of \citet{emsellem11}.

We should note that the spectral resolution of the current data, that corresponds to $\sigma_{inst}\sim$150 km s$^{-1}$, is not the optimal to derive these kinematics paramters, in particular in the regime of low velocity dispersion.

\subsection{Integrated, aperture limited and characteristic properties}
\label{sec:ana_int}

The analysis described so far provides with a set of parameters for
each individal spaxel, tessella, or as a function of the
galactocentric distance. Based on those distributions we estimate for
the extensive quantities (e.g., M$_\star$ or SFR) the values at
different apertures: integrated galaxy wide, the value within 1 Re, or
in the central aperture (1.5$\arcsec$/diameter). For the intensive quantities (e..g,
$\Sigma_\star$, the oxygen abundance or any of the analyzed stellar
indices), we estimate the azimuthal average radial distribution as
described in \citet{sanchez20}, \citet{sanchez21} and
\citet{jkbb23}. Essentially, we use the coordinates of the galaxy, the
position angle and ellipticity to trace a set of elliptical apertures
of 0.15 Re width, from 0 to 3.5 Re (or the edge of the FoV).  Then,
for each aperture we estimate the average value of the corresponding
parameter ($P_R$) and its standard deviation ($e\_P_R$). Then, for
each of those radial distributions we perform a linear regression to
estimate the value at the effective radius ($P_{Re}$) and the slope of
the gradient ($slope\_P$). As already discussed in previous studies, for many
parameters the value at the effective radius can be considered a good characterization
of the average value galaxy wide \citep[e.g.][and references therein]{sanchez20}.
For completeness, for some intensive parameters we derive the value in the central
aperture and the average value galaxy wide. 
%The complete list of parameters is included in Appendix \ref{ap:tab_par}.

\section{Results}
\label{sec:res}

Along this section we describe the results from the analysis outlined
in the previous section, including a description of the delivered
dataproducts, together with some examples that illustrate their
possible scientific use, highlighting the improvements introduced by
the new distributed dataset.

%%%%%%%%%%%%%%%%%%%%%%%%%%%%%%%%%%%%%%%%%%%%%%%%%%%%%%%%%%%%%%%%%%%%%%%5
\begin{figure*}
 \minipage{0.99\textwidth}
 \includegraphics[width=17.5cm]{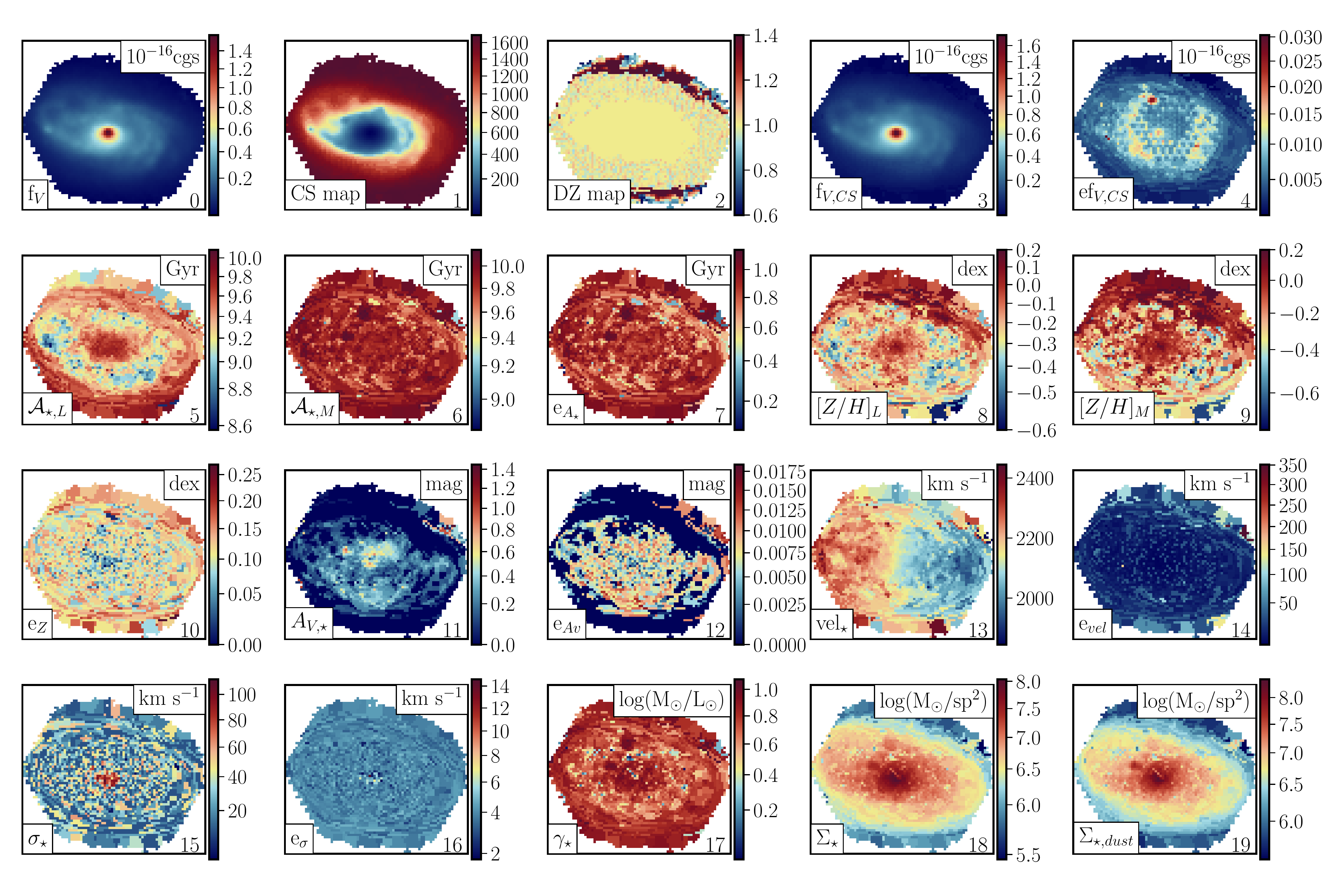}
 \endminipage
 \caption{Example of the content of the SSP extension in the Pipe3D fitsfile, corresponding to the galaxy NGC2906. Each panel shows a colour image with the property stored in the corresponding channel of the datacube, 
 as listed in Table \ref{tab:ssp}. The index, corresponding property, and units, are indicated in each panel as a label at the bottom-right, bottom-left and top-right of the figure.}
 \label{fig:SSP_map}
\end{figure*}
%%%%%%%%%%%%%%%%%%%%%%%%%%%%%%%%%%%%%%%%%%%%%%%%%%%%%%%%%%%%%%%%%%%%%%%5

\subsection{Pipe3D data model}
\label{sec:cubes}

The analysis described in Sec. \ref{sec:pyPipe3D} provides with a list of
parameters estimated at each individual spaxel or within a tessella
with their corresponding errors. Thus, for each datacube the analysis
provides with a set of maps or 2D arrays for each parameter, which
astrometry and dimensions correspond to that of the analyzed
datacube. Following \citet{pipe3d_ii,sanchez18,sanchez22} we group
those maps depending on the analysis that produce them and pack them
in a set of 3D arrays or cubes of dataproducts. This way, each channel
in the z-axis of each cube corresponds to the spatial distribution of
one particular parameter. Finally, each of those \pyp datacubes is stored as
an extension of the same FITS file for its distribution.

For the data analyzed along this study the Pipe3D files comprise nine
extensions for each derived FITS file: (i) ORG\_HDR, an initial
extension without data, containing all the metadata of the analyzed
datacube, in particular the astrometric solution (world coordinate
system); (ii) SSP, comprises the spatial distributions of the average
stellar properties derived by the stellar population fitting
(Sec. \ref{sec:st_fit}); (iii) SFH, includes the spatial distribution
of the coefficients $w_{\star,L}$ of the fitting of the stellar
population with the templates adopted SSP library (i.e.,
Eq. \ref{eq:dec}); (iv) INDICES, includes the spatial distribution of
the stellar indices (Sec. \ref{sec:st_ind}); (v) ELINES, comprises the
  properties of the strong emission lines derived by fitting them with
  a set of Gaussian functions (Sec. \ref{sec:el_fit}); (vi)
  FLUX\_ELINES and (vii) FLUX\_ELINES\_LONG comprising the parameters
  of the emission lines derived using a weighted-moment analysis
  (Sec. \ref{sec:el_fit}), for a two sub-sets of emission lines. In the first
  case we adopt the list of 54 emission lines adopted in previous analysis
  of the CALIFA data using Pipe3D \citep{pipe3d_ii}. In the second
  case we adopt a larger list of emission lines, with an updated value
  for the rest-frame wavelength, included in \citet{sanchez22}, adapted
  for the more limited wavelength range of the current data (with a total of 130 emission lines);
  (viii) GAIA\_MASK, mask of the field stars derived from the Gaia catalog described
  in Sec. \ref{sec:ana_mask}, and finally (ix) SELECT\_REG, mask of the regions with signal-to-noise
  larger than 3 in the stellar continuum ($V$-band), i.e., those regions where
  the analysis of the stellar population is reliable. A summary of the format of the FITS file
  is presented in Table \ref{tab:hdu}. For each analyzed galaxy ({\tt CUBENAME}.V500.drscube.fits.gz) we provide with one
  of those files, adopting the nomenclature {\tt CUBENAME}.Pipe3D.cube.fits.gz. The entire
  list of dataproducts files are distributed in the eCALIFA data release webpage \footnote{\url{http://ifs.astroscu.unam.mx/CALIFA/V500/v2.3/pyPipe3D/}}

  %\url{http://ifs.astroscu.unam.mx/CALIFA_WEB/public_html/} 

% --------------- Pipe3D.cube ------------------------%
\begin{table}
\begin{center}
\caption{Description of the Pipe3D file.}
\begin{tabular}{cll}\hline\hline
HDU	&  EXTENSION & Dimensions\\
\hline
  0 & ORG\_HDR            &()            \\
  1 & SSP                 & (NX, NY, 21)  \\
  2 & SFH                 & (NX, NY, 319) \\
  3 & INDICES             & (NX, NY, 70)  \\
  4 & ELINES              & (NX, NY, 11)  \\
  5 & FLUX\_ELINES        & (NX, NY, 432) \\
  6 & FLUX\_ELINES\_LONG  & (NX, NY, 1040)\\
  7 & GAIA\_MASK          & (NX, NY)      \\
  8 & SELECT\_REG         & (NX, NY)      \\
\hline
\end{tabular}\label{tab:hdu} 
\end{center}
NX and  NY may change from galaxy to galaxy, being $\sim$160 and $\sim$150 respectively.
\end{table}
% --------------- Pipe3D.cube ------------------------%

% --------------- SSP ------------------------%
\begin{table}
\begin{center}
\caption{Description of the SSP extension.}
\begin{tabular}{cll}\hline\hline
Channel	& Units	& Stellar index map\\
\hline
  0	&	10$^{-16}$ \flux	& Unbinned flux intensity \\
        & & at $\sim$5500\AA, f$_{V}$ \\
  1	&	none	                & Continuum segmentation \\
    & & index, $CS$\\
  2	&	none	                & Dezonification parameter,\\
   & & $DZ$\\
  3	&	10$^{-16}$ \flux	& Binned flux intensity at \\
   & & $\sim$5500\AA, f$_{V,CS}$\\
  4	&	10$^{-16}$ \flux	& StdDev of the flux at \\
  & & $\sim$5500\AA, ef$_{V,CS}$ \\
  5	&	{ log$_{10}$(yr)}& Lum. Weighted age, $\mathcal{A}_{\star,L}$,\\
  & &{ (log scale)}\\
%        &                                     & { in logarithmic scale}\\
6	&	{ log$_{10}$(yr)}& Mass Weighted age, $\mathcal{A}_{\star,M}$, \\
& & { (log scale)}\\
%        &                                     & { in logarithmic scale}\\
7	&	{ log$_{10}$(yr)}& Error of both $\mathcal{A}_{\star}$, \\ 
& & { (log scale)}\\
  8	&	dex	                & Lum. Weighted metallicity, \\                           
        &                       & $Z_{\star,L}$ in logarithmic scale,\\
        &                       & normalized to the solar \\
        &                       &value { ($Z_{\odot}=$0.017)}\\
9	&	dex	                & Lum. Weighted metallicity, \\
        &                       & $Z_{\star,M}$ in logarithmic scale, \\
        &                       &  normalized to the solar\\
        &                       & value{ ($Z_{\odot}=$0.017)}\\
10	&	dex	                & Error of both $Z_{\star}$\\
  11	&	mag	                & Dust extinction of the \\
         &                                    & stellar pop., A$_{V,\star}$\\
12	&	mag	                & Error of A$_{V,\star}$, e$_{{\rm A}_V}$\\
  13	&	km/s	                & Velocity of the \\
         &  & stellar pop., vel$_{\star}$\\
14	&	km/s	                & Error of the velocity, $e_{\rm vel}$ \\
  15	&	km/s	                & Velocity dispersion of the \\
        &    & stellar pop., $\sigma_{\star}$\\
16	&	km/s	                & Error of $\sigma_{\star}$, $e_\sigma$\\
  17	&	log$_{10}$(M$_\odot$/L$_\odot$)	& Stellar Mass-to-light ratio, \\
      &  & $\Upsilon_{\star}$\\
  18	&	log$_{10}$(M$_\odot$/sp$^2$)	& Stellar Mass density per \\
     &    & spaxel., $\Sigma_{\star}$\\
  19	&	log$_{10}$(M$_\odot$/sp$^2$)	& Dust corrected $\Sigma_{\star}$, \\
    & & defined as $\Sigma_{\star,dust}$\\
20	&	log$_{10}$(M$_\odot$/sp$^2$)	& error of $\Sigma_{\star}$\\
\hline
\end{tabular}\label{tab:ssp}
\end{center}
Channel indicates the Z-axis of the datacube starting from 0.			
$(1)$ measured along the entire wavelength range covered by the
spectroscopic data.
\end{table}
% --------------- SSP ------------------------%

\subsubsection{SSP extension}
\label{sec:ssp_cube}

This extension comprises the datacube in which are stored: (i) the
average properties of the stellar populations derived by \pyp based on
the decomposition of the stellar population on a set of SSP templates
(\ageLW, \ageMW, \metLW and \metMW), (ii) parameters as the stellar
mass and the $\Upsilon_\star$, (iii) the non-linear parameters
($v_\star$, $\sigma_\star$ and A$_{\rm V,\star}$), and (iv) additional
  parameters describing the light distribution, the binning pattern,
  the deszonification scaling applied during the fitting procedure.
  As the stellar population analysis performed on the spatial-binned datacube,
  the parameters stored in this extension reflect this binning. Thus, apart
  from the original light distribution and the deszonication scale all parameters
  are equal within each considered tessella.

  Table~\ref{tab:ssp} describes the parameter delivered in each
  channel (starting with 0), and its corresponding units (when
  required). This information is included in a set of header keywords
  labelled {\tt DESC\_N}, with {\tt N} corresponding to each channel
  in the z-axis. Figure \ref{fig:SSP_map} illustrates the content of
  this extension, showing for each channel the spatial distribution of
  the stored parameter corresponding to the galaxy NGC\,2906, an
  archetypal spiral galaxy included in the analyzed sample. Some known
  patterns known for this kind of galaxies, like radial gradients in
  the \ageLW and \metLW, the drop $\sigma_\star$ from the bulge
  outwards, and the clear rotational pattern seen in the $v_\star$
  map, are evident in this figure. Like in previous sections for more
  details on each parameter we refer the reader to \citet{sanchez22}.

%%%%%%%%%%%%%%%%%%%%%%%%%%%%%%%%%%%%%%%%%%%%%%%%%%%%%%%%%%%%%%%%%%%%%%%5
\begin{figure*}
 \minipage{0.99\textwidth}
 \includegraphics[width=17.5cm]{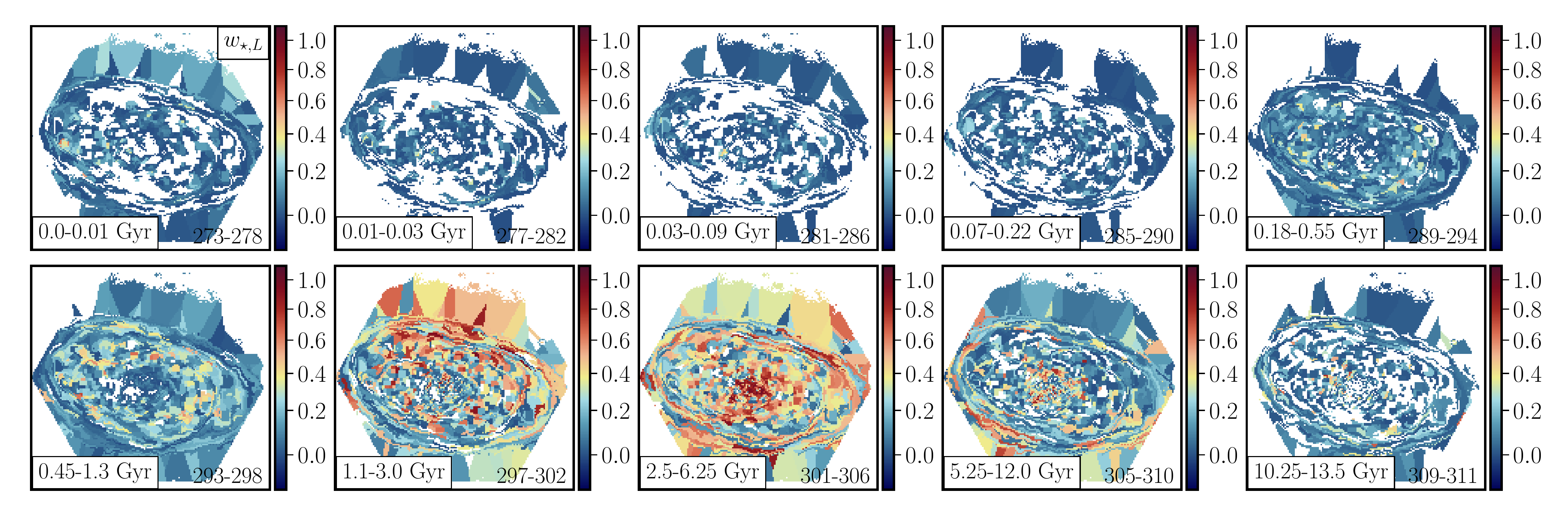}
 \endminipage
 \caption{Example of the content of the SFH extension in the Pipe3D fitsfile, corresponding to the galaxy NGC2906. Each panel shows a colour image with the fraction of light at the normalization wavelength ($w_{\star,L}$) for different ranges of age included in the SSP library. The range of ages is indicated in the bottom-left inbox, and the corresponding indices of the co-added channels in the SFH extension are shown in the bottom-right legend. { For instance, the top-left panel corresponds to the co-adding of all the maps in the SFH\_CUBE from the 273 to the 278, covering the age range between $<$0.01 Gyr.  As described in Sec. \ref{sec:sfh_cube} and shown in Table \ref{tab:sfh}, these channels run for the indices from 273 to 311, and they corresponds to light-weight maps at a fixed age (each channel) derived by co-adding all the light-weight maps for the seven different metallicities at that particular age.    }. For clarity we do not show the individual $w_{\star,L}$ maps included in the SFH extension{ , as this would require a plot with a total of 318 panels, that would not add any relevant information}.}
 \label{fig:SFH_map}
\end{figure*}
%%%%%%%%%%%%%%%%%%%%%%%%%%%%%%%%%%%%%%%%%%%%%%%%%%%%%%%%%%%%%%%%%%%%%%%5

%%%%%%%%%%%%%%%%%%%%%%%%%%%%%%%%%%%%%%%%%%%%%%%%%%%%%%%%%%%%%%%%%%%%%%%5
%\begin{figure*}
 %\minipage{0.99\textwidth}
 %\includegraphics[width=17.5cm]{figures/NGC2906.SFH_Z.pdf}
 %\endminipage
 %\caption{Similar figure as Fig. \ref{fig:SFH_map}, showing the $w_{\star,L}$ for the different metallicities included in the adopted SSP library (indicated in the bottom-left inset), corresponding to the channels of the SFH extension indicated in the bottom-right legend.}
 %\label{fig:SFH_Z_map}
%\end{figure*}
%%%%%%%%%%%%%%%%%%%%%%%%%%%%%%%%%%%%%%%%%%%%%%%%%%%%%%%%%%%%%%%%%%%%%%%5

% --------------- SFH ------------------------%
\begin{table}
\begin{center}
\caption{Description of the SFH extension.}
\begin{tabular}{cl}\hline\hline
Channel	&  Description of the map\\
\hline
0 &  $w_{\star,L}$ for ($\mathcal{A}_\star$,$Z_\star$)~=~(0.001 Gyr, 0.0001) \\
... &  ... \\
272 &  $w_{\star,L}$ for ($\mathcal{A}_\star$,$Z_\star$)~=~(13.5 Gyr, 0.04) \\
\hline  
273 &  $w_{\star,L}$ for $\mathcal{A}_\star$~=~0.001 Gyr\\
  ... &  ... \\
311 &  $w_{\star,L}$ for $\mathcal{A}_\star$~=~13.5 Gyr\\
\hline  
312 &  $w_{\star,L}$ for $Z_\star$~=~0.0001 \\
... &  ... \\
318 &  $w_{\star,L}$ for $Z_\star$~=~0.04 \\
\hline
\end{tabular}\label{tab:sfh}
\end{center}
Channel indicates the Z-axis of the datacube starting from 0, and $w_{\star,L}$ indicates the fraction (weight) of light at 5500\AA\ corresponding to: (i) an SSP of a certain age ($\mathcal{A}_\star$) and metallicity ($Z$)
(channels 0 to 272), 
(ii) a certain age, i.e., co-adding all $w_{\star,L}$ %the fractions of light 
corresponding to SSPs with the same age but different metallicity (channels 273 to 311), and 
(iii) a certain metallicity, i.e., co-adding all $w_{\star,L}$ %the fractions of light
corresponding to SSPs with the same metallicity but different age (channels 312 to 318). 
\end{table}
% --------------- SFH ------------------------%

\subsubsection{SFH extension}
\label{sec:sfh_cube}

The datacube stored in this extension includes the spatial
distribution of the coefficients of stellar decomposition for each SSP
template in the adopted library, i.e., $w_{\star,L}$(\age,\met) in
Eq.~\ref{eq:dec}. Thus, the first 273 channels covers the total range
of ages (39) and metallicities (7) included in the adopted SSP
temaplate ({\tt MaStar\_sLOG}). In addition it includes 39 additional
channels comprising the weights corresponding to each age, i.e., $w_{\star,L}$(\age), obtained by
co-adding all $w_{\star,L}$ corresponding to the same age covering the
seven different metallicites (channels from 273 to 311).  Finally, it
also includes 7 additional channels with the weights corresponding to
each metallicity, i.e., $w_{\star,L}$(\met) , obtained by co-adding all $w_{\star,L}$
corresponding to the same metallicity and covering the 39 different
ages (channels from 311 to 318). Table \ref{tab:sfh} lists the content
of each channel in this extension, information that it is included in a set headers keywords named DESC\_N, with N indicating the actual channel (running from 0 to 318)

Figure \ref{fig:SFH_map} illustrates the content of this extension,
showing the spatial distribution of the weights for each age,
$w_{\star,L}$(\age), re-binned in a set of 10 age ranges. This figure
illustrates how the age distribution of the stellar population
changes with the galactocentric distance, with a larger (smaller)
fraction of light corresponding to older (young) stellar populations,
i.e., \age$>$2.5Gyr ($<$300 Myr) is more concentrated in the inner
(outer) regions.

{ As described in \citep{pypipe3d} and summarized in Sec. \ref{sec:ana_error}, \pyp includes a procedure to estimate the errors of any derived quantity based on a MC procedure. However, for the current extension we decided not to distribute the errors as it would unnecessarily increase the size of this extension beyond the reasonable (due to the number of channels involved). Instead, we recommend the user to estimate the errors from the data themselves. The procedure to do so uses the covariance between adjacent spaxels, estimating the error as the standard deviation within in a box of the size of the PSF FWHM \citep[$\sim$1.5$\arcsec$ in average][]{edr} of the residuals once subtracted to the map corresponding to each channel an smoothed version with a Gaussian function of the size of the PSF too. This calculation relays on the fact that no parameter should change significantly within the size a resolution element (i.e., the PSF), and therefore, the observed variations spaxel by spaxel should be representative of the error. In average the typical error of $w_{\star,L}$(\age,\met) for the average S/N of our spectra is of the order of a 30-50\%, which corresponds to $\sim$20\% ($\sim$10\%) error for $w_{\age}$ ($w_{\met}$) \footnote{The procedure is included in the following notebook \url{https://github.com/sfsanchez72/califa_v2.3/blob/main/CALIFA_eDR_pyPipe3d.ipynb}, used to create figures in Sec. \ref{sec:cubes}.}.}

%Table \ref{tab:sfh} summarizes the information included in each channel of the SFH cube. This information is included in the header with a set of keywords named DESC\_N, indicating the content of channel N, with N running from 0 to 318. For completeness, the original file generated by \pyp from which the information in channel N was obtained,  has been listed in the header as FILE\_N.

%%%%%%%%%%%%%%%%%%%%%%%%%%%%%%%%%%%%%%%%%%%%%%%%%%%%%%%%%%%%%%%%%%%%%%%5
\begin{figure*}
 \minipage{0.99\textwidth}
 \includegraphics[width=17.5cm]{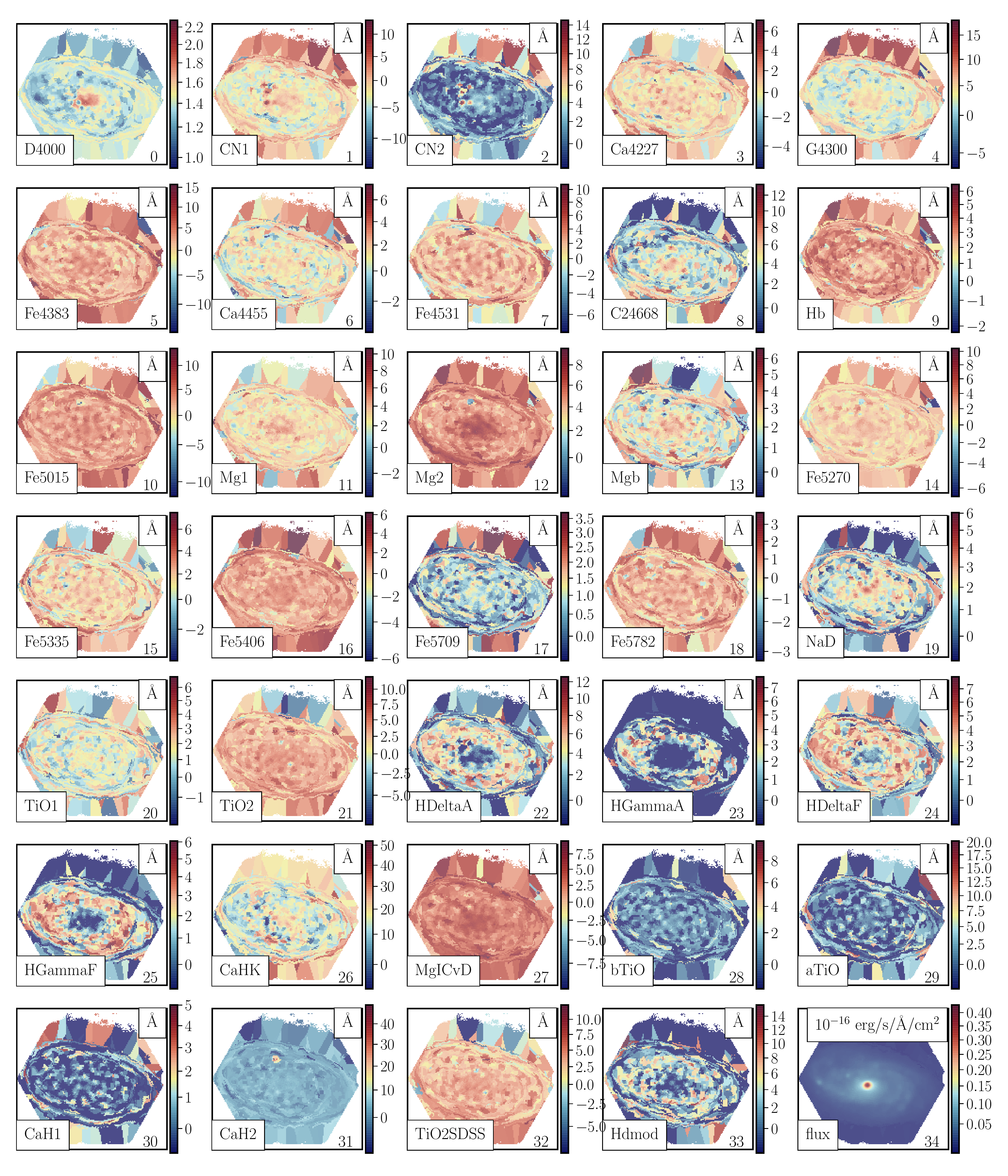}
 \endminipage
 \caption{Example of the content of the INDICES extension in the Pipe3D file, corresponding to the galaxy NGC2906. Each panel shows a color image with the content of a channel in this datacube. The actual content is indicated for each panel in the lower-left, the channel in the lower-right and the units of the represented quantity in the upper-right legend{. For the flux and flux error the units are \funitsI}.}
 \label{fig:IND_map}
\end{figure*}
%%%%%%%%%%%%%%%%%%%%%%%%%%%%%%%%%%%%%%%%%%%%%%%%%%%%%%%%%%%%%%%%%%%%%%%5

%--------------- INDICES ------------------------%
\begin{table*}
\begin{center}
\caption{Description of the INDICES extension.}
\begin{tabular}{lllccc}\hline\hline
ID & Channel	& Units	& Index $\lambda$ range (\AA) & blue $\lambda$ range (\AA) & red $\lambda$ range (\AA) \\
\hline
D4000      & 0/35 &  \AA  &   4050.000-4250.000 & 3750.000-3950.000 & \\
CN1	        & 1/36 &  \AA &   4142.125-4177.125 & 4080.125-4117.625 & 4244.125-4284.125\\
CN2	        & 2/37 &  \AA &   4142.125-4177.125 & 4083.875-4096.375 & 4244.125-4284.125\\
Ca4227     & 3/38 &  \AA   &  4222.250-4234.750 & 4211.000-4219.750 & 4241.000-4251.000\\
G4300       & 4/39 &  \AA  &   4281.375-4316.375 & 4266.375-4282.625 & 4318.875-4335.125\\
Fe4383     & 5/40 & \AA   &   4369.125-4420.375 & 4359.125-4370.375 & 4442.875-4455.375\\
Ca4455     & 6/41 & \AA &     4452.125-4474.625 & 4445.875-4454.625 &	4477.125-4492.125\\
Fe4531     & 7/42 & \AA &     4514.250-4559.250 & 4504.250-4514.250 &	4560.500-4579.250\\
C24668      & 8/43& \AA &     4634.000-4720.250 & 4611.500-4630.250 &	4742.750-4756.500\\
Hb	          & 9/44&  \AA &    4847.875-4876.625	& 4827.875-4847.875 &	4876.625-4891.625\\
Fe5015      &10/45& \AA &     4977.750-5054.000 & 4946.500-4977.750 &	5054.000-5065.250\\
Mg1	          &11/46& \AA &   5069.125-5134.125	&4895.125-4957.625 &	5301.125-5366.125\\
Mg2	          &12/47&  \AA &  5154.125-5196.625	& 4895.125-4957.625 &	5301.125-5366.125\\
Mgb	          &13/48&  \AA &    5160.125-5192.625	& 5142.625-5161.375 &	5191.375-5206.375\\
Fe5270      &14/49& \AA &     5245.650-5285.650	& 5233.150-5248.150 &	5285.650-5318.150\\
Fe5335      &15/50& \AA &     5312.125-5352.125	& 5304.625-5315.875 &	5353.375-5363.375\\
Fe5406      &16/51& \AA &     5387.500-5415.000	& 5376.250-5387.500 &	5415.000-5425.000\\
Fe5709      &17/52& \AA &     5696.625-5720.375	& 5672.875-5696.625 &	5722.875-5736.625\\
Fe5782       &18/53& \AA &    5776.625-5796.625	& 5765.375-5775.375 &	5797.875-5811.625\\
NaD	           &19/54& \AA &    5876.875-5909.375	& 5860.625-5875.625 &	5922.125-5948.125\\
TiO              &20/55& \AA &  5936.625-5994.125 & 5816.625-5849.125 &	6038.625-6103.625\\
TiO2            &21/56& \AA &  6189.625-6272.125 & 6066.625-6141.625 &	6372.625-6415.125\\
HDeltaA      &22/57& \AA &    4083.500-4122.250	& 4041.600-4079.750 &	4128.500-4161.000\\
HGammaA  &23/58& \AA &    4319.750-4363.500 & 4283.500-4319.750 & 4367.250-4419.750\\
HDeltaF	    &34/59& \AA &   4091.000-4112.250	& 4057.250-4088.500 &	4114.750-4137.250\\
HGammaF   &25/60& \AA &   4331.250-4352.250	& 4283.500-4319.750 & 4354.750-4384.750\\
CaHK	    &26/61& \AA &   3899.500-4003.500	& 3806.500-3833.800 &	4020.700-4052.400\\
MgICvD 	    &27/62& \AA &   5165.000-5220.000	& 5125.000-5165.000 &	5220.000-5260.000\\
bTiO	    &28/63& \AA & 4758.500-4800.000	& 4742.750-4756.500 &	4827.875-4847.875\\
aTiO	    &29/64& \AA & 5445.000-5600.000	& 5420.000-5442.000 &	5630.000-5655.000\\
CaH1	    &30/65& \AA & 6357.500-6401.750	& 6342.125-6356.500 &	6408.500-6429.750\\
CaH2	    &31/66& \AA & 6775.000-6900.000	& 6510.000-6539.250 &	7017.000-7064.000\\
TiO2SDSS    &32/67& \AA & 6189.625-6272.125	& 6066.625-6141.625 &	6422.000-6455.000\\
Hdmod        &33/68& \AA  &  4083.500-4122.250 & 4079.000-4083.000 & 4128.500-4161.000\\
\hline
SN               &34/69& 10$^{-16}$cgs & \multicolumn{3}{c}{Signal/Error across the entire wavelength range}\\
\hline
\end{tabular}\label{tab:index}
\end{center}
Channel indicates the Z-axis of the datacube starting from 0.			
\end{table*}

% --------------- INDICES ------------------------%

\subsubsection{INDICES extension}
\label{sec:ind_cube}

As described in Sec. \ref{sec:st_ind} \pyp estimates the spatial
distribution of a set of stellar indices, using the emission line
decontamined spectra obtained as a by product of the analysis
described in Sec. \ref{sec:st_fit}. This extension comprises the
results from this analysis, including in each channel the values
derived for each stellar index and the corresponding error.  In this
particular analysis we have updated the list of stellar indices
previously included in the analysis by Pipe3D, comprising a total of 33
indices, which actual definition and wavelength range where obtained
from the compilation incorporated in the MaNGA Data Analysis Pipeline
\citep{dap}\footnote{\url{https://www.sdss4.org/dr17/manga/manga-analysis-pipeline/}}.
In addition we include the D4000 parameter as defined by
\citet[][their Eq. 2]{gorgas99}.  We depart in this definition from the
more usual one that it is derived using the flux intensity in units of
frequency \citep[i.e., $F_\nu$][]{bruzual83}, as we find this former
definition more convenient to work with. Table \ref{tab:index}
describes the content of this extension, indicating the channel in
which each stellar index and its corresponding error are stored. In
addition, it is reproduced the wavelength range defining the stellar
index together with the blue and red wavelength ranges used to
estimate the adjacent continuum. Finally, we illustrate the
content of this extension in Figure \ref{fig:IND_map}, showing the
stellar indices derived for the galaxy NGC\, 2906. Note how some stellar indices
sensitive to the \age (\met) like D4000 (Mg2 or Mgb) present a
negative gradient from the inside-out, while other indices more
sensitive to the presence of young stellar populations (e.g., HdeltaA
o or Hb) present a clear increase near the location of the spiral arms
(i.e., where the younger stellar populations are found).

%%%%%%%%%%%%%%%%%%%%%%%%%%%%%%%%%%%%%%%%%%%%%%%%%%%%%%%%%%%%%%%%%%%%%%%5
\begin{figure}
 \minipage{0.99\textwidth}
 \includegraphics[width=8.5cm]{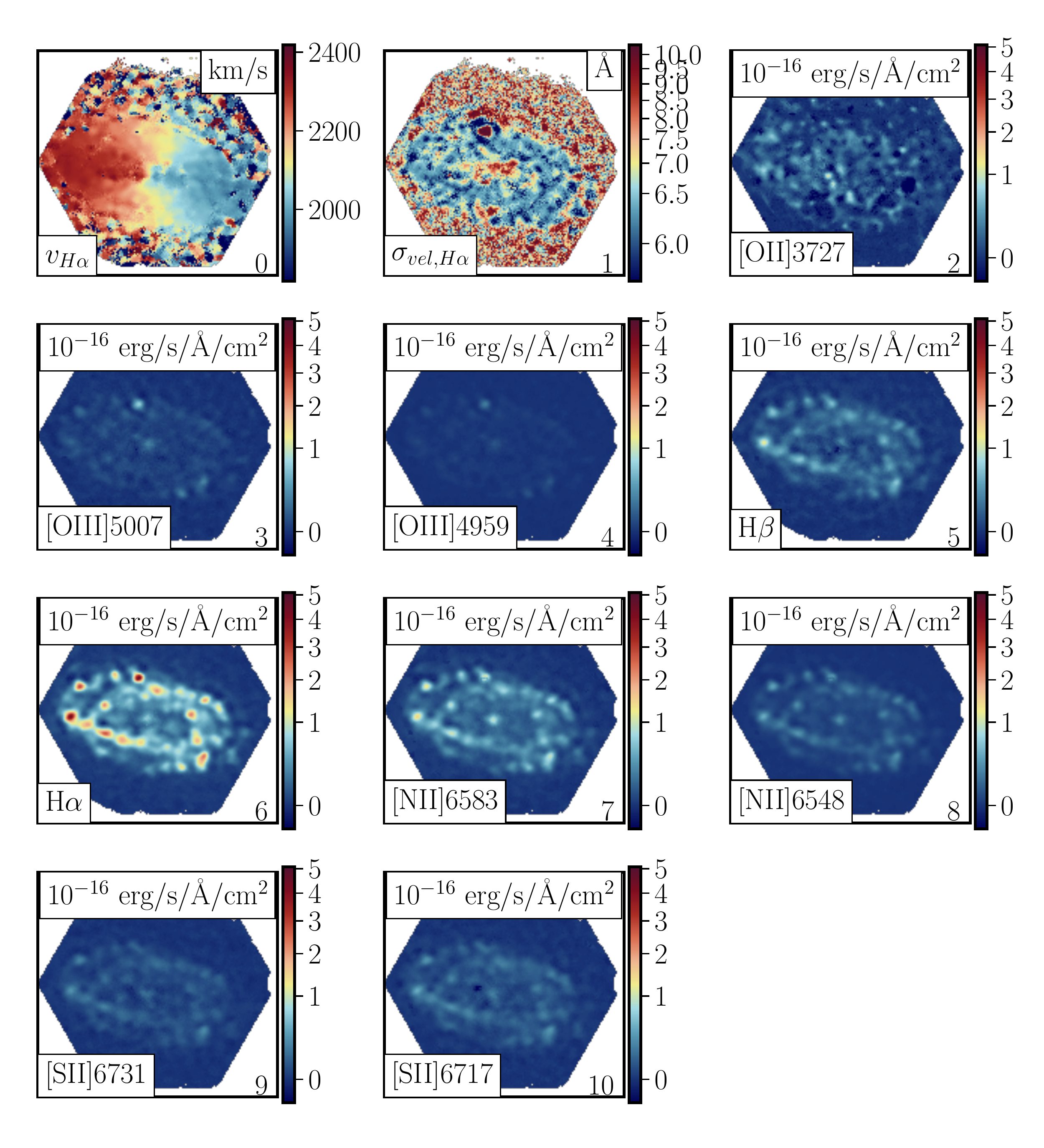}
 \endminipage
 \caption{Example of the content of the ELINES extension in the Pipe3D file, corresponding to the galaxy NGC2906. Each panel shows a color image with the content of a channel in this datacube. { The 1st panel corresponds to the velocity, in km/s, and the 2nd panel to \EWa, in \AA. The remaining panels represent the distribution of the flux intensities for the different analyzed emission lines (lower-left legend), in units of \funitsI}}
 \label{fig:ELINES_map}
\end{figure}
%%%%%%%%%%%%%%%%%%%%%%%%%%%%%%%%%%%%%%%%%%%%%%%%%%%%%%%%%%%%%%%%%%%%%%%5
%--------------- ELINES ------------------------%
\begin{table}
\begin{center}
\caption{Description of the ELINES extensions.}
\begin{tabular}{cll}\hline\hline
Channel	& Units	& Description of the map\\
\hline
0	& km/s	& H$\alpha$ velocity\\
1	& $\AA$	& H$\alpha$ velocity dispersion$^a$\\
2	& 10$^{-16}$ \flux &	[OII]3727 flux intensity \\
3	& 10$^{-16}$ \flux &	[OIII]5007 flux intensity \\
4	& 10$^{-16}$ \flux &	[OIII]4959 flux intensity \\
5	& 10$^{-16}$ \flux &	H$\beta$ flux intensity \\
6	& 10$^{-16}$ \flux &	H$\alpha$ flux intensity \\
7	& 10$^{-16}$ \flux &	[NII]6583 flux intensity \\
8	& 10$^{-16}$ \flux &	[NII]6548 flux intensity \\
9	& 10$^{-16}$ \flux &	[SII]6731 flux intensity \\
10	& 10$^{-16}$ \flux &	[SII]6717 flux intensity  \\
\hline
\end{tabular}\label{tab:e}
\end{center}
Channel indicates the Z-axis of the datacube starting from 0. 
$^a$\,FWHM, i.e., 2.354$\sigma$. The instrumental velocity dispersion
has not been removed. 

\end{table}
%--------------- ELINES ------------------------%

\subsubsection{ELINES extension}
\label{sec:elines_cube}

As described in Sec. \ref{sec:el_fit} the emission lines were analyzed
using two different methods.  In the first one a set of strong
emission lines were modelled with a Gaussian function. The spatial
results of this analysis are included in the ELINES extension, which
content is described in Table \ref{tab:e}. The information provided in this
table is stored in a set of header keywords, DESC\_{\tt N}, with {\tt N} being
the channel number. Figure \ref{fig:ELINES_map}
shows an example of the content included in this extension corresponding to the analysis
on the galaxy NGC\,2906. The 1st channel, that comprises the H$\alpha$ velocity map, shows a
typical rotational pattern for a spiral galaxy. The pattern is visible for those regions with clear
detection of the ionized gas (e.g., in H$\alpha$, 7th channel). For all emission lines the ionized gas
intensity maps present a clumpy structure tracing the location of the \HII\ regions and clusters,
which is the typical distribution for this kind of galaxies \citep[e.g., ][]{ARAA,sanchez20}.

%%%%%%%%%%%%%%%%%%%%%%%%%%%%%%%%%%%%%%%%%%%%%%%%%%%%%%%%%%%%%%%%%%%%%%%5
\begin{figure*}
 \minipage{0.99\textwidth}
 \includegraphics[width=17.5cm]{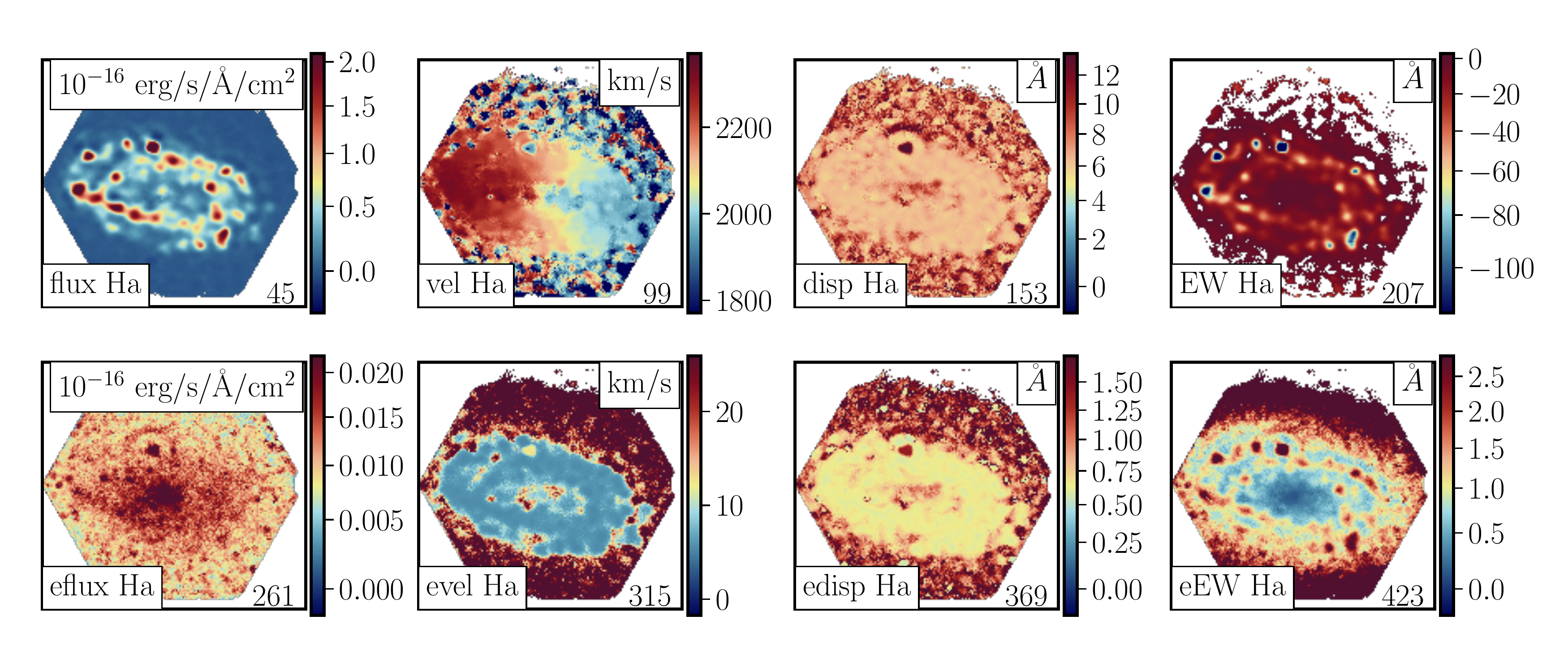}
 \endminipage
 \caption{Example of the content of the FLUX\_ELINES extension in the Pipe3D file, corresponding to the galaxy NGC2906. Each panel shows a color image with the content of a channel in this datacube. For each panel, the actual content is indicated in the lower-left, the channel number in the lower-right, and the units of the represented quantity in the upper-right legend. { For the flux and flux error the units are \funitsI}.}
 \label{fig:FELINES_map}
\end{figure*}
%%%%%%%%%%%%%%%%%%%%%%%%%%%%%%%%%%%%%%%%%%%%%%%%%%%%%%%%%%%%%%%%%%%%%%%5

%---------------------- FLUX_ELINES -------------------------- %
\begin{table}
\begin{center}
\caption{Description of the FLUX\_ELINES extensions.}
\begin{tabular}{cll}\hline\hline
Channel	& Units	& Description of the map\\
\hline
{\tt I}	    & 10$^{-16}$ \flux	& Flux intensity\\
{\tt I}+N	    &km/s	            & Velocity \\
{\tt I}+2N	&\AA	                & Velocity dispersion FWHM\\
{\tt I}+3N	&\AA	                & Equivalent width\\
{\tt I}+4N	&10$^{-16}$ \flux	& Flux error\\
{\tt I}+5N	&km/s	            & Velocity error\\
{\tt I}+6N	&\AA	            & Velocity dispersion error \\
{\tt I}+7N	&\AA	                & Equivalent width error\\
\hline
\hline
\end{tabular}\label{tab:fe}
\end{center}
{\tt I} is the index that identifies each of the analyzed emission lines, running
from 0 to {\tt N}, where {\tt N} is the total number of analyzed emission lines: (i) 53 for the
FLUX\_ELINES extension and (ii) 130 for the FLUX\_ELINES\_LONG one.
\end{table}
%---------------------- FLUX_ELINES ---------------------------- %

\subsubsection{FLUX\_ELINES and FLUX\_ELINES\_LONG extensions}
\label{sec:flux_elines_cube}

In addition to the Gaussian modelling the emission parameters are
estimated using a moment analysis, as described in
Sec. \ref{sec:el_fit}.  This second analysis is performed twice, using
two different list of emission lines, and their results are stored in
two different extensions of the \pyp data file: (i) the 53 emission
lines included in original version of the Pipe3D pipeline \citep[][Table 1, limited to $\lambda<$7400\AA]{pipe3d_ii}, included in the
FLUX\_ELINES extension, and (ii) a new set of 130 emission lines
extracted from the list published by \citet{snr_elines}, covering the
wavelength range between 3700\AA~and 7200\AA, in the rest-frame,
included in the FLUX\_ELINES\_LONG extension.  Details about this new
list of emission lines are included in Appendix \ref{ap:tab_fe}.

Table \ref{tab:fe} describes the format of both extensions. For each
analyzed emission line (defined by the running index {\tt I}), the
extension comprises four different parameters (flux intensity,
velocity, velocity dispersion and equivalent width)
and their corresponding errors. Thus, a total of eight channels, from
{\tt I} to {\tt I+7N}, correspond to the {\tt N} different parameters
derived for the same {\tt I} emission line, with 0$\le N<$8 and {\tt
  I} running from 0 to 53 (130) for the FLUX\_ELINES
(FLUX\_ELINES\_LONG) extension.

Figure \ref{fig:FELINES_map} shows an example of the content of these
extensions, corresponding to the four parameters derived for the
H$\alpha$ emission line, and their corresponding errors, stored in the
FLUX\_ELINES extension of the galaxy NGC\,2906. As expected the
distributions observed in the flux intensity, velocity and velocity
dispersion maps are very similar to those already seen in
Fig. \ref{fig:ELINES_map}, corresponding to the same emission line
derived adopting a Gaussian model. Finally, \EWa~map presents a clumpy
pattern that follows the intensity distribution, with clear negative
values at the location of the \HII\ regions. Much lower values, near
to 3\AA\ of below (in absolute values) are found at the locations
where the ionized gas presents a more smooth distribution, being
compatible with a diffuse ionized gas dominated by hot
evolved/post-AGB stars \citep[see Sec. 4 of ][for a more clear
description of this distribution]{sanchez21}.

As a final remark, we should highlight that in both the ELINES and FLUX\_ELINES extensions
the distributed velocity dispersion corresponds to the FWHM of the emission lines 
measured in units of \AA. Furthermore, the instrumental dispersion is not subtracted. As
described in \citet{sanchez22} it is needed to apply the following formula to estimate the
velocity dispersion in km s$^{-1}$:
\begin{equation}
    \sigma_{\AA}={\rm FWHM}/2.354, \ \ \ \sigma_{\rm km s^{-1}} = \frac{c}{\lambda}\sqrt{\sigma_{\AA}^2-\sigma_{\rm inst}^2},
\end{equation}
where $c$ is the speed of light in km s$^{-1}$, $\lambda$ is the wavelength of the emission line and $\sigma_{\rm inst}$ is the instrumental resolution ($\sim$2.6\AA).

\subsubsection{GAIA\_MASK extension SELECT\_REG extension}
\label{sec:masks}

These two extensions comprise the results of the two masking processes described in Sec. \ref{sec:ana_mask}: GAIA\_MASK includes the mask of the foreground stars identified using the Gaia DR3 catalog, while SELECT\_REG includes the mask generated using the signal-to-noise distribution for the continuum, by selecting a minimum value of S/N$>$1. The S/N ratio was derived by estimating the flux intensity and standard deviation for each individual spectrum within the wavelength range between 5589 and 5680 \AA. Although the main shape of the continuum is removed, the absorption features remains. Therefore, this S/N is an lower-limit to the real one.

\subsection{Catalog of individual parameters}
\label{sec:cat}

As described in Sec. \ref{sec:ana_int}, following \citet{sanchez22},
we derive a set of individual parameters for each galaxy, comprising
both integrated, aperture limited and characteristics parameters
(i.e., values at the effective radius), and, for a subset of them, the
slopes of their radial gradients. For doing so we make use of the
information included in each individual \pyp file described in
Sec. \ref{sec:cubes}, and the calculations described in
Sec. \ref{sec:ana_phy}. The final set of parameters is distributed as a
catalog that it is described in Appendix \ref{ap:tab_par}.

\subsection{NGC 2906: A show-case galaxy}
\label{sec:NGC2906}

Along the previous sections we have used the galaxy NGC\,2906 as a
show case to illustrate the content of each of the extensions in the
distributed \pyp files. NGC\,2906 is a rotational supported Sbc galaxy
with a stellar mass similar to that of the Milky-Way
($\sim$10$^{10}$M$_\odot$), being a star-forming galaxy located within
1$\sigma$ of the star-formation main sequence, and possible hosting a weak AGN \citep[e.g.][]{ang22}. Furthermore, it
presents a well defined bulge and two clear distinguishable spiral
arms. With these properties it can be considered an archetype
late-type galaxy, suitable to illustrate the content of the
distributed dataproducts. NGC\, 2906 was one of the observed by the
CALIFA survey the 17th of December 2012. Later on, it was observed
during the science verification of the MUSE instrument, being included
in the AMUSING++
compilation \citep{carlos20}. Therefore, it is also a suitable target to
explore the improvement of spatial resolution introduced by
the new reduction.

\subsubsection{NGC 2906: Resolved properties of the ionized gas}
\label{sec:res_NGC2906}

%%%%%%%%%%%%%%%%%%%%%%%%%%%%%%%%%%%%%%%%%%%%%%%%%%%%%%%%%%%%%%%%%%%%%%%5
\begin{figure}
 \minipage{0.99\textwidth}
 \includegraphics[width=0.49\textwidth]{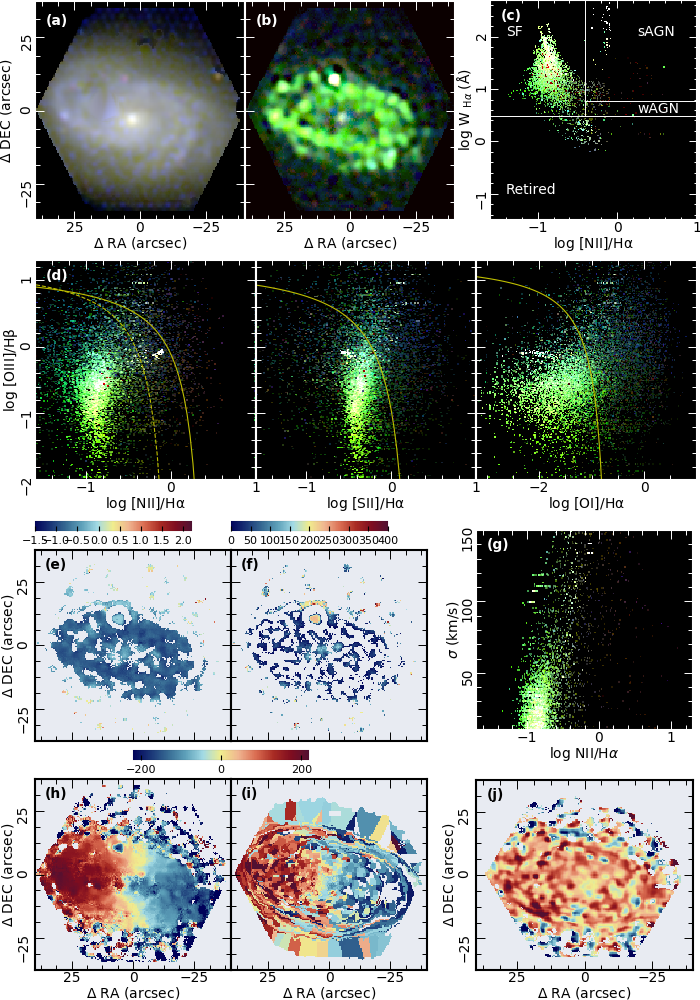}
 \endminipage
 \caption{Example of the information explored in the quality control process for the galaxy/cube NGC2906, extracted from the analysis presented by \citet{carlos20}. Each panel comprises the following information: (a) RGB three-color image created using the $u-$, $g-$ and $r$-band images synthetized from the reduced eCALIFA datacubes; (b) RGB color image created using the \oiii~(blue), \Ha~(green), and \nii~(red) intensity maps estimated by \pyp ; (c) spatially resolved WHAN diagnostic diagram, with each spaxel color-coded using the color-scheme shown in panel (b). The white solid lines indicate the boundaries proposed by \citet{cidfernandes11}, with the dominant ionizing source in each region included as labels; (d) spatially resolved diagnostic diagrams showing the distribution of [OIII]/H$\beta$ as a function of [NII]/H$\alpha$, [SII]/H$\alpha$ and [OI]/H$\alpha$, using a similar color scheme as panel $c$ \citep{baldwin81,veil01}. The yellow solid and dashed line in each panel corresponds to the demarcation line proposed to separate between SF and non-SF dominated ionization by \citet{kewley01} and \citet{kauff03}, respectively; (e) spatial distribution of the [NII]/H$\alpha$ line ratio; (f) spatial distribution of the H$\alpha$ velocity dispersion; (g) distribution of the 
 [NII]/H$\alpha$ line ratio as a function of the H$\alpha$ velocity dispersion across the extension of the galaxy; (h) H$\alpha$ velocity map (spaxel-wise); (i) stellar population velocity map (following the spatial-binning/tessallation); (j) difference between velocity maps shown in panels (h) and (i) {. Note that the scale of this later panel is not the same as those in panel (h) and (i).}}
 \label{fig:BPT_single}
\end{figure}
%%%%%%%%%%%%%%%%%%%%%%%%%%%%%%%%%%%%%%%%%%%%%%%%%%%%%%%%%%%%%%%%%%%%%%%

We already mentioned that one of the major improvements of the new analyzed
and distributed dataset is the increased spatial resolution. The
previous reduction provides with a spatial resolution that in average
is of the order of $\sim$2.5$\arcsec$/FWHM \citep[version
2.2;][]{sanchez16}. For the particular case of NGC\,2906, this FWHM was
estimated in 2.37$\arcsec$, according to the information provided by
the data reduction stored in the datacube header. The new reduction
improves the spatial resolution  with a final PSF FWHM near to the one
provided by the natural seeing \edrp. In the particular case of NGC\,2906 is
estimated in 1.1$\arcsec$/FWHM. When considering the offset between
the PSF FWHM estimated by the data-reduction itself and the real FWHM
of the natural seeing \edrp, a more realistic estimation would be
$\sim$1.5$\arcsec$. This is, in any case, a very significant improvement
in the spatial resolution with respect to the previous data reduction.

%The previous reduction provides with a spatial resolution of 2.5$\arcsec$/FWHM, 

A clear advantage of an improved spatial resolution is the ability to
distinguish between different structures within a galaxy. This is
particularly important, for instance, in the exploration of the
different ionizing sources and the understanding of the changes in the
physical conditions in the inter-stellar medium (ISM) across the
optical extension of galaxies. This has been highlighted in recent
reviews on the topic \citep{ARAA,sanchez21}, and the results by recent
projects aimed to explore the ionizing conditions using IFS with
unprecedented spatial resolutions \citep[e.g., PHANGS-MUSE;][]{emse22}.

%%%%%%%%%%%%%%%%%%%%%%%%%%%%%%%%%%%%%%%%%%%%%%%%%%%%%%%%%%%%%%%%%%%%%%%5
%\begin{figure*}
% \minipage{0.99\textwidth}
% \includegraphics[width=17.5cm]{figures/NGC2906.SFH_CM.pdf}
% \endminipage
% \caption{$\Sigma_\star$ derived from the weights included in the SFH extension using the mass-to-light ratio of each SSP and the flux intensity in each tessella (spatial bin) for different age ranges, indicated in the bottom-left inset, corresponding to the channels in the bottom-right legend. The first panel shows the actual $\Sigma_\star$, corresponding to integration along all look-back times. The subsequent panels show the cumulative $\Sigma_{\star,t}$ at the look-back time $t$ corresponding to the lower age of the considered range, without considering the redshift of the target. The age (\age) of the stellar population and the look-back time of the universe at which it was formed ($\tau_{lbt,\star}$) are related via the look-back time corresponding to the redshift at which it was observed ($\tau_{lbt,z}$): $\tau_{lbt,\star}$=$\tau_{lbt,z}$+\age. For simplicity, in this plot we consider $\tau_{lbt,z}\sim$0.}
% \label{fig:SFH_CM_map}
%\end{figure*}
%%%%%%%%%%%%%%%%%%%%%%%%%%%%%%%%%%%%%%%%%%%%%%%%%%%%%%%%%%%%%%%%%%%%%%%5

%\subsubsection{NGC2906: HII regions}
%\label{sec:HII_NGC2906}

%%%%%%%%%%%%%%%%%%%%%%%%%%%%%%%%%%%%%%%%%%%%%%%%%%%%%%%%%%%%%%%%%%%%%%%5
\begin{figure*}
 \minipage{0.99\textwidth}
 \includegraphics[width=17.5cm]{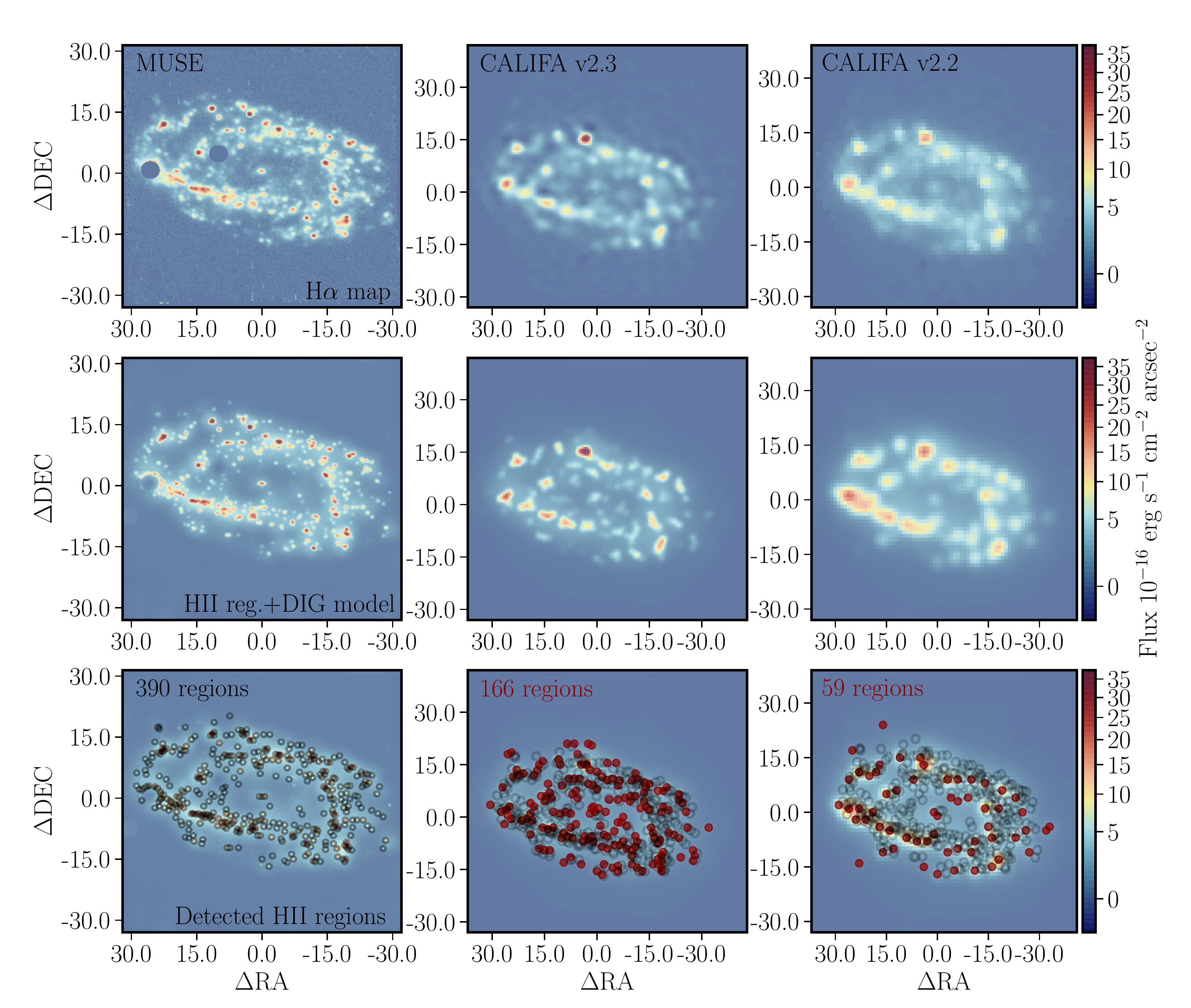}
 \endminipage
 \caption{\Ha\ intensity maps of NGC\ 2906 extracted using {\tt pyPipe3D} from (i) the MUSE datacube included in the AMUSING++ compilation \citep{carlos20} (top-left panel), (ii) the eCALIFA datacube corresponding to version 2.3 of the data reduction \edr (top-central panel), and (iii) the CALIFA datacube corresponding to version 2.2 of the data reduction (top-right panel). The best models for the \hii\ regions and diffuse ionized gas recovered by {\tt pyHIIExtractor} \citep{lugo22} for each of those maps shown in the top panels are included in the middle panels, following the same sequence from left-to-right (MUSE, eCALIFA v2.3 and v2.2). Finally, the bottom panels show each of these models together with the distribution of good-quality \hii\ regions detected using the MUSE data (black open-circles, included in all the bottom panels), and the regions detected using each of the eCALIFA data (solid red-circles). The number of  \hii\ regions detected using each dataset is included in a label in each of the bottom panels. Figure illustrate the effects of the degradation of the spatial resolution in the detectability of individual structures (e.g., \hii\ regions) in galaxies using this kind of data.}
 \label{fig:HII}
\end{figure*}
%%%%%%%%%%%%%%%%%%%%%%%%%%%%%%%%%%%%%%%%%%%%%%%%%%%%%%%%%%%%%%%%%%%%%%%5

%%%%%%%%%%%%%%%%%%%%%%%%%%%%%%%%%%%%%%%%%%%%%%%%%%%%%%%%%%%%%%%%%%%%%%%5
\begin{figure}
 \minipage{0.99\textwidth}
 \includegraphics[width=8.5cm]{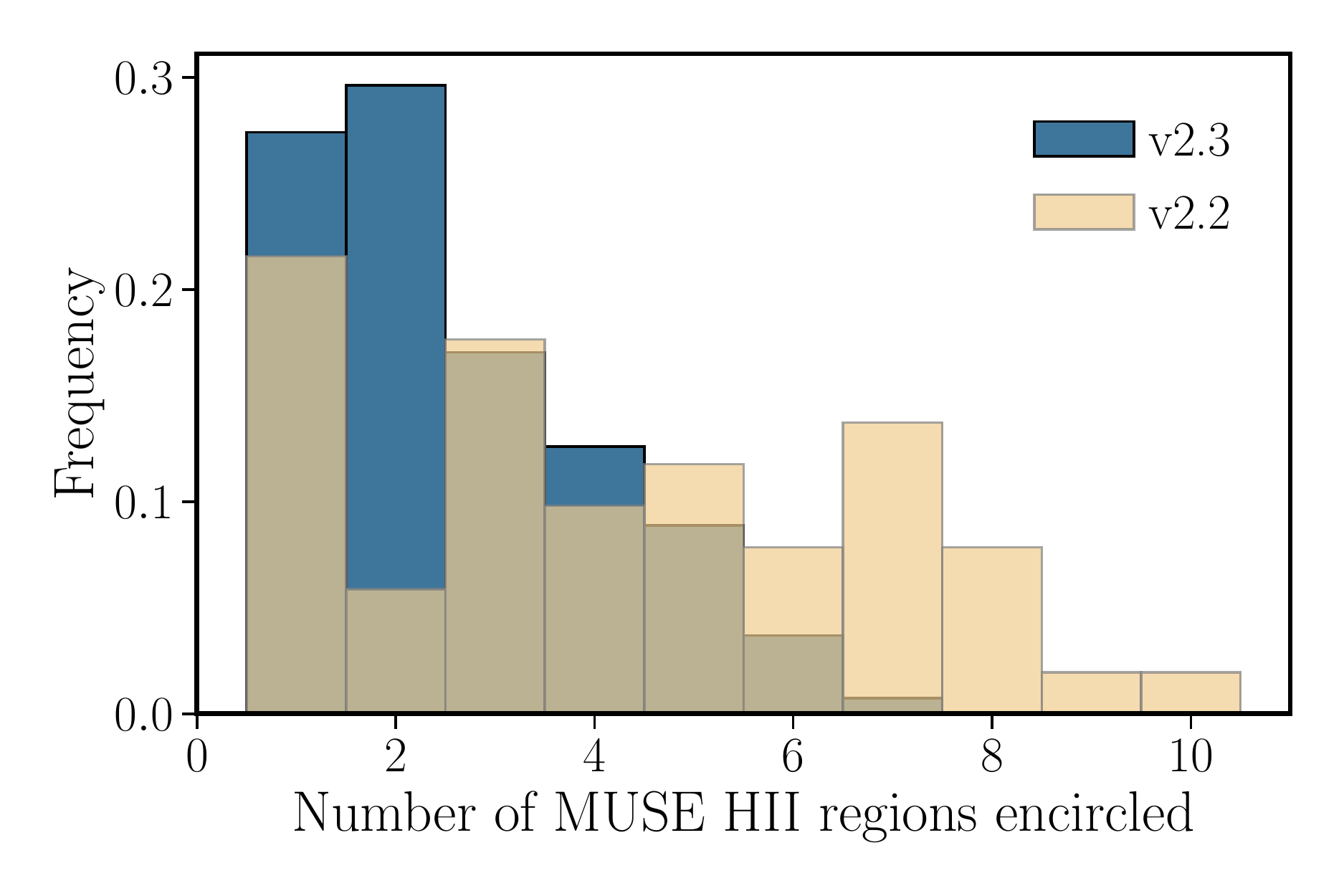}
 \endminipage
 \caption{Distribution of the frequency in which a certain number of \hii\ regions detected using the MUSE data shown in Fig. \ref{fig:HII} is encircled within the radius of an \hii\ region detected using the eCALIFA v2.3 dataset (blue histogram) or the  v2.2 one (yellow histogram). Figure illustrates the effect of the degradation of the spatial resolution in the clustering of \hii\ regions into a single blended structure.}
 \label{fig:HIIstats}
\end{figure}
%%%%%%%%%%%%%%%%%%%%%%%%%%%%%%%%%%%%%%%%%%%%%%%%%%%%%%%%%%%%%%%%%%%%%%%5
%

Figure \ref{fig:BPT_single} shows a comparison between the spatial
distribution of the continuum emission, created using the $u-$, $g-$
and $r-$band images synthetized from the reduced eCALIFA datacube,
dominated by the stellar population (panel a), and the ionized gas
emission lines, created using the \oiii~(blue), \Ha~(green) and \nii~(red) emission maps (panel b). The different morphological structures
within this galaxy are clearly appreciated in the continuum image:
(i) an evident (but small) bulge is observed in the center of the
galaxy, clearly dominated by a red/old stellar population; (ii) two
spiral arms, with some additional sub-arms, are traced by a blue/young
stellar component; in addition, (iii), a diffuse component dominated
by reddish/oldish stellar populations is appreciated beyond and in
between the spiral arms. The ionized gas map is dominated by a clumpy
structure, mostly green (i.e., with a dominant  \Ha\ flux over the other
two emission lines), that trace the location of the spiral arms seen
in the continuum image (i.e., associated with the presence of a
blue/young stellar population).

These clumps are most probably tracing \HII\ regions/clusters. This is
evident when explored their distribution in the different diagnostics
diagrams included in Fig. \ref{fig:BPT_single}, panels (c) and
(d). The vast majority of them are found in regions within the
diagnostic diagrams usually associated with classical \HII\ regions
\citep[e.g.][]{osterbrock89,kewley01,kauff03} or regions which
ionization is associated with recent star-formation activity
\citep[e.g.][]{cidfernandes11}, in agreement with the prescriptions
outlined by \citet{sanchez21}. Beside those clumps there is a
smooth/diffuse component that is observed at any location within the
galaxy, including the central regions and inter-arms, that most
probably is due to a combination of different ionizing sources such as
photons leaked from \HII\ regions and ionization due to the presence
of hot evolved/post-AGB stars
\citep[e.g.][]{binette94,rela12,singh13,weil2018,sanchez21,belfiore22}.
Finally, this object is a known host of a supernova remnant (SNR; Mart\'inez-Rodr\'iguez in prep.)
located at $\sim$10$\arcsec$ north and $\sim$2$\arcsec$ east from the
center of the galaxy. This SNR is observed as a white-roundish ionized
structure in panel (b), and as a cluster of white points in the
intermediate region\footnote{In between the \citet{kewley01} and
  \citet{kauff03} demarcation lines in the diagnostic diagram} in the
\oiii/\Hb~vs \nii/\Ha~diagram \citep[BPT diagram;][]{baldwin81},
included in panel (b).

Fig. \ref{fig:BPT_single} panel (e) shows the spatial distribution of
the N2 parameter, defined as log(\nii/\Ha). This parameter is
sensitive to oxygen abundance when the ionization is due to young
OB-stars \citep[e.g.][]{pettini04,marino13}. A mild
negative gradient is observed, with values of N2 slightly larger in the central
regions than in the outskirts, compatible with the known decline of
the oxygen abundance described for galaxies of this stellar mass and
morphology \citep[e.g., ][]{sanchez14,belf17,sanchez21,board23}.
Those regions ionized by other sources, different than young massive
stars, use to present larger values of N2. This is the reason why this
parameter is used to discriminate the ionizing source in the WHAN
diagram (panel c). In this particular galaxy we find anomalous larger
values of the N2 ratio compared to those at the same galactocentric
distances at the location of the reported SNR \citep[e.g.,][]{cid21}. We stress that this
N2-enhancement is a direct consequence of an extra ionizing source,
and should never be considered as an increase of the oxygen abundance.

If we compare the N2 distribution with the one shown in panel (f), the
H$\alpha$ velocity dispersion, we can appreciate that there is an
increase of this parameter at the location of the SNR too. As we
indicated before the spectral resolution of this data is not optimal
to measure the velocity dispersion, in particular in the disk of
late-type spirals, where this parameter is expected to be well below
the instrumental resolution.  This is the reason why in many
locations across the extension of this galaxy the velocity dispersion
is masked. For those regions for which a reliable estimation is
obtained we appreciate that a decline in the velocity dispersion from
value near $\sim$150 \kms in the central regions to values as low as
$\sim$30 \kms, in the outer disk.  However, at the location of the SNR
the velocity dispersion rises up to $\sim$200 \kms. Recent
explorations have proposed that the velocity dispersion should be
considered as an extra parameter to discriminate between different
ionizing sources \citep{dagos19,carlos20,law21}. Indeed, the N2
parameter usually follows the velocity dispersion, as appreciated in
panel (g) of Fig. \ref{fig:BPT_single}. Due to their association with
the gas disk and their relatively low internal pressure \HII/SF
regions are usually associated with regions of low velocity dispersion
\citep[e.g.][]{law21}.  On the contrary, regions ionized by shocks use
to present larger velocity dispersions
\citep[e.g.][]{dagos19,carlos20}. 

The last three panels of Fig. \ref{fig:BPT_single} show the H$\alpha$
and stellar velocity maps (panels h and i), and finally the difference
between both of them (panel j). The well defined rotational pattern
appreciated in both velocity maps, together with the low velocity
dispersion seen in panel (f), suggests that this is a rotational
supported galaxy. A possible perturbation on the rotational pattern is
appreciated in the inner regions of the H$\alpha$ velocity map,
although a more detailed analysis would be required to drawn any
conclusion in this regards, which is beyond the scope of the current analysis.  %Beside that the difference between both velocity maps do not show any clear pattern, beside a general offset.
Apart from a general offset, there is no discernible pattern in the difference between both velocity maps. { The nature of the offset is a known effect in the calibration of the zero-velocity associated with the measurements using an SSP template, a miss-match that we have already reported in previous articles \citep[e.g.][]{sanchez22}, that it is associated with the definition of the central wavelength at the edge or at centre of the spectral pixel. It produces an offset of half of the spectral pixel (in velocities), times the expansion factor introduced by the redshift, i.e., 1+z. This offset should be corrected prior to any detailed comparison between the two velocity maps (for all the delivered dataset). A more detailed exploration, matching the offset in velocity in the central regions is required to explore possible patterns associated with known effects, such as the asymmetric drift or the possible presence of non-radial motions that may affect differently the gas and the stellar populations. }
Like in the case of the N2 and velocity dispersion maps the strongest
deviation appreciated in the gas velocity maps is at the location of
the SNR, where it is observed a clear blue-shift. This kind of
asymmetries or multiple kinematic components are frequent in the
presence of shock ionization \citep[e.g.][]{veil01,carlos16}.

\subsubsection{NGC 2906: Ability to segretate \HII\ regions}
\label{sec:HII_NGC2906}

The content of Fig. \ref{fig:BPT_single}, together with the
distributions shown in Figs. \ref{fig:SSP_map} to \ref{fig:FELINES_map}
clearly illustrates the kind of information extracted by \pyp from the
new reduced datacubes. It also shows in a qualitative way how the new
spatial resolution allows to distinguish between different structures
within the galaxy. In this section we present a more quantitative
statement of this improvement, by comparing the ability to detect and
segregate \HII\ regions (and associations) using three IFS datasets on
the galaxy NGC\,2906: (i) the MUSE data analyzed by Pipe3D as part of
the AMUSING++ compilation \citep{carlos20}; (ii) the eCALIFA v2.3 data
analyzed by \pyp discussed in this article and (iii) the CALIFA v2.2
data analyzed by \pyp distributed as part of the DR3
\citep{sanchez16}.

Figure \ref{fig:HII}, top panels, show the H$\alpha$ intensity
map extracted from the three datasets. As expected the MUSE data
present the richest content of sub-structures/clumpy regions (i.e.,
\HII\ regions), due to its better spatial resolution
(FWHM$\sim$0.6$\arcsec$). In comparison, the eCALIFA v2.3 data
(FWHM$\sim$1.1-1.5$\arcsec$) still traces the brightest clumpy
structures seen in the MUSE data. However, it presents a clear
degradation of the image quality, bluring and/or grouping into a
single structure several clumps clearly visible and/or segregated in
the MUSE data. Finally, the lowest number of substructures is
appreciated in the CALIFA v2.2 data (FWHM$\sim$2.37$\arcsec$), with
the largest degree of clumpy regions lost or grouped into a single
unresolved or barely resolved structure.

To quantify these differences we apply the recently developed {\tt
  pyHIIExtractor} code \citep{lugo22} to the three datasets. This code
comprises an algorithm that detects clumpy regions in emission line
maps, deriving both their flux intensities and sizes. It also
segragates the detected regions from the diffuse ionized gas (DIG),
generating a model for both components. Being developed to explore the
\HII\ regions of the galaxies within the AMUSING++ compilation (Lugo
Aranda et al., in prep.), its capabilities has been tested with a wide
range of simulations. Fig. \ref{fig:HII}, middle-column panels, show
the combined models (\HII\ regions+DIG) generated by this code when
applied to the H$\alpha$ maps shown in the upper panels, using the
same input parameters: (i) a minimum size of 30 pc
($\sim$0.6$\arcsec$ at the redshift of the object), (ii) a maximum
size of 300 pc, (iii) a overall detection limit of 3$\sigma$, and
(iv) a minimum contrast of a 50\% with respect to the DIG in the
detection of a clumpy structure. A direct comparison between the upper
and middle-column panels illustrates how well the code recovers the
original light distribution. It also highlights its limits: it is unable to recover some of the faintest/smaller clumpy regions in the MUSE data. We should note that the most limiting factor is the contrast with respect to the level of the adjacent DIG, followed by the imposed minimum size.

The three bottom panels in Fig \ref{fig:HII} show again the same H$\alpha$
models, together with the distribution of clumpy structures detected
by {\tt pyHIIExtractor} using the MUSE data, that is repeated in the
three panels for comparison purposes. In addition, it is shown the
distribution of regions detected by this code using the eCALIFA v2.3
(central panel), and CALIFA v.2.2 (right panel), respectively.  In the
case of the MUSE data it is detected a total of 390 clumpy regions, a
number that decreases to 166 for the eCALIFA v2.3 data and to just 59
for the CALIFA v2.2 data, respectively. Thus, the resolution clearly affects the ability to detect individual structures such as \HII\ regions. Furthermore the new reduced and analyzed dataset presents a clear improvement with respect to the previous dataset. 

As discussed previously, \HII\ regions are lost due to the bluring effect
introduced by the degradation of the resolution (i.e., they are not
distinguished from the background), but also due to an aggregation
with nearby clumps. To quantify both effects we determine how many
regions of the ones detected in the MUSE data are not included within
the radius of the clumpy structures detected in the other two
datasets.  This will tell us the number of regions that are totally
lost. Finally, we determine the number of regions detected in the MUSE
encircled within the radius of the regions detected using the other
two datasets. This will quantify the frequency in which an original
region is aggregated to form a new structure in the lower resolution
datasets. For doing so we make use of the {\tt cKDthree} algorithm
included in the {\tt scipy.spatial} python package to look for the
nearest neighbours of each of the regions detected in the MUSE data in
the catalog of regions derived using the other two datasets.  Then, we
determine whether a region is included within any nearest neighbour by
comparing their radii with the distance.

Of the 390 regions detected in the MUSE data, 351 are included within
one of the 166 regions detected in the eCALIFA data. Thus, most of the
regions are not really lost ($\sim$90\%), but they are rather
aggregated into larger clumps, that in average comprises $\sim$2 of
the former regions. On the contrary, in the case of the CALIFA v2.2
data, only 218 of the original 390 regions are included within one of
the detected 59 regions. Therefore, the ``recovery'' rate is much
lower ($\sim$55\%). Furthermore, the new clumps comprise a larger
number of the original regions ($\sim$3-4). Therefore, the improved
resolution achieved by the new reduction increases the recovery rate
of \HII\ regions and decreases the number of agregated/grouped regions
by a factor two. A more detailed exploration on how the original \HII\
regions are aggregated into larger structures is presented in Figure
\ref{fig:HIIstats}, where it is shown the frequency at which a certain
number of the former regions is aggregated to a region detected in the
other datasets. In the case of the eCALIFA v2.3 dataset most of the
detected regions comprises one or two MUSE-detected regions
($\sim$60\%), with less than $\sim$20\% encircling more than four of
those regions. On the contrary, in the case of the CALIFA v2.2 data,
only a $\sim$25\% of the regions comprises one or two MUSE-detected
regions, while more than $\sim$50\% aggregates more than four of those
regions.

In summary, this exploration clearly indicates that the new reduced and
analyzed data have significantly improved out the ability to detect and
study sub-structures in the observed galaxies.

%improved spatial resolution
%provides with a much 
% For the discussion!
%
%\citet{mast14} already demonstrated that the degradation of the
%resolution and the aggregation of adjacent \HII\ regions leads to a
%significant modification of the derived physical parameters and their
%radial gradients, in particular the line ratios and the estimated
%oxygen abundances.

%u

%390 390 351 166 136
%[1 2 3 4 5 6 7] [37 40 23 17 12  5  1] 135
%390 390 218 59 54

%390 166 144
%390 59 56

\subsection{Ionization conditions across the optical extension of galaxies}
\label{sec:ion_nat}

%%%%%%%%%%%%%%%%%%%%%%%%%%%%%%%%%%%%%%%%%%%%%%%%%%%%%%%%%%%%%%%%%%%%%%%5
\begin{figure*}
 \minipage{0.99\textwidth}
 \includegraphics[width=17.5cm]{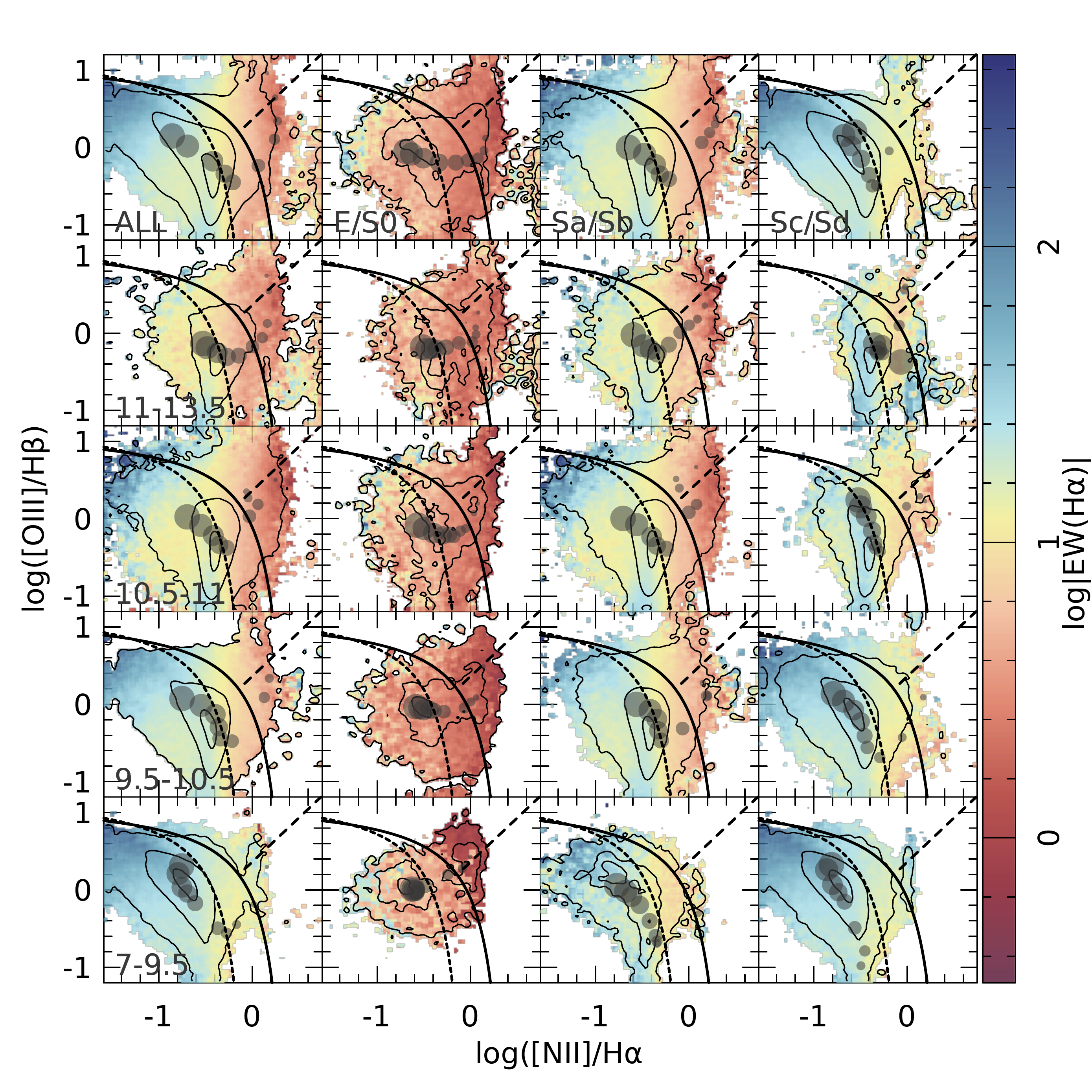}
 \endminipage
 \caption{\EWa\ across the classical [OIII/H$\beta$]~vs.~[NII]/H$\alpha$  BPT diagram \citep{baldwin81} for $\sim$9 million individual spaxels extracted from a sub-sample of 660 face-on ($a/b<$0.85), well resolved (Re$>$2$\arcsec$) and redshift limited (0.005$<z<$0.05) galaxies. Each panel shows the distribution for a sub-set of galaxies with different morphologies (columns) and stellar masses (rows), and the top-left panel includes the complete sub-sample, using the same symbols and color schemes: (i) the density of spaxels is traced by the successive contours encircling 90, 50 and 10\% of the individual points; (ii) the average values at different galactocentric distances, ranging from 0.1 to 2.1 Re ($\Delta$Re=0.2), are represented by the grey circles, which size increases with the distance; (iii) the classical demarcation lines defined by \citet{kewley01} and \citet{kauff03} are represented by a solid and short-dashed lines, respectively; finally (iv) the demarcation line segregating between Seyferts and LINES defined by \citet{kewley01} is shown as a long-dashed line.}
 \label{fig:ex_BPT}
\end{figure*}
%%%%%%%%%%%%%%%%%%%%%%%%%%%%%%%%%%%%%%%%%%%%%%%%%%%%%%%%%%%%%%%%%%%%%%%5

In previous sections we have illutrated the impact of the intrinsic
properties of the data (i.e., spatial coverage and sampling) and the
improved spatial resolution of the new reduced dataset, in the
exploration of the ionizing conditions and ISM properties on one
particular galaxy NGC\,2906. In this section we illustrate their
impact in the exploration of the general patterns of those properties
described for the full population of galaxies. For doing so we
replicate the explorations presented by \citet{ARAA} and
\citet{sanchez22}, shown in their Fig. 5 an Fig. 10, respectively.
There it was presented the spaxel-wise spatial resolved distribution
of values across the classical BPT diagnostic diagram, [OIII]/H$\beta$ vs. [NII]/H$\alpha$ \citep[BPT;][]{baldwin81} for two
different samples of galaxies, segregated by stellar mass and
morphology.  In both cases it was claimed that the selected samples
are representative of the galaxy population in the nearby Universe:
(i) \citet{ARAA} explored an heterogeneous compilation of galaxies
observed using different IFS instruments by different surveys, and
(ii) \citet{sanchez22} used the final DR of the MaNGA IFS galaxy
survey.

Following both studies we select a refined sub-sample of 660 galaxies
from the original dataset, excluding edge-on galaxies ($a/b>$0.85),
galaxies poorly resolved (Re$<$2$\arcsec$), and limiting the redshift
range (0.005$<z<$0.05). This way we  (i) avoid projection issues,
providing with a cleaner interpretation of the galactocentric
distances, (ii) exclude galaxies that are not resolved by the data,
and (iii) limit the effects of the wide range of physical resolutions
($\sim$500 pc, in average, ranging from 150 pc to 1 kpc) in the
interpretation of the results. We should highlight that this selection
biased the sample excluding (i) the most prominent/evident galactic
outflows, shock ionized and frequently observed as biconical,
cononical or filamentary structures of gas extending along the
vertical axis in edge-on galaxies
\citep[e.g.][]{bland95,heckman90,carlos20}; and (ii) the weak
extra-planar diffuse ionized gas that is observed as part of the thick
disk in some edge-on galaxies too
\citep[e.g.][]{floresfajardo11,levy18}.

For each galaxy we obtain the distribution of the average \EWa~and the
density of spaxels across the considered BPT diagram. Then, for the
full sample and for any sub-sample selected by stellar mass and/or
morphology, we average both distributions. This way each ionized
spaxel and each galaxy weights the same, irrespectively of the flux
intensity of the emission lines. Therefore, by construction the
possible contribution of strong but spatially concentrated
ionizing sources, like AGNs, is somehow diluted. Together with the
bias against strong shock ionized structures, we acknowledge that most
of the patterns that emerge from this exploration are related to the
dominant (in terms of spatial extend) ionizing sources in galaxies:
i.e., stars.

Figure \ref{fig:ex_BPT} shows, for each panel, the average
distributions along the BPT diagram for a different subset of
galaxies. Top-left panel includes the diagram for the full
sub-sample. Then, each column corresponds to galaxies of different
morphologies: (i) early-type galaxies (E/S0), (ii) early-spirals or
spirals with prominent bulge (Sa/Sb), and (iii) late-spirals, or
spirals with weak or no bulge (Sc/Sd). Finally, each row corresponds
to galaxies with a different stellar mass, from high-mass
(10$^{11-13.5}$M$_\odot$) to low-mass (10$^{7-9.5}$M$_\odot$). This
way, any panel at a certain column and row represents the BPT diagram
for galaxies with a particular morphology and stellar mass. In
addition it is included in each panel the location of the azimuthal
average [OIII]/H$\beta$ and [NII]/H$\alpha$ value at different
galactocentic distances.

Figure \ref{fig:ex_BPT} shows very similar patterns and distributions as
those already reported by \citet{ARAA} and \citet{sanchez22}. For the
full sub-sample (top-left panel) the three studies show the highest
densities along the known location of either classical \HII\ regions
\citep[e.g.][]{osterbrock89} or SF galaxies
\citep[e.g.][]{kauff03}. The highest values of \EWa ~($>$10\AA) are
found at these locations too, corresponding to galactocentric
distances $>$0.7 Re (i.e., to the disk of galaxies). The distribution
presents an extension of much lower density towards the right-side of
the diagram, where the \EWa~presents the lowest average values
($\sim$1\AA), corresponding to the inner regions of galaxies ($<$0.5
Re, i.e., the bulges). This pattern fits with our current
understanding of the dominant ionizing processes in galaxies
\citep[e.g.][]{sanchez21}. While the ionization in the disk is
dominated by the contribution of OB young massive stars, tracers of
recent SF processes, the inner region is dominated by a LINER-like
ionization \citep{heckman87} produced most probably by Hot-Evoled
Low-Mass/post-AGB
stars \citep[HOLMES;][]{binette94,floresfajardo11,singh13,belfiore17a,lacerda18}.
Thus, the radial pattern in line ratios is primarily dominated by a
change in the dominant ionizing source. Beside that, there is an
additional change in the line ratios from the inside-out within the
disk regions of galaxies, spaning from the bottom-right (for
R$\sim$0.6 Re) towards the upper-left (for R$\sim$2 Re) regions of
higher density. This additional pattern is induced by a general decline of the
metal abundances from the inside-out that it observed in the average population
of spiral galaxies \citep[e.g.][]{sanchez14,ARAA}.

The main result from this exploration is that the dominant ionizing
source is traced by the dominant stellar
population. Therefore, as the star-formation history, final stellar
composition, and inside-out distribution within galaxies depend on the
current morphology and accumulated stellar mass
\citep[e.g.][]{rosa14,rgb17,ARAA}, and thus the described patterns will be
strongly modulated by those two parameters. This is evident in
Fig. \ref{fig:ex_BPT}. Massive and early-type galaxies (E/S0 and
M$_\star \sim 10^{11-13.5}$ M$_\odot$), those presenting the older and
less diverse stellar population, present the lowest and more
homogeneous distribution of \EWa, with most of the ionized regions
covering a regime between the center and the right-side of the BPT
diagram. We should stress that contrary to the usual perception their
ionized regions are not all restricted to the regime above the
classical demarcation lines adopted to separate between SF and non-SF
ionization \citep[e.g.][]{kewley01,kauff03}, despite the fact that no
star-formation (and therefore, no young OB-stars) are present in these
galaxies. As indicated before, their ionization is dominated by either
HOLMES/post-AGB stars \citep[the presence of a significant
  contribution of low-/mid-velocity shocks cannot be excluded
  too; and see][]{dopita96}. Early-type of lower stellar masses present a
larger fraction of ionization in the \HII/SF regime of the diagram,
although the presence of LINER-like ionization is observed at any
stellar mass. Rejuvenation induced by the capture of less massive, gas
rich, galaxies or the effect of the slow dimming of a disk remant
could explain this ionization \citep[][]{gomes15c,oyarzun19}.

On the other hand, low massive and late-type galaxies (Sc/Sd and
M$_\star \sim 10^{7-9.5}$ M$_\odot$), those presenting the youngest
stellar populations, present the highest \EWa~values, with a
significant variation from the inside-out (from 10\AA ~to
100\AA). Their ionized regions are essentially restricted to the
classical location of \HII\ regions, well below the \citet{kewley01}
demarcation line, and in most of the cases even below the more
restrictive \citet{kauff03} one. We discussed before that their line
ratios change from the inside-out following a negative galactocentric
abundance gradient. Following the same trend described for eary-type
galaxies (E/S0) discussed before, the ionizatoin of late-type galaxies
of higher stellar, being still dominated by the presence of young
massive stars, present a clear increase of harder ionizations, with
the distribution steadily shifting towards the so-called intermediate
and LINER-like region of the diagram (i.e., present an increase in the
\nii/\Ha\ line ratio).

Finally, early-spirals with MW-like stellar masses (Sa/Sb
M$_\star \sim 10^{10.5-11}$ M$_\odot$), those with the strongest
gradient in the stellar populations from the inside-out, present the
highest range of \EWa\ too (from $\sim$1\AA\ to $\sim$100\AA).  Their
ionized regions cover the widest range of possible line ratios, with a
location similar to those of massive/early-type galaxies (for their
central regions), and similar to those of less-massive/late-type
galaxies (for the outer regions). This highlights again the strong
connection between the dominant stellar population and the observed
properties of the ionized gas (i.e., line ratios). Like in the two
previous cases (E/S0 and Sc/Sb), the low mass early-type spirals
present an ionization more dominated by young massive star, while the
high mass ones present a stronger component of intermediate/LINER-like
ionization.

As indicated before, all these patterns are very similar to those
described in \citet{ARAA} and \citet{sanchez22}. However, there are
subtlet but relevant differences, most of them related to the average
radial distribution (traced by the grey solid circles in
Fig. \ref{fig:ex_BPT}).  In the two previous explorations the radial
trends were well defined for all morphological subsamples for stellar
masses above 10$^{10.5}$M$_\odot$. These trends are very similar to
the one reported here, describing a shift from the right-side in the
central regions (more to the LINER-like in the presene of a bulge or
for early-type galaxies) towards the left-side in the outer ones (more
prominent and following the loci of classical \HII\ regions for disk
dominated galaxies). However, below that mass in both previous studies
the trends are less clear, or they directly bend or reverse from this
average one observed in the current dataset. This is more evident for
late-type/low-mass galaxies (bottom-right panel,
Fig. \ref{fig:ex_BPT}). As this radial trend, for this particular
sub-sample, is determined by a decrease (increase) in the oxygen
abundance (ionization strength) from the inside-out, we consider that
the distributions observed when using the current dataset fit better
with our current understanding of the chemical structure of disk
galaxies than the previously reported ones.

We consider that most probably the source of the discrepancy is in the
analyzed data themselves. In the case of \citet{ARAA} it was used an
heterogeneous compilation of IFS data from different surveys,
including data of high spatial resolution
\citep[AMUSING++;][]{carlos20}, but also data from CALIFA v2.2, MaNGA
\citep{manga} and SAMI \citep{sami}, that indeed dominates the
statistics. Despite the effort in that review to homogenize the
dataset and select only the well resolved data, as we show in
Sec. \ref{sec:HII_NGC2906}, the improved spatial resolution resolution
have a clear effect in the exploration of the properties of the
ionized gas. The fact that the results presented in \citet{sanchez22},
using the MaNGA data with lower spatial resolution (even lower in
physical terms, due to the average redshift of that sample), present
the same discrepancies, reinforce that suspicion.

\section{Summary and conclusions}
\label{sec:dis}

Along this article we present one of the largest and better quality
distributions of spatial resolved spectrocopic properties of galaxies
obtained using IFS data. Comparing with the most recent distributions
of similar products, like the results from the analysis using \pyp on
the final DR of the MaNGA IFS galaxy survey \citep{sanchez22}, we
acknowledge that this later distribution covers a much larger number of
galaxies ($\sim$10,000). However, the number of independent spectra
provided by that survey ($\sim$2 millions) is only a factor two larger
than the one provided here. This implies that, in average, eCALIFA
provides with nearly five times more sampled spatial elements per
galaxy. In addition, the MaNGA covers up to 1.5 Re for $\sim$2/3 of
the sample, while eCALIFA covers more than 2 Re for $\sim$85\% of the
galaxies (\citealt{ARAA}, \edrnp). Finally, the new reduction introduced by
\edr~improved the spatial resolution of the original CALIFA data from
2.4$\arcsec$/FWHM ($\sim$700 pc) to $\sim$1.0-1.5$\arcsec$/FWHM
($\sim$300 pc). In summary, despite the more limited number of
galaxies, the spatial sampling and coverage of the optical extension
of galaxies is significantly better, in comparison to other IFS surveys with larger galaxy samples.

We describe along this article the analysis performed on this
dataset, making a particular emphasis in the description of the
changes introduced in the analysis with respect to previous similar
explorations. For each galaxy we deliver a single FITS file comprising
different extensions in which each of them includes the spatial
distributions of the different physical and observational quantities
derived by the analysis. We present a detail description of each of
these extensions, illustrating their content using the results for the
galaxy NGC\,2906 as a showcase. Finally, we extract for each galaxy a
set of integrated and/or characteristics parameters and, when
required, the slopes of their radial gradients. 
We provide an additional catalog containing over 550 derived quantities for each object in the dataset.
%This dataset comprises a catalog with more than 550 quantities derived for each object that we distribute as an additional catalog. 

We illustrated the content of this new set of dataproducts by
exploring (i) the properties of the ionized gas in the archetypal
galaxy NGC\,2906, and (ii) the distribution across the classical BPT
diagnostic diagram of the spatial resolved ionized gas for the full
sample, segregated by morphology and stellar mass. In both cases, it is
appreciated the effects of the improved spatial resolution, in
particular in the ability to detect \HII\ regions and in the
recovering radial trends that were less evident or
directly not observed when using data of more coarse resolution.

The complete set of dataproducts and the catalog of
invidual quantities is freely distributed for its use by the communitty as part of the
eCALIFA data release
\footnote{\url{http://ifs.astroscu.unam.mx/CALIFA_WEB/public_html/}}.

% Acknowledgements
\section*{ACKNOWLEDGMENTS}

{ We thanks the referee for his/her suggestions that have improved the content of the current manuscript}

S.F.S. thanks the PAPIIT-DGAPA AG100622 project. J.K.B.B. and S.F.S. acknowledge support from the CONACYT grant CF19-39578. R.G.B. acknowledges financial support from the grants CEX2021-001131-S funded by MCIN/AEI/10.13039/501100011033, SEV-2017-0709, and to PID2019-109067-GB100. L.G. acknowledges financial support from the Spanish Ministerio de Ciencia e Innovaci\'on (MCIN), the Agencia Estatal de Investigaci\'on (AEI) 10.13039/501100011033, and the European Social Fund (ESF) "Investing in your future" under the 2019 Ram\'on y Cajal program RYC2019-027683-I and the PID2020-115253GA-I00 HOSTFLOWS project, from Centro Superior de Investigaciones Cient\'ificas (CSIC) under the PIE project 20215AT016, and the program Unidad de Excelencia Mar\'ia de Maeztu CEX2020-001058-M.

This study uses data provided by the Calar Alto Legacy
Integral Field Area (CALIFA) survey (http://califa.caha.es/). Based on
observations collected at the Centro Astron\'omico Hispano Alem\'an
(CAHA) at Calar Alto, operated jointly by the Max-Planck-Institut
f\"ur Astronomie and the Instituto de Astrof\'isica de Andaluc\'ia
(CSIC). 

This research made use of
Astropy,\footnote{http://www.astropy.org} a community-developed core
Python package for Astronomy \citep{astropy:2013, astropy:2018}.

% References
%
% Margin notes within bibliography
%\section*{LITERATURE\ CITED}

%To download the appropriate bibliography style file, please see \url{http://www.annualreviews.org/page/authors/author-instructions/preparing/latex}. \\
%\noindent
%Please see the Style Guide document for instructions on preparing your Literature Cited.
%The citations should be listed in alphabetical order, with no titles. For example:

%\citep{sanchez2012}

\bibliographystyle{ar-style2}
\bibliography{my_bib}
%\bibliography{CALIFAI,Alenka,Califa8_SFH,references-VAR,CALIFA_DR3,rgb_califa_ML}

%\bibliography{CALIFAI,Alenka,Califa8_SFH,references-VAR}
%\bibliography{CALIFAI,Alenka,Califa8_SFH,references-VAR,CALIFA_DR3,rgb_califa_ML}
%\bibliography{CALIFAI,Alenka,Califa8_SFH,references-VAR,CALIFA_DR3.bib}

\appendix

\section{List of emission lines using the moment analysis}
\label{ap:tab_fe}

As indicated before, we perform the moment analysis described in Sec
\ref{sec:el_fit} over two different list of emission lines, which
results are distributed in the FLUX\_ELINES and FLUX\_ELINES\_LONG
extensions (Sec. \ref{sec:cubes}). In the case of the FLUX\_ELINES
extension we use the list of emission lines distributed in
\citet{pipe3d_ii}, Tab. 1 of that article. For the FLUX\_ELINES\_LONG
extension it was analyzed the list of emission lines included in Table
\ref{tab:fe_long_list}. For each emission line we include in this table a running index {\tt I} that defines the channel in which each parameter is included (as described
in Tab. \ref{tab:fe}), together with its rest-frame wavelength and an {\tt Id} to identify each line. The reported wavelegnths of these emission lines were extracted from \citet{snr_elines}.

%----------------------------------------------------------------

\begin{table*}
\begin{center}
\caption{FLUX\_ELINES\_LONG extension: Analyzed emission lines}
%\hskip-2.8cm
\hspace*{-3em}
\begin{tabular}{lll|lll|lll|lll|lll}\hline\hline
{\tt \#I} &$\lambda$ (\AA) & Id &
{\tt \#I} &$\lambda$ (\AA) & Id &
{\tt \#I} &$\lambda$ (\AA) & Id &
{\tt \#I} &$\lambda$ (\AA) & Id &
{\tt \#I} & $\lambda$ (\AA) & Id \\
  \hline
  0 &  3726.03  &   [OII] $^*$  & 26 &  4287.4  &   [FeII]  & 52 &  4754.83  &   [FeIII]  & 78 &  5158.0  &   [FeII]  & 104 &  5577.34  &   [OI]  \\ 
1 &  3728.82  &   [OII] $^*$ & 27 &  4340.49  &   H$\gamma$$^*$  & 53 &  4769.6  &   [FeIII]  & 79 &  5158.9  &   [FeVII]  & 105 &  5631.1  &   [FeVI]  \\ 
2 &  3734.37  &   HI$^*$  & 28 &  4358.1  &   [FeII]  & 54 &  4774.74  &   [FeII]  & 80 &  5176.0  &   [FeVI]  & 106 &  5677.0  &   [FeVI]  \\ 
3 &  3750.15  &   HI  & 29 &  4358.37  &   [FeII]  & 55 &  4777.88  &   [FeIII]  & 81 &  5184.8  &   [FeII]  & 107 &  5720.7  &   [FeVII]  \\ 
4 &  3758.9  &   [FeVII]  & 30 &  4359.34  &   [FeII]  & 56 &  4813.9  &   [FeIII]  & 82 &  5191.82  &   [ArIII]  & 108 &  5754.59  &   [NII]  \\ 
5 &  3770.63  &   HI  & 31 &  4363.21  &   [OIII]  & 57 &  4814.55  &   [FeII]  & 83 &  5197.9  &   [NI]$^*$  & 109 &  5876.0  &   HeI$^*$  \\ 
6 &  3797.9  &   HI$^*$  & 32 &  4413.78  &   [FeII]  & 58 &  4861.36  &   H$\beta$$^*$  & 84 &  5200.26  &   [NI]$^*$  & 110 &  5889.95  &   NaI$^*$  \\ 
7 &  3819.61  &   HeI  & 33 &  4414.45  &   [FeII]  & 59 &  4881.11  &   [FeIII]  & 85 &  5220.06  &   [FeII]  & 111 &  5895.92  &   NaI$^*$  \\ 
8 &  3835.38  &   HI  & 34 &  4416.27  &   [FeII]  & 60 &  4889.63  &   [FeII]  & 86 &  5261.61  &   [FeII]  & 112 &  6087.0  &   [FeVII]  \\ 
9 &  3868.75  &   [NeIII]  & 35 &  4452.11  &   [FeII]  & 61 &  4893.4  &   [FeVII]  & 87 &  5268.88  &   [FeII]  & 113 &  6300.3  &   [OI]$^*$  \\ 
10 &  3888.65  &   HeI$^*$  & 36 &  4457.95  &   [FeII]  & 62 &  4905.35  &   [FeII]  & 88 &  5270.3  &   [FeIII]  & 114 &  6312.06  &   [SIII]  \\ 
11 &  3889.05  &   HI$^*$  & 37 &  4470.29  &   [FeII]  & 63 &  4921.93  &   HeI  & 89 &  5273.38  &   [FeII]  & 115 &  6363.78  &   [OI]  \\ 
12 &  3933.66  &   CaII  & 38 &  4471.48  &   HeI  & 64 &  4924.5  &   [FeIII]  & 90 &  5277.8  &   [FeVI]  & 116 &  6374.51  &   [FeX]  \\ 
13 &  3964.73  &   HeI$^*$  & 39 &  4474.91  &   [FeII]  & 65 &  4930.5  &   [FeIII]  & 91 &  5296.84  &   [FeII]  & 117 &  6435.1  &   [ArV]  \\ 
14 &  3967.46  &   [NeIII]$^*$ & 40 &  4485.21  &   [NiII]  & 66 &  4942.5  &   [FeVII]  & 92 &  5302.86  &   [FeXIV]  & 118 &  6548.05  &   [NII]$^*$  \\ 
15 &  3968.47  &   CaII $^*$ & 41 &  4562.48  &   [MgI]  & 67 &  4958.91  &   [OIII]$^*$  & 93 &  5309.18  &   [CaV]  & 119 &  6562.85  &   H$\alpha$$^*$  \\ 
16 &  3970.07  &   H$\epsilon$$^*$  & 42 &  4571.1  &   MgI]  & 68 &  4972.5  &   [FeVI]  & 94 &  5333.65  &   [FeII]  & 120 &  6583.45  &   [NII]$^*$  \\ 
17 &  4026.19  &   HeI  & 43 &  4632.27  &   [FeII]  & 69 &  4973.39  &   [FeII]  & 95 &  5335.2  &   [FeVI]  & 121 &  6678.15  &   HeI  \\ 
18 &  4068.6  &   [SII]  & 44 &  4658.1  &   [FeIII]  & 70 &  4985.9  &   [FeIII]  & 96 &  5376.47  &   [FeII]  & 122 &  6716.44  &   [SII]$^*$  \\ 
19 &  4076.35  &   [SII]  & 45 &  4685.68  &   HeII  & 71 &  5006.84  &   [OIII]$^*$  & 97 &  5411.52  &   HeII  & 123 &  6730.82  &   [SII]$^*$  \\ 
20 &  4101.77  &   H$\delta$$^*$  & 46 &  4701.62  &   [FeIII]  & 72 &  5015.68  &   HeI$^*$  & 98 &  5412.64  &   [FeII]  & 124 &  6855.18  &   FeI  \\ 
21 &  4120.81  &   HeI  & 47 &  4711.33  &   [ArIV]  & 73 &  5039.1  &   [FeII]  & 99 &  5424.2  &   [FeVI]  & 125 &  7005.67  &   [ArV]  \\ 
22 &  4177.21  &   [FeII]  & 48 &  4713.14  &   HeI  & 74 &  5072.4  &   [FeII]  & 100 &  5426.6  &   [FeVI]  & 126 &  7065.19  &   HeI  \\ 
23 &  4227.2  &   [FeV]  & 49 &  4724.17  &   [NeIV]  & 75 &  5107.95  &   [FeII]  & 101 &  5484.8  &   [FeVI]  & 127 &  7135.8  &   [ArIII]$^*$  \\ 
24 &  4243.98  &   [FeII]  & 50 &  4733.93  &   [FeIII]  & 76 &  5111.63  &   [FeII]  & 102 &  5517.71  &   [ClIII]  & 128 &  7155.14  &   [FeII]  \\ 
25 &  4267.0  &   CII  & 51 &  4740.2  &   [ArIV]  & 77 &  5145.8  &   [FeVI]  & 103 &  5527.33  &   [FeII]  & 129 &  7171.98  &   [FeII]  \\ 
\end{tabular}\label{tab:fe_long_list}
\end{center}
{\tt \#I} is the running index described in Tab. \ref{tab:fe}, and {\tt Id} is a label to identify the emission lines the  is the label to each emission line
included in the FLUX\_ELINES\_LONG extension. $^*$ emission lines most frequently detected in galaxies, according to \citet{sanchez22}, for which we distribute the characteristic value (value at the effective radius) and the slope of its radial gradient (Sec. \ref{sec:ana_int}).
\end{table*}
%----------------------------------------------------------------

\section{List of Integrated and characteristic parametrers}
\label{ap:tab_par}

Table \ref{tab:cat} lists the complete set of integrated, characteristic parameters and radial gradient slopes distributed for each analyzed datacube,
described in Sec. \ref{sec:ana_int} and included in the {\tt eCALIFA.pyPipe3D.fits} FITs file\footnote{\url{http://ifs.astroscu.unam.mx/CALIFA/V500/v2.3/tables/}}. For each parameter we include the column in which it is stored the information in the FITs file table, the name of the parameter, its units (when required) and a brief description.

\begin{table*}
\begin{center}
\caption{Integrated and characteristic parameters delivered for each analyzed datacube.}
\begin{tabular}{llll}
\hline\hline
\#  & parameter name & units & Description  \\ 
\hline
0 & ID &  --  &  None \\ 
1 & cubename &  --  &  eCALIFA name of the datacube \\ 
2 & galaxy &  --  &  eCALIFA name of the galay \\ 
3 & Re\_arc &  arcsec  &  Adopted effective radius in arcsec   \\ 
4 & FoV &  --  &  Ratio between the diagonal radius of the cube and Re  \\ 
5 & redshift &  --  &  Redshift of the galaxy derived by \pyp \\ 
6 & DL &  { Mpc}  &  Adopted luminosity distance   \\ 
7 & DA &  { Mpc}  &  Adopted angular-diameter distance   \\ 
8 & PA &  degrees  &  Adopted position angle in degrees   \\ 
9 & ellip &  --  &  Adopted ellipticity   \\ 
10 & Re\_kpc &  kpc  &  Effective radius in kpc   \\ 
11 & log\_Mass &  log(M$_\odot$)  &  Integrated stellar mass in units of the solar mass in logarithm scale   \\ 
12 & e\_log\_Mass &  log(M$_\odot$)  &  Error of the integrated stellar mass { in logratihm-scale}  \\ 
13 & log\_SFR\_Ha &  --  &  -- \\ 
14 & e\_log\_SFR\_Ha &  --  &  -- \\ 
15 & log\_SFR\_ssp &  log(M$_\odot$ yr$^{-1}$)  &  Integrated SFR derived from the SSP analysis t$$<$$32Myr   \\ 
16 & e\_log\_SFR\_ssp &  --  &  -- \\ 
17 & log\_NII\_Ha\_cen &  --  &  Logarithm of the [NII]6583/H$\alpha$ line ratio in the central aperture   \\ 
18 & e\_log\_NII\_Ha\_cen &  --  &  Error in the logarithm of the [NII]6583/H$\alpha$ line ratio   \\ 
19 & log\_OIII\_Hb\_cen &  --  &  Logarithm of the [OIII]5007/H$\beta$ line ratio in the central aperture   \\ 
20 & e\_log\_OIII\_Hb\_cen &  --  &  Error in the logarithm of the [OIII]5007/H$\beta$ line ratio    \\ 
21 & log\_SII\_Ha\_cen &  --  &  Logarithm of the [SII]6717+6731/H$\alpha$ line ratio in the central aperture  \\ 
22 & e\_log\_SII\_Ha\_cen &  --  &  Error in the logarithm of the [SII]6717/H$\alpha$ line ratio    \\ 
23 & log\_OII\_Hb\_cen &  --  &  Logarithm of the [OII]3727/H$\beta$ line ratio in the central aperture \\ 
24 & e\_log\_OII\_Hb\_cen &  --  &  Error in the logarithm of the [OII]3727/H$\beta$ line ratio   \\ 
25 & EW\_Ha\_cen &  \AA\  &  EW of H$\alpha$ in the central aperture   \\ 
26 & e\_EW\_Ha\_cen &  \AA\  &  Error of the EW of H$\alpha$ in the central aperture   \\ 
27 & ZH\_LW\_Re\_fit &  dex  &  { LW metallicity of the stellar population at Re, normalized to the solar value, in logarithm scales}   \\ 
28 & e\_ZH\_LW\_Re\_fit &  dex  &  Error in the luminosity weighted metallicity of the stellar population   \\ 
29 & alpha\_ZH\_LW\_Re\_fit &  dex/Re  &  Slope of the gradient of the LW metallicity of the stellar population   \\ 
30 & e\_alpha\_ZH\_LW\_Re\_fit &  dex/Re  &  Error of the slope of the gradient of the LW log-metallicity   \\ 
31 & ZH\_MW\_Re\_fit &  dex  &  { MW metallicity of the stellar population at Re, normalized to the solar value, in logarithm scales}   \\ 
32 & e\_ZH\_MW\_Re\_fit &  dex  &  Error in the Mass weighted metallicity of the stellar population  \\ 
33 & alpha\_ZH\_MW\_Re\_fit &  dex/Re  &  Slope of the gradient of the MW log-metallicity of the stellar population   \\ 
34 & e\_alpha\_ZH\_MW\_Re\_fit &  dex/Re  &  Error of the slope of the gradient of the MW log-metallicity   \\ 
\hline 
\end{tabular}\label{tab:cat}
\end{center}
\end{table*}

\addtocounter{table}{-1}

\begin{table*}
\begin{center}
\caption{Integrated and characteristic parameters delivered for each analyzed datacube.}
\begin{tabular}{llll}
\hline\hline
\#  & parameter name & units & Description  \\ 
\hline
35 & Age\_LW\_Re\_fit &  log(yr)  &  Luminosity weighted age of the stellar population in logarithm scale   \\ 
36 & e\_Age\_LW\_Re\_fit &  log(yr)  &  Error in the luminosity weighted age of the stellar population   \\ 
37 & alpha\_Age\_LW\_Re\_fit &  log(yr)/Re  &  Slope of the gradient of the LW log-age of the stellar population   \\ 
38 & e\_alpha\_Age\_LW\_Re\_fit &  log(yr)/Re  &  Error of the slope of the gradient of the LW log-age of the stellar population  \\ 
39 & Age\_MW\_Re\_fit &  log(yr)  &  Mass weighted age of the stellar population in logarithm   \\ 
40 & e\_Age\_MW\_Re\_fit &  log(yr)  &  Error in the Mass weighted age of the stellar population   \\ 
41 & alpha\_Age\_MW\_Re\_fit &  { log(yr)}/Re  &  Slope of the gradient of the MW log-age of the stellar population   \\ 
42 & e\_alpha\_Age\_MW\_Re\_fit &  { log(yr)}/Re  &  Error of the slope of the gradient of the MW log-age of the stellar population   \\ 
43 & log\_Mass\_gas &  log(M$_\odot$)  &  Integrated gas mass in units of the solar mass in logarithm scale   \\ 
44 & vel\_sigma\_Re &  --  &  Velocity/dispersion ratio for the stellar populations within 1.5 Re   \\ 
45 & e\_vel\_sigma\_Re &  --  &  Error in the velocity/dispersion ratio for the stellar population   \\ 
46 & log\_SFR\_SF &  log(M$_\odot$ yr$^{-1}$)  &  Integrated SFR using only the spaxels compatible with SF   \\ 
47 & log\_SFR\_D\_C &  log(M$_\odot$ yr$^{-1}$)  &  Integrated SFR diffuse corrected   \\ 
48 & OH\_O3N2\_cen &  dex  &  Oxygen abundance using the calibrator O3N2 at the central aperture \\ 
49 & e\_OH\_O3N2\_cen &  dex  &  Error in Oxygen abundance using the calibrator O3N2 at the central aperture   \\ 
50 & OH\_N2\_cen &  dex  &  Oxygen abundance using the calibrator N2 at the central aperture \\ 
51 & e\_OH\_N2\_cen &  dex  &  Error in Oxygen abundance using the calibrator N2 at the central aperture   \\ 
52 & OH\_ONS\_cen &  dex  &  Oxygen abundance using the calibrator ONS at the central aperture \\ 
53 & e\_OH\_ONS\_cen &  dex  &  Error in Oxygen abundance using the calibrator ONS at the central aperture   \\ 
54 & OH\_R23\_cen &  dex  &  Oxygen abundance using the calibrator R23 at the central aperture \\ 
55 & e\_OH\_R23\_cen &  dex  &  Error in Oxygen abundance using the calibrator R23 at the central aperture   \\ 
56 & OH\_pyqz\_cen &  dex  &  Oxygen abundance using the calibrator pyqz at the central aperture \\ 
57 & e\_OH\_pyqz\_cen &  dex  &  Error in Oxygen abundance using the calibrator pyqz at the central aperture   \\ 
58 & OH\_t2\_cen &  dex  &  Oxygen abundance using the calibrator t2 at the central aperture \\ 
59 & e\_OH\_t2\_cen &  dex  &  Error in Oxygen abundance using the calibrator t2 at the central aperture   \\ 
60 & OH\_M08\_cen &  dex  &  Oxygen abundance using the calibrator M08 at the central aperture \\ 
61 & e\_OH\_M08\_cen &  dex  &  Error in Oxygen abundance using the calibrator M08 at the central aperture   \\ 
62 & OH\_T04\_cen &  dex  &  Oxygen abundance using the calibrator T04 at the central aperture \\ 
63 & e\_OH\_T04\_cen &  dex  &  Error in Oxygen abundance using the calibrator T04 at the central aperture   \\ 
64 & OH\_dop\_cen &  dex  &  Oxygen abundance using the calibrator dop at the central aperture \\ 
65 & e\_OH\_dop\_cen &  dex  &  Error in Oxygen abundance using the calibrator dop at the central aperture   \\ 
66 & OH\_O3N2\_EPM09\_cen &  dex  &  Oxygen abundance using the calibrator O3N2\_EPM09 at the central aperture \\ 
67 & e\_OH\_O3N2\_EPM09\_cen &  dex  &  Error in Oxygen abundance using the calibrator O3N2\_EPM09 at the central aperture   \\ 
68 & log\_OI\_Ha\_cen &  dex  &  Logarithm of the [OI]6301/H$\alpha$ within a central aperture \\ 
69 & e\_log\_OI\_Ha\_cen &  dex  &  Error in the logarithm of the [OI]6301/H$\alpha$ within a central aperture    \\ 
\hline 
\end{tabular}\label{tab:cat}
\end{center}
\end{table*}

\addtocounter{table}{-1}

\begin{table*}
\begin{center}
\caption{Integrated and characteristic parameters delivered for each analyzed datacube.}
\begin{tabular}{llll}
\hline\hline
\#  & parameter name & units & Description  \\ 
\hline
70 & Ha\_Hb\_cen &  --  &  Ratio between the flux of Ha$\alpha$ and H$\beta$ within a central aperture \\ 
71 & e\_Ha\_Hb\_cen &  --  &  Error of Ha\_Hb\_cen   \\ 
72 & log\_NII\_Ha\_Re &  dex  &  Logarithm of the [NII]6583/H$\alpha$ line ratio at 1Re \\ 
73 & e\_log\_NII\_Ha\_Re &  dex  &  Error in the logarithm of the [NII]6583/H$\alpha$ ratio at 1Re   \\ 
74 & log\_OIII\_Hb\_Re &  dex  &  Logarithm of the [OIII]5007/H$\beta$ line ratio at 1Re \\ 
75 & e\_log\_OIII\_Hb\_Re &  dex  &  Error in the logarithm of the [OIII]5007/H$\beta$ ratio at 1Re   \\ 
76 & log\_SII\_Ha\_Re &  dex  &  Logarithm of the [SII]6717+6731/H$\alpha$ ratio at 1Re \\ 
77 & e\_log\_SII\_Ha\_Re &  dex  &  Error in the logarithm of the [SII]6717/H$\alpha$ ratio at 1Re   \\ 
78 & log\_OII\_Hb\_Re &  dex  &  Logarithm of the [OII]3727/H$\beta$ ratio at 1Re \\ 
79 & e\_log\_OII\_Hb\_Re &  dex  &  Error in the logarithm of the [OII]3727/H$\beta$ at 1Re   \\ 
80 & log\_OI\_Ha\_Re &  dex  &  Logarithm of the [OI]6301/H$\alpha$ ratio at 1Re \\ 
81 & e\_log\_OI\_Ha\_Re &  dex  &  Error in the logarithm of the [OI]6301/H$\alpha$ ratio at 1Re   \\ 
82 & EW\_Ha\_Re &  \AA\  &  EW of H$\alpha$ at 1Re \\ 
83 & e\_EW\_Ha\_Re &  \AA\  &  Error of the EW of H$\alpha$ at 1Re   \\ 
84 & Ha\_Hb\_Re &  --  &  Ratio between the flux of H$\alpha$ and H$\beta$ within at 1Re \\ 
85 & e\_Ha\_Hb\_Re &  --  &  Error of Ha\_Hb\_Re   \\ 
86 & log\_NII\_Ha\_ALL &  dex  &  Logarithm of the [NII]6583/H$\alpha$ line ratio within the entire FoV \\ 
87 & e\_log\_NII\_Ha\_ALL &  dex  &  Error in the logarithm of the [NII]6583/$\alpha$ ratio within the entire FoV   \\ 
88 & log\_OIII\_Hb\_ALL &  dex  &  Logarithm of the [OIII]5007/H$\beta$ line ratio within the FoV \\ 
89 & e\_log\_OIII\_Hb\_ALL &  dex  &  Error in the logarithm of the [OIII]5007/H$\beta$ ratio within the FoV   \\ 
90 & log\_SII\_Ha\_ALL &  dex  &  Logarithm of the [SII]6717+6731/H$\alpha$ ratio within the entire FoV \\ 
91 & e\_log\_SII\_Ha\_ALL &  dex  &  Error in the logarithm of the [SII]6717/H$\alpha$ ratio within the FoV   \\ 
92 & log\_OII\_Hb\_ALL &  dex  &  Logarithm of the [OII]3727/H$\beta$ ratio within the entire FoV \\ 
93 & e\_log\_OII\_Hb\_ALL &  dex  &  Error in the logarithm of the [OII]3727/H$\beta$ ratio within the FoV   \\ 
94 & log\_OI\_Ha\_ALL &  dex  &  Logarithm of the [OI]6301/H$\alpha$ ratio within the entire FoV \\ 
95 & e\_log\_OI\_Ha\_ALL &  dex  &  Error in the logarithm of the [OI]6301/H$\beta$ ratio within the FoV   \\ 
96 & EW\_Ha\_ALL &  \AA\  &  EW of H$\alpha$ within the entire FoV \\ 
97 & e\_EW\_Ha\_ALL &  \AA\  &  Error of the EW of H$\alpha$ within the FoV   \\ 
98 & Ha\_Hb\_ALL &  --  &  Ratio between the flux of H$\alpha$ and H$\beta$ within the entire FoV \\ 
99 & Sigma\_Mass\_cen &  log(M$_\odot$ pc$^{-2}$)  &  Stellar Mass surface density in the central aperture   \\ 
100 & e\_Sigma\_Mass\_cen &  log(M$_\odot$ pc$^{-2}$)  &  Error in Stellar Mass density in the central aperture \\ 
101 & Sigma\_Mass\_Re &  log(M$_\odot$ pc$^{-2}$)  &  Stellar Mass surface density at 1Re   \\ 
102 & e\_Sigma\_Mass\_Re &  log(M$_\odot$ pc$^{-2}$)  &  Error in Stellar Mass density at 1Re \\ 
103 & Sigma\_Mass\_ALL &  log(M$_\odot$ pc$^{-2}$)  &  Average Stellar Mass surface density within the entire FoV   \\ 
104 & e\_Sigma\_Mass\_ALL &  log(M$_\odot$ pc$^{-2}$)  &  Error in the average Stellar Mass surface density within the FoV \\ 
\hline 
\end{tabular}\label{tab:cat}
\end{center}
\end{table*}

\addtocounter{table}{-1}

\begin{table*}
\begin{center}
\caption{Integrated and characteristic parameters delivered for each analyzed datacube.}
\begin{tabular}{llll}
\hline\hline
\#  & parameter name & units & Description  \\ 
\hline
105 & T30 &  Gyr  &  Lookback time at which the galaxy has formed 30\% of its stellar mass   \\ 
106 & ZH\_T30 &  dex  &  Stellar metallicity, normalized by the solar value, in logarithm-scale at T30   time \\ 
107 & ZH\_Re\_T30 &  dex  &  Stellar metallicity, normalized by the solar value, in logarithm-scale at Re at T30   time   \\ 
108 & a\_ZH\_T30 &  dex/Re  &  Slope of the ZH gradient at T30   time \\ 
109 & T40 &  Gyr  &  Lookback time at which the galaxy has formed 40\% of its stellar mass   \\ 
110 & ZH\_T40 &  dex  &  Stellar metallicity, normalized by the solar value, in logarithm-scale at T40   time \\ 
111 & ZH\_Re\_T40 &  dex  &  Stellar metallicity, normalized by the solar value, in logarithm-scale at Re at T40   time   \\ 
112 & a\_ZH\_T40 &  dex/Re  &  Slope of the ZH gradient at T40   time \\ 
113 & T50 &  Gyr  &  Lookback time at which the galaxy has formed 50\% of its stellar mass   \\ 
114 & ZH\_T50 &  dex  &  Stellar metallicity, normalized by the solar value, in logarithm-scale at T50   time \\ 
115 & ZH\_Re\_T50 &  dex  &  Stellar metallicity, normalized by the solar value, in logarithm-scale at Re at T50   time   \\ 
116 & a\_ZH\_T50 &  dex/Re  &  Slope of the ZH gradient at T50   time \\ 
117 & T60 &  Gyr  &  Lookback time at which the galaxy has formed 60\% of its stellar mass   \\ 
118 & ZH\_T60 &  dex  &  Stellar metallicity, normalized by the solar value, in logarithm-scale at T60   time \\ 
119 & ZH\_Re\_T60 &  dex  &  Stellar metallicity, normalized by the solar value, in logarithm-scale at Re at T60   time   \\ 
120 & a\_ZH\_T60 &  dex/Re  &  Slope of the ZH gradient at T60   time \\ 
121 & T70 &  Gyr  &  Lookback time at which the galaxy has formed 70\% of its stellar mass   \\ 
122 & ZH\_T70 &  dex  &  Stellar metallicity, normalized by the solar value, in logarithm-scale at T70   time \\ 
123 & ZH\_Re\_T70 &  dex  &  Stellar metallicity, normalized by the solar value, in logarithm-scale at Re at T70   time   \\ 
124 & a\_ZH\_T70 &  dex/Re  &  Slope of the ZH gradient at T70   time \\ 
125 & T80 &  Gyr  &  Lookback time at which the galaxy has formed 80\% of its stellar mass   \\ 
126 & ZH\_T80 &  dex  &  Stellar metallicity, normalized by the solar value, in logarithm-scale at T80   time \\ 
127 & ZH\_Re\_T80 &  dex  &  Stellar metallicity, normalized by the solar value, in logarithm-scale at Re at T80   time   \\ 
128 & a\_ZH\_T80 &  dex/Re  &  Slope of the ZH gradient at T80   time \\ 
129 & T90 &  Gyr  &  Lookback time at which the galaxy has formed 90\% of its stellar mass   \\ 
130 & ZH\_T90 &  dex  &  Stellar metallicity, normalized by the solar value, in logarithm-scale at T90   time \\ 
131 & ZH\_Re\_T90 &  dex  &  Stellar metallicity, normalized by the solar value, in logarithm-scale at Re at T90   time   \\ 
132 & a\_ZH\_T90 &  dex/Re  &  Slope of the ZH gradient at T90   time \\ 
133 & T95 &  Gyr  &  Lookback time at which the galaxy has formed 95\% of its stellar mass   \\ 
134 & ZH\_T95 &  dex  &  Stellar metallicity, normalized by the solar value, in logarithm-scale at T95   time \\ 
135 & ZH\_Re\_T95 &  dex  &  Stellar metallicity, normalized by the solar value, in logarithm-scale at Re at T95   time   \\ 
136 & a\_ZH\_T95 &  dex/Re  &  Slope of the ZH gradient at T95   time \\ 
137 & T99 &  Gyr  &  Lookback time at which the galaxy has formed 99\% of its stellar mass   \\ 
138 & ZH\_T99 &  dex  &  Stellar metallicity, normalized by the solar value, in logarithm-scale at T99   time \\ 
139 & ZH\_Re\_T99 &  dex  &  Stellar metallicity, normalized by the solar value, in logarithm-scale at Re at T99   time   \\ 
\hline 
\end{tabular}\label{tab:cat}
\end{center}
\end{table*}

\addtocounter{table}{-1}

\begin{table*}
\begin{center}
\caption{Integrated and characteristic parameters delivered for each analyzed datacube.}
\begin{tabular}{llll}
\hline\hline
\#  & parameter name & units & Description  \\ 
\hline
140 & a\_ZH\_T99 &  dex/Re  &  Slope of the ZH gradient at T99   time \\ 
141 & log\_Mass\_gas\_Av\_gas\_OH &  log(M$_\odot$)  &  { Integrated gas mass using the calibrator log\_Mass\_gas\_Av\_gas\_OH, in logarithm-scale}  \\ 
142 & log\_Mass\_gas\_Av\_ssp\_OH &  log(M$_\odot$)  &  { Integrated gas mass using the calibrator log\_Mass\_gas\_Av\_ssp\_OH, in logarithm-scale} \\ 
143 & vel\_ssp\_2 &  km s$^{-1}$  &  Stellar velocity at 2 Re   \\ 
144 & e\_vel\_ssp\_2 &  km/s  &  Error in vel\_ssp\_2 \\ 
145 & vel\_Ha\_2 &  km s$^{-1}$  &  H$\alpha$ velocity at 2 Re   \\ 
146 & e\_vel\_Ha\_2 &  km/s  &  Error in vel\_Ha\_2 \\ 
147 & vel\_ssp\_1 &  km s$^{-1}$  &  Stellar velocity at 1 Re   \\ 
148 & e\_vel\_ssp\_1 &  km/s  &  Error of vel\_ssp\_1 \\ 
149 & vel\_Ha\_1 &  km/s  &  H$\alpha$ velocity at 1 Re   \\ 
150 & e\_vel\_Ha\_1 &  km/s  &  Error of e\_vel\_Ha\_1 \\ 
151 & log\_SFR\_ssp\_100Myr &  log(M$_\odot$ yr$^{-1}$)  &  Integrated SFR derived from the SSP analysis for t$$<$$100Myr   \\ 
152 & log\_SFR\_ssp\_10Myr &  log(M$_\odot$ yr$^{-1}$)  &  Integrated SFR derived from the SSP analysis for t$$<$$10Myr \\ 
153 & vel\_disp\_Ha\_cen &  km/s  &  H$\alpha$ Velocity dispersion at the central aperture   \\ 
154 & vel\_disp\_ssp\_cen &  km/s  &  Stellar velocity dispersion at the central aperture \\ 
155 & vel\_disp\_Ha\_1Re &  km/s  &  H$\alpha$ velocity dispersion at 1 Re   \\ 
156 & vel\_disp\_ssp\_1Re &  km/s  &  Stellar velocity at 1 Re  159  \\ 
157 & log\_Mass\_in\_Re &  --  &  -- \\ 
158 & ML\_int &  M$_\odot$/L$_\odot$  &  V-band mass-to-light ratio from integrated quantities \\ 
159 & ML\_avg &  M$_\odot$/L$_\odot$  &  V-band mass-to-light ratio averaged across the FoV   \\ 
160 & F\_Ha\_cen &  10$^{-16}$ \flux  &  Flux intensity of H$\alpha$ in the central aperture \\ 
161 & e\_F\_Ha\_cen &  10$^{-16}$ \flux  &  Error of F\_Ha\_cen   \\ 
162 & R50\_kpc\_V &  kpc  &  { Radius at which it is integrated half of the light in the V-band within the FoV} \\ 
163 & e\_R50\_kpc\_V &  kpc  &  Error of R50\_kpc\_V   \\ 
164 & R50\_kpc\_Mass &  kpc  &  { Radius at which is integrated half of the stellar mass within the FoV } \\ 
165 & e\_R50\_kpc\_Mass &  kpc  &  Error of R50\_kpc\_Mass   \\ 
166 & log\_Mass\_corr\_in\_R50\_V &  log(M$_\odot$)  &  Integrated stellar mass within { R50\_kpc\_V} \\ 
167 & e\_log\_Mass\_corr\_in\_R50\_V &  log(M$_\odot$)  &  Error of log\_Mass\_corr\_in\_R50\_V   \\ 
168 & log\_Mass\_gas\_Av\_gas\_log\_log &  log(M$_\odot$)  &  Molecular gas mass derived from the dust extinction { using the log-log calibration} \\ 
169 & Av\_gas\_Re &  mag  &  Dust extinction in the V-band derived from the H$\alpha$/H$\beta$ line ratio { at 1 Re}   \\ 
170 & e\_Av\_gas\_Re &  mag  &  Error of Av\_gas\_Re \\ 
171 & Av\_ssp\_Re &  mag  &  Dust extinction in the V-band derived from the analysis of the stellar population   \\ 
172 & e\_Av\_ssp\_Re &  mag  &  Error of the Av\_ssp\_Re \\ 
173 & Lambda\_Re &  --  &  Specific angular momentum ($\lambda$-parameter) for the stellar populations { within 1 Re}   \\ 
174 & e\_Lambda\_Re &  --  &  { Error of the Lambda\_Re parameter} \\ 
\hline 
\end{tabular}\label{tab:cat}
\end{center}
\end{table*}

\addtocounter{table}{-1}

\begin{table*}
\begin{center}
\caption{Integrated and characteristic parameters delivered for each analyzed datacube.}
\begin{tabular}{llll}
\hline\hline
\#  & parameter name & units & Description  \\ 
\hline
175 & flux\_[OII]3726.03\_Re\_fit &  10$^{-16}$ \flux   &  Flux intensity of [OII]3726.03 at 1Re \\ 
176 & e\_flux\_[OII]3726.03\_Re\_fit &  10$^{-16}$ \flux  &  Error in the flux intensity of [OII]3726.03 at 1Re   \\ 
177 & flux\_[OII]3726.03\_alpha\_fit &  10$^{-16}$ \flux  &  Flux intensity of [OII]3726.03 at 1Re \\ 
178 & e\_flux\_[OII]3726.03\_alpha\_fit &  10$^{-16}$ \flux  &  Error in the flux intensity of [OII]3726.03 at 1Re   \\ 
179 & flux\_[OII]3728.82\_Re\_fit &  10$^{-16}$ \flux  &  Flux intensity of [OII]3728.82 at 1Re \\ 
180 & e\_flux\_[OII]3728.82\_Re\_fit &  10$^{-16}$ \flux  &  Error in the flux intensity of [OII]3728.82 at 1Re   \\ 
181 & flux\_[OII]3728.82\_alpha\_fit &  10$^{-16}$ \flux  &  Flux intensity of [OII]3728.82 at 1Re \\ 
182 & e\_flux\_[OII]3728.82\_alpha\_fit &  10$^{-16}$ \flux  &  Error in the flux intensity of [OII]3728.82 at 1Re   \\ 
183 & flux\_HI3734.37\_Re\_fit &  10$^{-16}$ \flux  &  Flux intensity of HI3734.37 at 1Re \\ 
184 & e\_flux\_HI3734.37\_Re\_fit &  10$^{-16}$ \flux  &  Error in the flux intensity of HI3734.37 at 1Re   \\ 
185 & flux\_HI3734.37\_alpha\_fit &  10$^{-16}$ \flux  &  Flux intensity of HI3734.37 at 1Re \\ 
186 & e\_flux\_HI3734.37\_alpha\_fit &  10$^{-16}$ \flux  &  Error in the flux intensity of HI3734.37 at 1Re   \\ 
187 & flux\_HI3797.9\_Re\_fit &  10$^{-16}$ \flux  &  Flux intensity of HI3797.9 at 1Re \\ 
188 & e\_flux\_HI3797.9\_Re\_fit &  10$^{-16}$ \flux  &  Error in the flux intensity of HI3797.9 at 1Re   \\ 
189 & flux\_HI3797.9\_alpha\_fit &  10$^{-16}$ \flux  &  Flux intensity of HI3797.9 at 1Re \\ 
190 & e\_flux\_HI3797.9\_alpha\_fit &  10$^{-16}$ \flux  &  Error in the flux intensity of HI3797.9 at 1Re   \\ 
191 & flux\_HeI3888.65\_Re\_fit &  10$^{-16}$ \flux  &  Flux intensity of HeI3888.65 at 1Re \\ 
192 & e\_flux\_HeI3888.65\_Re\_fit &  10$^{-16}$ \flux  &  Error in the flux intensity of HeI3888.65 at 1Re   \\ 
193 & flux\_HeI3888.65\_alpha\_fit &  10$^{-16}$ \flux  &  Flux intensity of HeI3888.65 at 1Re  \\ 
194 & e\_flux\_HeI3888.65\_alpha\_fit &  10$^{-16}$ \flux  &  Error in the flux intensity of HeI3888.65 at 1Re   \\ 
195 & flux\_HI3889.05\_Re\_fit &  10$^{-16}$ \flux  &  Flux intensity of HI3889.05 at 1Re \\ 
196 & e\_flux\_HI3889.05\_Re\_fit &  10$^{-16}$ \flux  &  Error in the flux intensity of HI3889.05 at 1Re   \\ 
197 & flux\_HI3889.05\_alpha\_fit &  10$^{-16}$ \flux  &  Flux intensity of HI3889.05 at 1Re \\ 
198 & e\_flux\_HI3889.05\_alpha\_fit &  10$^{-16}$ \flux  &  Error in the flux intensity of HI3889.05 at 1Re   \\ 
199 & flux\_HeI3964.73\_Re\_fit &  10$^{-16}$ \flux  &  Flux intensity of HeI3964.73 at 1Re \\ 
200 & e\_flux\_HeI3964.73\_Re\_fit &  10$^{-16}$ \flux  &  Error in the flux intensity of HeI3964.73 at 1Re   \\ 
201 & flux\_HeI3964.73\_alpha\_fit &  10$^{-16}$ \flux  &  Flux intensity of HeI3964.73 at 1Re \\ 
202 & e\_flux\_HeI3964.73\_alpha\_fit &  10$^{-16}$ \flux  &  Error in the flux intensity of HeI3964.73 at 1Re   \\ 
203 & flux\_[NeIII]3967.46\_Re\_fit &  10$^{-16}$ \flux  &  Flux intensity of [NeIII]3967.46 at 1Re \\ 
204 & e\_flux\_[NeIII]3967.46\_Re\_fit &  10$^{-16}$ \flux  &  Error in the flux intensity of [NeIII]3967.46 at 1Re   \\ 
205 & flux\_[NeIII]3967.46\_alpha\_fit &  10$^{-16}$ \flux  &  Flux intensity of [NeIII]3967.46 at 1Re \\ 
206 & e\_flux\_[NeIII]3967.46\_alpha\_fit &  10$^{-16}$ \flux  &  Error in the flux intensity of [NeIII]3967.46 at 1Re   \\ 
207 & flux\_CaII3968.47\_Re\_fit &  10$^{-16}$ \flux  &  Flux intensity of CaII3968.47 at 1Re \\ 
208 & e\_flux\_CaII3968.47\_Re\_fit &  10$^{-16}$ \flux  &  Error in the flux intensity of CaII3968.47 at 1Re   \\ 
209 & flux\_CaII3968.47\_alpha\_fit &  10$^{-16}$ \flux  &  Flux intensity of CaII3968.47 at 1Re \\ 
\hline 
\end{tabular}\label{tab:cat}
\end{center}
\end{table*}

\addtocounter{table}{-1}

\begin{table*}
\begin{center}
\caption{Integrated and characteristic parameters delivered for each analyzed datacube.}
\begin{tabular}{llll}
\hline\hline
\#  & parameter name & units & Description  \\ 
\hline
210 & e\_flux\_CaII3968.47\_alpha\_fit &  10$^{-16}$ \flux  &  Error in the flux intensity of CaII3968.47 at 1Re   \\ 
211 & flux\_Hepsilon3970.07\_Re\_fit &  10$^{-16}$ \flux  &  Flux intensity of H$\epsilon$3970.07 at 1Re \\ 
212 & e\_flux\_Hepsilon3970.07\_Re\_fit &  10$^{-16}$ \flux  &  Error in the flux intensity of H$\epsilon$3970.07 at 1Re   \\ 
213 & flux\_Hepsilon3970.07\_alpha\_fit &  10$^{-16}$ \flux  &  Flux intensity of H$\epsilon$3970.07 at 1Re \\ 
214 & e\_flux\_Hepsilon3970.07\_alpha\_fit &  10$^{-16}$ \flux  &  Error in the flux intensity of H$\epsilon$3970.07 at 1Re   \\ 
215 & flux\_Hdelta4101.77\_Re\_fit &  10$^{-16}$ \flux  &  Flux intensity of H$\delta$4101.77 at 1Re \\ 
216 & e\_flux\_Hdelta4101.77\_Re\_fit &  10$^{-16}$ \flux  &  Error in the flux intensity of H$\delta$4101.77 at 1Re   \\ 
217 & flux\_Hdelta4101.77\_alpha\_fit &  10$^{-16}$ \flux  &  Flux intensity of H$\delta$4101.77 at 1Re \\ 
218 & e\_flux\_Hdelta4101.77\_alpha\_fit &  10$^{-16}$ \flux  &  Error in the flux intensity of H$\delta$4101.77 at 1Re   \\ 
219 & flux\_Hgamma4340.49\_Re\_fit &  10$^{-16}$ \flux  &  Flux intensity of H$\gamma$4340.49 at 1Re \\ 
220 & e\_flux\_Hgamma4340.49\_Re\_fit &  10$^{-16}$ \flux  &  Error in the flux intensity of H$\gamma$4340.49 at 1Re   \\ 
221 & flux\_Hgamma4340.49\_alpha\_fit &  10$^{-16}$ \flux  &  Flux intensity of H$\gamma$4340.49 at 1Re \\ 
222 & e\_flux\_Hgamma4340.49\_alpha\_fit &  10$^{-16}$ \flux  &  Error in the flux intensity of H$\gamma$4340.49 at 1Re   \\ 
223 & flux\_Hbeta4861.36\_Re\_fit &  10$^{-16}$ \flux  &  Flux intensity of H$\beta$4861.36 at 1Re \\ 
224 & e\_flux\_Hbeta4861.36\_Re\_fit &  10$^{-16}$ \flux  &  Error in the flux intensity of H$\beta$4861.36 at 1Re   \\ 
225 & flux\_Hbeta4861.36\_alpha\_fit &  10$^{-16}$ \flux  &  Flux intensity of H$\beta$4861.36 at 1Re \\ 
226 & e\_flux\_Hbeta4861.36\_alpha\_fit &  10$^{-16}$ \flux  &  Error in the flux intensity of H$\beta$4861.36 at 1Re   \\ 
227 & flux\_[OIII]4958.91\_Re\_fit &  10$^{-16}$ \flux  &  Flux intensity of [OIII]4958.91 at 1Re \\ 
228 & e\_flux\_[OIII]4958.91\_Re\_fit &  10$^{-16}$ \flux  &  Error in the flux intensity of [OIII]4958.91 at 1Re   \\ 
229 & flux\_[OIII]4958.91\_alpha\_fit &  10$^{-16}$ \flux  &  Flux intensity of [OIII]4958.91 at 1Re \\ 
230 & e\_flux\_[OIII]4958.91\_alpha\_fit &  10$^{-16}$ \flux  &  Error in the flux intensity of [OIII]4958.91 at 1Re   \\ 
231 & flux\_[OIII]5006.84\_Re\_fit &  10$^{-16}$ \flux  &  Flux intensity of [OIII]5006.84 at 1Re \\ 
232 & e\_flux\_[OIII]5006.84\_Re\_fit &  10$^{-16}$ \flux  &  Error in the flux intensity of [OIII]5006.84 at 1Re   \\ 
233 & flux\_[OIII]5006.84\_alpha\_fit &  10$^{-16}$ \flux  &  Flux intensity of [OIII]5006.84 at 1Re \\ 
234 & e\_flux\_[OIII]5006.84\_alpha\_fit &  10$^{-16}$ \flux  &  Error in the flux intensity of [OIII]5006.84 at 1Re   \\ 
235 & flux\_HeI5015.68\_Re\_fit &  10$^{-16}$ \flux  &  Flux intensity of HeI5015.68 at 1Re \\ 
236 & e\_flux\_HeI5015.68\_Re\_fit &  10$^{-16}$ \flux  &  Error in the flux intensity of HeI5015.68 at 1Re   \\ 
237 & flux\_HeI5015.68\_alpha\_fit &  10$^{-16}$ \flux  &  Flux intensity of HeI5015.68 at 1Re \\ 
238 & e\_flux\_HeI5015.68\_alpha\_fit &  10$^{-16}$ \flux  &  Error in the flux intensity of HeI5015.68 at 1Re   \\ 
239 & flux\_[NI]5197.9\_Re\_fit &  10$^{-16}$ \flux  &  Flux intensity of [NI]5197.9 at 1Re \\ 
240 & e\_flux\_[NI]5197.9\_Re\_fit &  10$^{-16}$ \flux  &  Error in the flux intensity of [NI]5197.9 at 1Re   \\ 
241 & flux\_[NI]5197.9\_alpha\_fit &  10$^{-16}$ \flux  &  Flux intensity of [NI]5197.9 at 1Re \\ 
242 & e\_flux\_[NI]5197.9\_alpha\_fit &  10$^{-16}$ \flux  &  Error in the flux intensity of [NI]5197.9 at 1Re   \\ 
243 & flux\_[NI]5200.26\_Re\_fit &  10$^{-16}$ \flux  &  Flux intensity of [NI]5200.26 at 1Re \\ 
244 & e\_flux\_[NI]5200.26\_Re\_fit &  10$^{-16}$ \flux  &  Error in the flux intensity of [NI]5200.26 at 1Re   \\ 
\hline 
\end{tabular}\label{tab:cat}
\end{center}
\end{table*}

\addtocounter{table}{-1}

\begin{table*}
\begin{center}
\caption{Integrated and characteristic parameters delivered for each analyzed datacube.}
\begin{tabular}{llll}
\hline\hline
\#  & parameter name & units & Description  \\ 
\hline
245 & flux\_[NI]5200.26\_alpha\_fit &  10$^{-16}$ \flux  &  Flux intensity of [NI]5200.26 at 1Re \\ 
246 & e\_flux\_[NI]5200.26\_alpha\_fit &  10$^{-16}$ \flux  &  Error in the flux intensity of [NI]5200.26 at 1Re   \\ 
247 & flux\_HeI5876.0\_Re\_fit &  10$^{-16}$ \flux  &  Flux intensity of HeI5876.0 at 1Re \\ 
248 & e\_flux\_HeI5876.0\_Re\_fit &  10$^{-16}$ \flux  &  Error in the flux intensity of HeI5876.0 at 1Re   \\ 
249 & flux\_HeI5876.0\_alpha\_fit &  10$^{-16}$ \flux  &  Flux intensity of HeI5876.0 at 1Re \\ 
250 & e\_flux\_HeI5876.0\_alpha\_fit &  10$^{-16}$ \flux  &  Error in the flux intensity of HeI5876.0 at 1Re   \\ 
251 & flux\_NaI5889.95\_Re\_fit &  10$^{-16}$ \flux  &  Flux intensity of NaI5889.95 at 1Re \\ 
252 & e\_flux\_NaI5889.95\_Re\_fit &  10$^{-16}$ \flux  &  Error in the flux intensity of NaI5889.95 at 1Re   \\ 
253 & flux\_NaI5889.95\_alpha\_fit &  10$^{-16}$ \flux  &  Flux intensity of NaI5889.95 at 1Re \\ 
254 & e\_flux\_NaI5889.95\_alpha\_fit &  10$^{-16}$ \flux  &  Error in the flux intensity of NaI5889.95 at 1Re   \\ 
255 & flux\_NaI5895.92\_Re\_fit &  10$^{-16}$ \flux  &  Flux intensity of NaI5895.92 at 1Re \\ 
256 & e\_flux\_NaI5895.92\_Re\_fit &  10$^{-16}$ \flux  &  Error in the flux intensity of NaI5895.92 at 1Re   \\ 
257 & flux\_NaI5895.92\_alpha\_fit &  10$^{-16}$ \flux  &  Flux intensity of NaI5895.92 at 1Re \\ 
258 & e\_flux\_NaI5895.92\_alpha\_fit &  10$^{-16}$ \flux  &  Error in the flux intensity of NaI5895.92 at 1Re   \\ 
259 & flux\_[OI]6300.3\_Re\_fit &  10$^{-16}$ \flux  &  Flux intensity of [OI]6300.3 at 1Re \\ 
260 & e\_flux\_[OI]6300.3\_Re\_fit &  10$^{-16}$ \flux  &  Error in the flux intensity of [OI]6300.3 at 1Re   \\ 
261 & flux\_[OI]6300.3\_alpha\_fit &  10$^{-16}$ \flux  &  Flux intensity of [OI]6300.3 at 1Re \\ 
262 & e\_flux\_[OI]6300.3\_alpha\_fit &  10$^{-16}$ \flux  &  Error in the flux intensity of [OI]6300.3 at 1Re   \\ 
263 & flux\_[NII]6548.05\_Re\_fit &  10$^{-16}$ \flux  &  Flux intensity of [NII]6548.05 at 1Re \\ 
264 & e\_flux\_[NII]6548.05\_Re\_fit &  10$^{-16}$ \flux  &  Error in the flux intensity of [NII]6548.05 at 1Re   \\ 
265 & flux\_[NII]6548.05\_alpha\_fit &  10$^{-16}$ \flux  &  Flux intensity of [NII]6548.05 at 1Re \\ 
266 & e\_flux\_[NII]6548.05\_alpha\_fit &  10$^{-16}$ \flux  &  Error in the flux intensity of [NII]6548.05 at 1Re   \\ 
267 & flux\_Halpha6562.85\_Re\_fit &  10$^{-16}$ \flux  &  Flux intensity of H$\alpha$6562.85 at 1Re \\ 
268 & e\_flux\_Halpha6562.85\_Re\_fit &  10$^{-16}$ \flux  &  Error in the flux intensity of H$\alpha$6562.85 at 1Re   \\ 
269 & flux\_Halpha6562.85\_alpha\_fit &  10$^{-16}$ \flux  &  Flux intensity of H$\alpha$6562.85 at 1Re \\ 
270 & e\_flux\_Halpha6562.85\_alpha\_fit &  10$^{-16}$ \flux  &  Error in the flux intensity of H$\alpha$6562.85 at 1Re   \\ 
271 & flux\_[NII]6583.45\_Re\_fit &  10$^{-16}$ \flux  &  Flux intensity of [NII]6583.45 at 1Re \\ 
272 & e\_flux\_[NII]6583.45\_Re\_fit &  10$^{-16}$ \flux  &  Error in the flux intensity of [NII]6583.45 at 1Re   \\ 
273 & flux\_[NII]6583.45\_alpha\_fit &  10$^{-16}$ \flux  &  Flux intensity of [NII]6583.45 at 1Re \\ 
274 & e\_flux\_[NII]6583.45\_alpha\_fit &  10$^{-16}$ \flux  &  Error in the flux intensity of [NII]6583.45 at 1Re   \\ 
275 & flux\_[SII]6716.44\_Re\_fit &  10$^{-16}$ \flux  &  Flux intensity of [SII]6716.44 at 1Re \\ 
276 & e\_flux\_[SII]6716.44\_Re\_fit &  10$^{-16}$ \flux  &  Error in the flux intensity of [SII]6716.44 at 1Re   \\ 
277 & flux\_[SII]6716.44\_alpha\_fit &  10$^{-16}$ \flux  &  Flux intensity of [SII]6716.44 at 1Re \\ 
278 & e\_flux\_[SII]6716.44\_alpha\_fit &  10$^{-16}$ \flux  &  Error in the flux intensity of [SII]6716.44 at 1Re   \\ 
279 & flux\_[SII]6730.82\_Re\_fit &  10$^{-16}$ \flux  &  Flux intensity of [SII]6730.82 at 1Re \\ 
\hline 
\end{tabular}\label{tab:cat}
\end{center}
\end{table*}

\addtocounter{table}{-1}

\begin{table*}
\begin{center}
\caption{Integrated and characteristic parameters delivered for each analyzed datacube.}
\begin{tabular}{llll}
\hline\hline
\#  & parameter name & units & Description  \\ 
\hline
280 & e\_flux\_[SII]6730.82\_Re\_fit &  10$^{-16}$ \flux  &  Error in the flux intensity of [SII]6730.82 at 1Re   \\ 
281 & flux\_[SII]6730.82\_alpha\_fit &  10$^{-16}$ \flux  &  Flux intensity of [SII]6730.82 at 1Re \\ 
282 & e\_flux\_[SII]6730.82\_alpha\_fit &  10$^{-16}$ \flux  &  Error in the flux intensity of [SII]6730.82 at 1Re   \\ 
283 & flux\_[ArIII]7135.8\_Re\_fit &  10$^{-16}$ \flux  &  Flux intensity of [ArIII]7135.8 at 1Re \\ 
284 & e\_flux\_[ArIII]7135.8\_Re\_fit &  10$^{-16}$ \flux  &  Error in the flux intensity of [ArIII]7135.8 at 1Re   \\ 
285 & flux\_[ArIII]7135.8\_alpha\_fit &  10$^{-16}$ \flux  &  Flux intensity of [ArIII]7135.8 at 1Re \\ 
286 & e\_flux\_[ArIII]7135.8\_alpha\_fit &  10$^{-16}$ \flux  &  Error in the flux intensity of [ArIII]7135.8 at 1Re   \\ 
287 & OH\_Mar13\_N2\_Re\_fit &  dex  &  Oxygen abundance using the calibrator Mar13\_N2 at 1Re \\ 
288 & e\_OH\_Mar13\_N2\_Re\_fit &  dex  &  Error in Oxygen abundance using the calibrator Mar13\_N2 at 1Re   \\ 
289 & OH\_Mar13\_N2\_alpha\_fit &  dex/Re  &  Slope of the O/H gradient using the calibrator Mar13\_N2 \\ 
290 & e\_OH\_Mar13\_N2\_alpha\_fit &  dex/Re  &  Error in the slope of the O/H gradient using the calibrator Mar13\_N2   \\ 
291 & OH\_Mar13\_O3N2\_Re\_fit &  dex  &  Oxygen abundance using the calibrator Mar13\_O3N2 at 1Re \\ 
292 & e\_OH\_Mar13\_O3N2\_Re\_fit &  dex  &  Error in Oxygen abundance using the calibrator Mar13\_O3N2 at 1Re   \\ 
293 & OH\_Mar13\_O3N2\_alpha\_fit &  dex/Re  &  Slope of the O/H gradient using the calibrator Mar13\_O3N2 \\ 
294 & e\_OH\_Mar13\_O3N2\_alpha\_fit &  dex/Re  &  Error in the slope of the O/H gradient using the calibrator Mar13\_O3N2   \\ 
295 & OH\_T04\_Re\_fit &  dex  &  Oxygen abundance using the calibrator T04 at 1Re \\ 
296 & e\_OH\_T04\_Re\_fit &  dex  &  Error in Oxygen abundance using the calibrator T04 at 1Re   \\ 
297 & OH\_T04\_alpha\_fit &  dex/Re  &  Slope of the O/H gradient using the calibrator T04 \\ 
298 & e\_OH\_T04\_alpha\_fit &  dex/Re  &  Error in the slope of the O/H gradient using the calibrator T04   \\ 
299 & OH\_Pet04\_N2\_lin\_Re\_fit &  dex  &  Oxygen abundance using the calibrator Pet04\_N2\_lin at 1Re \\ 
300 & e\_OH\_Pet04\_N2\_lin\_Re\_fit &  dex  &  Error in Oxygen abundance using the calibrator Pet04\_N2\_lin at 1Re   \\ 
301 & OH\_Pet04\_N2\_lin\_alpha\_fit &  dex/Re  &  Slope of the O/H gradient using the calibrator Pet04\_N2\_lin \\ 
302 & e\_OH\_Pet04\_N2\_lin\_alpha\_fit &  dex/Re  &  Error in the slope of the O/H gradient using the calibrator Pet04\_N2\_lin   \\ 
303 & OH\_Pet04\_N2\_poly\_Re\_fit &  dex  &  Oxygen abundance using the calibrator Pet04\_N2\_poly at 1Re \\ 
304 & e\_OH\_Pet04\_N2\_poly\_Re\_fit &  dex  &  Error in Oxygen abundance using the calibrator Pet04\_N2\_poly at 1Re   \\ 
305 & OH\_Pet04\_N2\_poly\_alpha\_fit &  dex/Re  &  Slope of the O/H gradient using the calibrator Pet04\_N2\_poly \\ 
306 & e\_OH\_Pet04\_N2\_poly\_alpha\_fit &  dex/Re  &  Error in the slope of the O/H gradient using the calibrator Pet04\_N2\_poly   \\ 
307 & OH\_Pet04\_O3N2\_Re\_fit &  dex  &  Oxygen abundance using the calibrator Pet04\_O3N2 at 1Re \\ 
308 & e\_OH\_Pet04\_O3N2\_Re\_fit &  dex  &  Error in Oxygen abundance using the calibrator Pet04\_O3N2 at 1Re   \\ 
309 & OH\_Pet04\_O3N2\_alpha\_fit &  dex/Re  &  Slope of the O/H gradient using the calibrator Pet04\_O3N2 \\ 
310 & e\_OH\_Pet04\_O3N2\_alpha\_fit &  dex/Re  &  Error in the slope of the O/H gradient using the calibrator Pet04\_O3N2   \\ 
311 & OH\_Kew02\_N2O2\_Re\_fit &  dex  &  Oxygen abundance using the calibrator Kew02\_N2O2 at 1Re \\ 
312 & e\_OH\_Kew02\_N2O2\_Re\_fit &  dex  &  Error in Oxygen abundance using the calibrator Kew02\_N2O2 at 1Re   \\ 
313 & OH\_Kew02\_N2O2\_alpha\_fit &  dex/Re  &  Slope of the O/H gradient using the calibrator Kew02\_N2O2 \\ 
314 & e\_OH\_Kew02\_N2O2\_alpha\_fit &  dex/Re  &  Error in the slope of the O/H gradient using the calibrator Kew02\_N2O2   \\ 
\hline 
\end{tabular}\label{tab:cat}
\end{center}
\end{table*}

\addtocounter{table}{-1}

\begin{table*}
\begin{center}
\caption{Integrated and characteristic parameters delivered for each analyzed datacube.}
\begin{tabular}{llll}
\hline\hline
\#  & parameter name & units & Description  \\ 
\hline
315 & OH\_Pil10\_ONS\_Re\_fit &  dex  &  Oxygen abundance using the calibrator Pil10\_ONS at 1Re \\ 
316 & e\_OH\_Pil10\_ONS\_Re\_fit &  dex  &  Error in Oxygen abundance using the calibrator Pil10\_ONS at 1Re   \\ 
317 & OH\_Pil10\_ONS\_alpha\_fit &  dex/Re  &  Slope of the O/H gradient using the calibrator Pil10\_ONS \\ 
318 & e\_OH\_Pil10\_ONS\_alpha\_fit &  dex/Re  &  Error in the slope of the O/H gradient using the calibrator Pil10\_ONS   \\ 
319 & OH\_Pil10\_ON\_Re\_fit &  dex  &  Oxygen abundance using the calibrator Pil10\_ON at 1Re \\ 
320 & e\_OH\_Pil10\_ON\_Re\_fit &  dex  &  Error in Oxygen abundance using the calibrator Pil10\_ON at 1Re   \\ 
321 & OH\_Pil10\_ON\_alpha\_fit &  dex/Re  &  Slope of the O/H gradient using the calibrator Pil10\_ON \\ 
322 & e\_OH\_Pil10\_ON\_alpha\_fit &  dex/Re  &  Error in the slope of the O/H gradient using the calibrator Pil10\_ON   \\ 
323 & OH\_Pil11\_NS\_Re\_fit &  dex  &  Oxygen abundance using the calibrator Pil11\_NS at 1Re \\ 
324 & e\_OH\_Pil11\_NS\_Re\_fit &  dex  &  Error in Oxygen abundance using the calibrator Pil11\_NS at 1Re   \\ 
325 & OH\_Pil11\_NS\_alpha\_fit &  dex/Re  &  Slope of the O/H gradient using the calibrator Pil11\_NS \\ 
326 & e\_OH\_Pil11\_NS\_alpha\_fit &  dex/Re  &  Error in the slope of the O/H gradient using the calibrator Pil11\_NS   \\ 
327 & OH\_Cur20\_RS32\_Re\_fit &  dex  &  Oxygen abundance using the calibrator Cur20\_RS32 at 1Re \\ 
328 & e\_OH\_Cur20\_RS32\_Re\_fit &  dex  &  Error in Oxygen abundance using the calibrator Cur20\_RS32 at 1Re   \\ 
329 & OH\_Cur20\_RS32\_alpha\_fit &  dex/Re  &  Slope of the O/H gradient using the calibrator Cur20\_RS32 \\ 
330 & e\_OH\_Cur20\_RS32\_alpha\_fit &  dex/Re  &  Error in the slope of the O/H gradient using the calibrator Cur20\_RS32   \\ 
331 & OH\_Cur20\_R3\_Re\_fit &  dex  &  Oxygen abundance using the calibrator Cur20\_R3 at 1Re \\ 
332 & e\_OH\_Cur20\_R3\_Re\_fit &  dex  &  Error in Oxygen abundance using the calibrator Cur20\_R3 at 1Re   \\ 
333 & OH\_Cur20\_R3\_alpha\_fit &  dex/Re  &  Slope of the O/H gradient using the calibrator Cur20\_R3 \\ 
334 & e\_OH\_Cur20\_R3\_alpha\_fit &  dex/Re  &  Error in the slope of the O/H gradient using the calibrator Cur20\_R3   \\ 
335 & OH\_Cur20\_O3O2\_Re\_fit &  dex  &  Oxygen abundance using the calibrator Cur20\_O3O2 at 1Re \\ 
336 & e\_OH\_Cur20\_O3O2\_Re\_fit &  dex  &  Error in Oxygen abundance using the calibrator Cur20\_O3O2 at 1Re   \\ 
337 & OH\_Cur20\_O3O2\_alpha\_fit &  dex/Re  &  Slope of the O/H gradient using the calibrator Cur20\_O3O2 \\ 
338 & e\_OH\_Cur20\_O3O2\_alpha\_fit &  dex/Re  &  Error in the slope of the O/H gradient using the calibrator Cur20\_O3O2   \\ 
339 & OH\_Cur20\_S2\_Re\_fit &  dex  &  Oxygen abundance using the calibrator Cur20\_S2 at 1Re \\ 
340 & e\_OH\_Cur20\_S2\_Re\_fit &  dex  &  Error in Oxygen abundance using the calibrator Cur20\_S2 at 1Re   \\ 
341 & OH\_Cur20\_S2\_alpha\_fit &  dex/Re  &  Slope of the O/H gradient using the calibrator Cur20\_S2 \\ 
342 & e\_OH\_Cur20\_S2\_alpha\_fit &  dex/Re  &  Error in the slope of the O/H gradient using the calibrator Cur20\_S2   \\ 
343 & OH\_Cur20\_R2\_Re\_fit &  dex  &  Oxygen abundance using the calibrator Cur20\_R2 at 1Re \\ 
344 & e\_OH\_Cur20\_R2\_Re\_fit &  dex  &  Error in Oxygen abundance using the calibrator Cur20\_R2 at 1Re   \\ 
345 & OH\_Cur20\_R2\_alpha\_fit &  dex/Re  &  Slope of the O/H gradient using the calibrator Cur20\_R2 \\ 
346 & e\_OH\_Cur20\_R2\_alpha\_fit &  dex/Re  &  Error in the slope of the O/H gradient using the calibrator Cur20\_R2   \\ 
347 & OH\_Cur20\_N2\_Re\_fit &  dex  &  Oxygen abundance using the calibrator Cur20\_N2 at 1Re \\ 
348 & e\_OH\_Cur20\_N2\_Re\_fit &  dex  &  Error in Oxygen abundance using the calibrator Cur20\_N2 at 1Re   \\ 
349 & OH\_Cur20\_N2\_alpha\_fit &  dex/Re  &  Slope of the O/H gradient using the calibrator Cur20\_N2 \\ 
\hline 
\end{tabular}\label{tab:cat}
\end{center}
\end{table*}

\addtocounter{table}{-1}

\begin{table*}
\begin{center}
\caption{Integrated and characteristic parameters delivered for each analyzed datacube.}
\begin{tabular}{llll}
\hline\hline
\#  & parameter name & units & Description  \\ 
\hline
350 & e\_OH\_Cur20\_N2\_alpha\_fit &  dex/Re  &  Error in the slope of the O/H gradient using the calibrator Cur20\_N2   \\ 
351 & OH\_Cur20\_R23\_Re\_fit &  dex  &  Oxygen abundance using the calibrator Cur20\_R23 at 1Re \\ 
352 & e\_OH\_Cur20\_R23\_Re\_fit &  dex  &  Error in Oxygen abundance using the calibrator Cur20\_R23 at 1Re  374  \\ 
353 & OH\_Cur20\_R23\_alpha\_fit &  dex  &  Slope of the O/H gradient using the calibrator Cur20\_R23 \\ 
354 & e\_OH\_Cur20\_R23\_alpha\_fit &  dex/Re  &  Error in the slope of the O/H gradient using the calibrator Cur20\_R23   \\ 
355 & OH\_Cur20\_O3N2\_Re\_fit &  dex  &  Oxygen abundance using the calibrator Cur20\_O3N2 at 1Re \\ 
356 & e\_OH\_Cur20\_O3N2\_Re\_fit &  dex  &  Error in Oxygen abundance using the calibrator Cur20\_O3N2 at 1Re   \\ 
357 & OH\_Cur20\_O3N2\_alpha\_fit &  dex/Re  &  Slope of the O/H gradient using the calibrator Cur20\_O3N2 \\ 
358 & e\_OH\_Cur20\_O3N2\_alpha\_fit &  dex/Re  &  Error in the slope of the O/H gradient using the calibrator Cur20\_O3N2   \\ 
359 & OH\_Cur20\_O3S2\_Re\_fit &  dex  &  Oxygen abundance using the calibrator Cur20\_O3S2 at 1Re \\ 
360 & e\_OH\_Cur20\_O3S2\_Re\_fit &  dex  &  Error in Oxygen abundance using the calibrator Cur20\_O3S2 at 1Re   \\ 
361 & OH\_Cur20\_O3S2\_alpha\_fit &  dex/Re  &  Slope of the O/H gradient using the calibrator Cur20\_O3S2 \\ 
362 & e\_OH\_Cur20\_O3S2\_alpha\_fit &  dex/Re  &  Error in the slope of the O/H gradient using the calibrator Cur20\_O3S2   \\ 
363 & OH\_KK04\_Re\_fit &  dex  &  Oxygen abundance using the calibrator KK04 at 1Re \\ 
364 & e\_OH\_KK04\_Re\_fit &  dex  &  Error in Oxygen abundance using the calibrator KK04 at 1Re   \\ 
365 & OH\_KK04\_alpha\_fit &  dex/Re  &  Slope of the O/H gradient using the calibrator KK04 \\ 
366 & e\_OH\_KK04\_alpha\_fit &  dex/Re  &  Error in the slope of the O/H gradient using the calibrator KK04   \\ 
367 & OH\_Pil16\_R\_Re\_fit &  dex  &  Oxygen abundance using the calibrator Pil16\_R at 1Re \\ 
368 & e\_OH\_Pil16\_R\_Re\_fit &  dex  &  Error in Oxygen abundance using the calibrator Pil16\_R at 1Re   \\ 
369 & OH\_Pil16\_R\_alpha\_fit &  dex/Re  &  Slope of the O/H gradient using the calibrator Pil16\_R \\ 
370 & e\_OH\_Pil16\_R\_alpha\_fit &  dex/Re  &  Error in the slope of the O/H gradient using the calibrator Pil16\_R   \\ 
371 & OH\_Pil16\_S\_Re\_fit &  dex  &  Oxygen abundance using the calibrator Pil16\_S at 1Re \\ 
372 & e\_OH\_Pil16\_S\_Re\_fit &  dex  &  Error in Oxygen abundance using the calibrator Pil16\_S at 1Re   \\ 
373 & OH\_Pil16\_S\_alpha\_fit &  dex/Re  &  Slope of the O/H gradient using the calibrator Pil16\_S \\ 
374 & e\_OH\_Pil16\_S\_alpha\_fit &  dex/Re  &  Error in the slope of the O/H gradient using the calibrator Pil16\_S   \\ 
375 & OH\_Ho\_Re\_fit &  dex  &  Oxygen abundance using the calibrator Ho at 1Re \\ 
376 & e\_OH\_Ho\_Re\_fit &  dex  &  Error in Oxygen abundance using the calibrator Ho at 1Re   \\ 
377 & OH\_Ho\_alpha\_fit &  dex/Re  &  Slope of the O/H gradient using the calibrator Ho \\ 
378 & e\_OH\_Ho\_alpha\_fit &  dex/Re  &  Error in the slope of the O/H gradient using the calibrator Ho   \\ 
379 & U\_Dors\_O32\_Re\_fit &  dex  &   log(U), ionization parameter using the calibrator Dors\_O32 at 1Re \\ 
380 & e\_U\_Dors\_O32\_Re\_fit &  dex  &  Error in log(U) using the calibrator Dors\_O32 at 1Re   \\ 
381 & U\_Dors\_O32\_alpha\_fit &  dex/Re  &  Slope of the log(U) gradient using the calibrator Dors\_O32 \\ 
382 & e\_U\_Dors\_O32\_alpha\_fit &  dex/Re  &  Error in the slope of the log(U) gradient using the calibrator Dors\_O32   \\ 
383 & U\_Dors\_S\_Re\_fit &  dex  &  log(U) ionization parameter using the calibrator Dors\_S at 1Re \\ 
384 & e\_U\_Dors\_S\_Re\_fit &  dex  &  Error in log(U) using the calibrator Dors\_S at 1Re   \\ 
\hline 
\end{tabular}\label{tab:cat}
\end{center}
\end{table*}

\addtocounter{table}{-1}

\begin{table*}
\begin{center}
\caption{Integrated and characteristic parameters delivered for each analyzed datacube.}
\begin{tabular}{llll}
\hline\hline
\#  & parameter name & units & Description  \\ 
\hline
385 & U\_Dors\_S\_alpha\_fit &  dex/Re  &  Slope of the log(U) gradient using the calibrator Dors\_S \\ 
386 & e\_U\_Dors\_S\_alpha\_fit &  dex/Re  &  Error in the slope of the log(U) gradient using the calibrator Dors\_S   \\ 
387 & U\_Mor16\_O23\_fs\_Re\_fit &  dex  &   log(U) ionization parameter using the calibrator Mor16\_O23\_fs at 1Re \\ 
388 & e\_U\_Mor16\_O23\_fs\_Re\_fit &  dex  &  Error in log(U) using the calibrator Mor16\_O23\_fs at 1Re   \\ 
389 & U\_Mor16\_O23\_fs\_alpha\_fit &  dex/Re  &  Slope of the log(U) gradient using the calibrator Mor16\_O23\_fs \\ 
390 & e\_U\_Mor16\_O23\_fs\_alpha\_fit &  dex/Re  &  Error in the slope of the log(U) gradient using the calibrator Mor16\_O23\_fs   \\ 
391 & U\_Mor16\_O23\_ts\_Re\_fit &  dex  &   log(U) ionization parameter using the calibrator Mor16\_O23\_ts at 1Re \\ 
392 & e\_U\_Mor16\_O23\_ts\_Re\_fit &  dex  &  Error in log(U) using the calibrator Mor16\_O23\_ts at 1Re   \\ 
393 & U\_Mor16\_O23\_ts\_alpha\_fit &  dex/Re  &  Slope of the log(U) gradient using the calibrator Mor16\_O23\_ts \\ 
394 & e\_U\_Mor16\_O23\_ts\_alpha\_fit &  dex/Re  &  Error in the slope of the log(U) gradient using the calibrator Mor16\_O23\_ts   \\ 
395 & NH\_Pil16\_R\_Re\_fit &  dex  &  Nitrogen abundance using the calibrator Pil16\_R at 1Re \\ 
396 & e\_NH\_Pil16\_R\_Re\_fit &  dex  &  Error in Nitrogen abundance using the calibrator Pil16\_R at 1Re   \\ 
397 & NH\_Pil16\_R\_alpha\_fit &  dex/Re  &  Slope of the N/H gradient using the calibrator Pil16\_R \\ 
398 & e\_NH\_Pil16\_R\_alpha\_fit &  dex/Re  &  Error in the slope of the N/H gradient using the calibrator Pil16\_R   \\ 
399 & NO\_Pil16\_R\_Re\_fit &  dex  &  N/O abundance using the calibrator Pil16\_R at 1Re \\ 
400 & e\_NO\_Pil16\_R\_Re\_fit &  dex  &  Error in N/O abundance using the calibrator Pil16\_R at 1Re   \\ 
401 & NO\_Pil16\_R\_alpha\_fit &  dex/Re  &  Slope of the N/O gradient using the calibrator Pil16\_R \\ 
402 & e\_NO\_Pil16\_R\_alpha\_fit &  dex/Re  &  Error in the slope of the N/O gradient using the calibrator Pil16\_R   \\ 
403 & NO\_Pil16\_Ho\_R\_Re\_fit &  dex  &  N/O abundance using the calibrator Pil16\_Ho\_R at 1Re \\ 
404 & e\_NO\_Pil16\_Ho\_R\_Re\_fit &  dex  &  Error in N/O abundance using the calibrator Pil16\_Ho\_R at 1Re   \\ 
405 & NO\_Pil16\_Ho\_R\_alpha\_fit &  dex/Re  &  Slope of the N/O gradient using the calibrator Pil16\_Ho\_R \\ 
406 & e\_NO\_Pil16\_Ho\_R\_alpha\_fit &  dex/Re  &  Error in the slope of the N/O gradient using the calibrator Pil16\_Ho\_R   \\ 
407 & NO\_Pil16\_N2\_R2\_Re\_fit &  dex  &  N/O abundance using the calibrator Pil16\_N2\_R2 at 1Re \\ 
408 & e\_NO\_Pil16\_N2\_R2\_Re\_fit &  dex  &  Error in N/O abundance using the calibrator Pil16\_N2\_R2 at 1Re   \\ 
409 & NO\_Pil16\_N2\_R2\_alpha\_fit &  dex/Re  &  Slope of the N/O gradient using the calibrator Pil16\_N2\_R2 \\ 
410 & e\_NO\_Pil16\_N2\_R2\_alpha\_fit &  dex/Re  &  Error in the slope of the N/O gradient using the calibrator Pil16\_N2\_R2   \\ 
411 & Ne\_Oster\_S\_Re\_fit &  dex  &  electron density using the Oster\_S estimator at 1Re \\ 
412 & e\_Ne\_Oster\_S\_Re\_fit &  dex  &  Error in n\_e using the Oster\_S estimator at 1Re   \\ 
413 & Ne\_Oster\_S\_alpha\_fit &  dex/Re  &  Slope of the n\_e gradient using the Oster\_S estimator \\ 
414 & e\_Ne\_Oster\_S\_alpha\_fit &  dex/Re  &  Error in the slope of n\_e gradient using the Oster\_S estimator   \\ 
415 & C24668\_Re\_fit &  Angstrom  &  Value of the C24668 stellar index at 1Re \\ 
416 & C24668\_alpha\_fit &  --  &  Slope of the gradient of the C24668 index \\ 
417 & CN1\_Re\_fit &  Angstrom  &  Value of the CN1 stellar index at 1Re \\ 
418 & CN1\_alpha\_fit &  --  &  Slope of the gradient of the CN1 index \\ 
419 & CN2\_Re\_fit &  Angstrom  &  Value of the CN2 stellar index at 1Re \\ 
\hline 
\end{tabular}\label{tab:cat}
\end{center}
\end{table*}

\addtocounter{table}{-1}

\begin{table*}
\begin{center}
\caption{Integrated and characteristic parameters delivered for each analyzed datacube.}
\begin{tabular}{llll}
\hline\hline
\#  & parameter name & units & Description  \\ 
\hline
420 & CN2\_alpha\_fit &  --  &  Slope of the gradient of the CN2 index \\ 
421 & Ca4227\_Re\_fit &  Angstrom  &  Value of the Ca4227 stellar index at 1Re \\ 
422 & Ca4227\_alpha\_fit &  --  &  Slope of the gradient of the Ca4227 index \\ 
423 & Ca4455\_Re\_fit &  Angstrom  &  Value of the Ca4455 stellar index at 1Re \\ 
424 & Ca4455\_alpha\_fit &  --  &  Slope of the gradient of the Ca4455 index \\ 
425 & CaH1\_Re\_fit &  Angstrom  &  Value of the CaH1 stellar index at 1Re \\ 
426 & CaH1\_alpha\_fit &  --  &  Slope of the gradient of the CaH1 index \\ 
427 & CaH2\_Re\_fit &  Angstrom  &  Value of the CaH2 stellar index at 1Re \\ 
428 & CaH2\_alpha\_fit &  --  &  Slope of the gradient of the CaH2 index \\ 
429 & CaHK\_Re\_fit &  Angstrom  &  Value of the CaHK stellar index at 1Re \\ 
430 & CaHK\_alpha\_fit &  --  &  Slope of the gradient of the CaHK index \\ 
431 & Fe4383\_Re\_fit &  Angstrom  &  Value of the Fe4383 stellar index at 1Re \\ 
432 & Fe4383\_alpha\_fit &  --  &  Slope of the gradient of the Fe4383 index \\ 
433 & Fe4531\_Re\_fit &  Angstrom  &  Value of the Fe4531 stellar index at 1Re \\ 
434 & Fe4531\_alpha\_fit &  --  &  Slope of the gradient of the Fe4531 index \\ 
435 & Fe5015\_Re\_fit &  Angstrom  &  Value of the Fe5015 stellar index at 1Re \\ 
436 & Fe5015\_alpha\_fit &  --  &  Slope of the gradient of the Fe5015 index \\ 
437 & Fe5270\_Re\_fit &  \AA\  &  Value of the Fe5270 stellar index at 1Re \\ 
438 & Fe5270\_alpha\_fit &  \AA/Re  &  Slope of the gradient of the Fe5270 index \\ 
439 & Fe5335\_Re\_fit &  \AA\  &  Value of the Fe5335 stellar index at 1Re \\ 
440 & Fe5335\_alpha\_fit &  \AA/Re  &  Slope of the gradient of the Fe5335 index \\ 
441 & Fe5406\_Re\_fit &  Angstrom  &  Value of the Fe5406 stellar index at 1Re \\ 
442 & Fe5406\_alpha\_fit &  --  &  Slope of the gradient of the Fe5406 index \\ 
443 & Fe5709\_Re\_fit &  Angstrom  &  Value of the Fe5709 stellar index at 1Re \\ 
444 & Fe5709\_alpha\_fit &  --  &  Slope of the gradient of the Fe5709 index \\ 
445 & Fe5782\_Re\_fit &  Angstrom  &  Value of the Fe5782 stellar index at 1Re \\ 
446 & Fe5782\_alpha\_fit &  --  &  Slope of the gradient of the Fe5782 index \\ 
447 & G4300\_Re\_fit &  Angstrom  &  Value of the G4300 stellar index at 1Re \\ 
448 & G4300\_alpha\_fit &  --  &  Slope of the gradient of the G4300 index \\ 
449 & HDeltaA\_Re\_fit &  Angstrom  &  Value of the HDeltaA stellar index at 1Re \\ 
450 & HDeltaA\_alpha\_fit &  --  &  Slope of the gradient of the HDeltaA index \\ 
451 & HDeltaF\_Re\_fit &  Angstrom  &  Value of the HDeltaF stellar index at 1Re \\ 
452 & HDeltaF\_alpha\_fit &  --  &  Slope of the gradient of the HDeltaF index \\ 
453 & HGammaA\_Re\_fit &  Angstrom  &  Value of the HGammaA stellar index at 1Re \\ 
454 & HGammaA\_alpha\_fit &  --  &  Slope of the gradient of the HGammaA index \\ 
\hline 
\end{tabular}\label{tab:cat}
\end{center}
\end{table*}

\addtocounter{table}{-1}

\begin{table*}
\begin{center}
\caption{Integrated and characteristic parameters delivered for each analyzed datacube.}
\begin{tabular}{llll}
\hline\hline
\#  & parameter name & units & Description  \\ 
\hline
455 & HGammaF\_Re\_fit &  Angstrom  &  Value of the HGammaF stellar index at 1Re \\ 
456 & HGammaF\_alpha\_fit &  --  &  Slope of the gradient of the HGammaF index \\ 
457 & Hb\_Re\_fit &  \AA\  &  Value of the H$\beta$ stellar index at 1Re \\ 
458 & Hb\_alpha\_fit &  \AA/Re  &  Slope of the gradient of the H$\beta$ index \\ 
459 & Hdmod\_Re\_fit &  \AA\  &  Value of the Hdmod stellar index at 1Re \\ 
460 & Hdmod\_alpha\_fit &  \AA/Re  &  Slope of the gradient of the Hdmod index \\ 
461 & Mg1\_Re\_fit &  Angstrom  &  Value of the Mg1 stellar index at 1Re \\ 
462 & Mg1\_alpha\_fit &  --  &  Slope of the gradient of the Mg1 index \\ 
463 & Mg2\_Re\_fit &  Angstrom  &  Value of the Mg2 stellar index at 1Re \\ 
464 & Mg2\_alpha\_fit &  --  &  Slope of the gradient of the Mg2 index \\ 
465 & MgICvD\_Re\_fit &  Angstrom  &  Value of the MgICvD stellar index at 1Re \\ 
466 & MgICvD\_alpha\_fit &  --  &  Slope of the gradient of the MgICvD index \\ 
467 & Mgb\_Re\_fit &  \AA\  &  Value of the Mgb stellar index at 1Re \\ 
468 & Mgb\_alpha\_fit &  \AA/Re  &  Slope of the gradient of the Mgb index \\ 
469 & NaD\_Re\_fit &  Angstrom  &  Value of the NaD stellar index at 1Re \\ 
470 & NaD\_alpha\_fit &  --  &  Slope of the gradient of the NaD index \\ 
471 & SN\_Re\_fit &  --  &  Value of the SN stellar index at 1Re \\ 
472 & SN\_alpha\_fit &  --  &  Slope of the gradient of the SN index \\ 
473 & TiO1\_Re\_fit &  Angstrom  &  Value of the TiO1 stellar index at 1Re \\ 
474 & TiO1\_alpha\_fit &  --  &  Slope of the gradient of the TiO1 index \\ 
475 & TiO2SDSS\_Re\_fit &  Angstrom  &  Value of the TiO2SDSS stellar index at 1Re \\ 
476 & TiO2SDSS\_alpha\_fit &  --  &  Slope of the gradient of the TiO2SDSS index \\ 
477 & TiO2\_Re\_fit &  Angstrom  &  Value of the TiO2 stellar index at 1Re \\ 
478 & TiO2\_alpha\_fit &  --  &  Slope of the gradient of the TiO2 index \\ 
479 & aTiO\_Re\_fit &  Angstrom  &  Value of the aTiO stellar index at 1Re \\ 
480 & aTiO\_alpha\_fit &  --  &  Slope of the gradient of the aTiO index \\ 
481 & bTiO\_Re\_fit &  Angstrom  &  Value of the bTiO stellar index at 1Re \\ 
482 & bTiO\_alpha\_fit &  --  &  Slope of the gradient of the bTiO index \\ 
483 & e\_C24668\_Re\_fit &  Angstrom  &  Error of the C24668 stellar index at 1Re \\ 
484 & e\_C24668\_alpha\_fit &  --  &  Error in the slope of the gradient of the C24668 index \\ 
485 & e\_CN1\_Re\_fit &  Angstrom  &  Error of the CN1 stellar index at 1Re \\ 
486 & e\_CN1\_alpha\_fit &  --  &  Error in the slope of the gradient of the CN1 index \\ 
487 & e\_CN2\_Re\_fit &  Angstrom  &  Error of the CN2 stellar index at 1Re \\ 
488 & e\_CN2\_alpha\_fit &  --  &  Error in the slope of the gradient of the CN2 index \\ 
489 & e\_Ca4227\_Re\_fit &  Angstrom  &  Error of the Ca4227 stellar index at 1Re \\ 
\hline 
\end{tabular}\label{tab:cat}
\end{center}
\end{table*}

\addtocounter{table}{-1}

\begin{table*}
\begin{center}
\caption{Integrated and characteristic parameters delivered for each analyzed datacube.}
\begin{tabular}{llll}
\hline\hline
\#  & parameter name & units & Description  \\ 
\hline
490 & e\_Ca4227\_alpha\_fit &  --  &  Error in the slope of the gradient of the Ca4227 index \\ 
491 & e\_Ca4455\_Re\_fit &  Angstrom  &  Error of the Ca4455 stellar index at 1Re \\ 
492 & e\_Ca4455\_alpha\_fit &  --  &  Error in the slope of the gradient of the Ca4455 index \\ 
493 & e\_CaH1\_Re\_fit &  Angstrom  &  Error of the CaH1 stellar index at 1Re \\ 
494 & e\_CaH1\_alpha\_fit &  --  &  Error in the slope of the gradient of the CaH1 index \\ 
495 & e\_CaH2\_Re\_fit &  Angstrom  &  Error of the CaH2 stellar index at 1Re \\ 
496 & e\_CaH2\_alpha\_fit &  --  &  Error in the slope of the gradient of the CaH2 index \\ 
497 & e\_CaHK\_Re\_fit &  Angstrom  &  Error of the CaHK stellar index at 1Re \\ 
498 & e\_CaHK\_alpha\_fit &  --  &  Error in the slope of the gradient of the CaHK index \\ 
499 & e\_Fe4383\_Re\_fit &  Angstrom  &  Error of the Fe4383 stellar index at 1Re \\ 
500 & e\_Fe4383\_alpha\_fit &  --  &  Error in the slope of the gradient of the Fe4383 index \\ 
501 & e\_Fe4531\_Re\_fit &  Angstrom  &  Error of the Fe4531 stellar index at 1Re \\ 
502 & e\_Fe4531\_alpha\_fit &  --  &  Error in the slope of the gradient of the Fe4531 index \\ 
503 & e\_Fe5015\_Re\_fit &  Angstrom  &  Error of the Fe5015 stellar index at 1Re \\ 
504 & e\_Fe5015\_alpha\_fit &  --  &  Error in the slope of the gradient of the Fe5015 index \\ 
505 & e\_Fe5270\_Re\_fit &  \AA\  &  Error of the Fe5270 stellar index at 1Re   \\ 
506 & e\_Fe5270\_alpha\_fit &  \AA/Re  &  Error of the slope of the gradient of the Fe5270 index   \\ 
507 & e\_Fe5335\_Re\_fit &  \AA\  &  Error of the Fe5335 stellar index at 1Re   \\ 
508 & e\_Fe5335\_alpha\_fit &  \AA/Re  &  Error of the slope of the gradient of the Fe5335 index   \\ 
509 & e\_Fe5406\_Re\_fit &  Angstrom  &  Error of the Fe5406 stellar index at 1Re \\ 
510 & e\_Fe5406\_alpha\_fit &  --  &  Error in the slope of the gradient of the Fe5406 index \\ 
511 & e\_Fe5709\_Re\_fit &  Angstrom  &  Error of the Fe5709 stellar index at 1Re \\ 
512 & e\_Fe5709\_alpha\_fit &  --  &  Error in the slope of the gradient of the Fe5709 index \\ 
513 & e\_Fe5782\_Re\_fit &  Angstrom  &  Error of the Fe5782 stellar index at 1Re \\ 
514 & e\_Fe5782\_alpha\_fit &  --  &  Error in the slope of the gradient of the Fe5782 index \\ 
515 & e\_G4300\_Re\_fit &  Angstrom  &  Error of the G4300 stellar index at 1Re \\ 
516 & e\_G4300\_alpha\_fit &  --  &  Error in the slope of the gradient of the G4300 index \\ 
517 & e\_HDeltaA\_Re\_fit &  Angstrom  &  Error of the HDeltaA stellar index at 1Re \\ 
518 & e\_HDeltaA\_alpha\_fit &  --  &  Error in the slope of the gradient of the HDeltaA index \\ 
519 & e\_HDeltaF\_Re\_fit &  Angstrom  &  Error of the HDeltaF stellar index at 1Re \\ 
520 & e\_HDeltaF\_alpha\_fit &  --  &  Error in the slope of the gradient of the HDeltaF index \\ 
521 & e\_HGammaA\_Re\_fit &  Angstrom  &  Error of the HGammaA stellar index at 1Re \\ 
522 & e\_HGammaA\_alpha\_fit &  --  &  Error in the slope of the gradient of the HGammaA index \\ 
523 & e\_HGammaF\_Re\_fit &  Angstrom  &  Error of the HGammaF stellar index at 1Re \\ 
524 & e\_HGammaF\_alpha\_fit &  --  &  Error in the slope of the gradient of the HGammaF index \\ 
\hline 
\end{tabular}\label{tab:cat}
\end{center}
\end{table*}

\addtocounter{table}{-1}

\begin{table*}
\begin{center}
\caption{Integrated and characteristic parameters delivered for each analyzed datacube.}
\begin{tabular}{llll}
\hline\hline
\#  & parameter name & units & Description  \\ 
\hline
525 & e\_Hb\_Re\_fit &  \AA\  &  Error of the H$\beta$ stellar index at 1Re   \\ 
526 & e\_Hb\_alpha\_fit &  \AA/Re  &  Error of the slope of the gradient of the H$\beta$index   \\ 
527 & e\_Hdmod\_Re\_fit &  \AA\  &  Error of the Hdmod stellar index at 1Re   \\ 
528 & e\_Hdmod\_alpha\_fit &  \AA/Re  &  Error of the slope of the gradient of the Hdmod index   \\ 
529 & e\_Mg1\_Re\_fit &  Angstrom  &  Error of the Mg1 stellar index at 1Re \\ 
530 & e\_Mg1\_alpha\_fit &  --  &  Error in the slope of the gradient of the Mg1 index \\ 
531 & e\_Mg2\_Re\_fit &  Angstrom  &  Error of the Mg2 stellar index at 1Re \\ 
532 & e\_Mg2\_alpha\_fit &  --  &  Error in the slope of the gradient of the Mg2 index \\ 
533 & e\_MgICvD\_Re\_fit &  Angstrom  &  Error of the MgICvD stellar index at 1Re \\ 
534 & e\_MgICvD\_alpha\_fit &  --  &  Error in the slope of the gradient of the MgICvD index \\ 
535 & e\_Mgb\_Re\_fit &  \AA\  &  Error of the Mgb stellar index at 1Re   \\ 
536 & e\_Mgb\_alpha\_fit &  \AA/Re  &  Error of the slope of the gradient of the Mgb index   \\ 
537 & e\_NaD\_Re\_fit &  Angstrom  &  Error of the NaD stellar index at 1Re \\ 
538 & e\_NaD\_alpha\_fit &  --  &  Error in the slope of the gradient of the NaD index \\ 
539 & e\_SN\_Re\_fit &  --  &  Error of the SN stellar index at 1Re \\ 
540 & e\_SN\_alpha\_fit &  --  &  Error in the slope of the gradient of the SN index \\ 
541 & e\_TiO1\_Re\_fit &  Angstrom  &  Error of the TiO1 stellar index at 1Re \\ 
542 & e\_TiO1\_alpha\_fit &  --  &  Error in the slope of the gradient of the TiO1 index \\ 
543 & e\_TiO2SDSS\_Re\_fit &  Angstrom  &  Error of the TiO2SDSS stellar index at 1Re \\ 
544 & e\_TiO2SDSS\_alpha\_fit &  --  &  Error in the slope of the gradient of the TiO2SDSS index \\ 
545 & e\_TiO2\_Re\_fit &  Angstrom  &  Error of the TiO2 stellar index at 1Re \\ 
546 & e\_TiO2\_alpha\_fit &  --  &  Error in the slope of the gradient of the TiO2 index \\ 
547 & e\_aTiO\_Re\_fit &  Angstrom  &  Error of the aTiO stellar index at 1Re \\ 
548 & e\_aTiO\_alpha\_fit &  --  &  Error in the slope of the gradient of the aTiO index \\ 
549 & e\_bTiO\_Re\_fit &  Angstrom  &  Error of the bTiO stellar index at 1Re \\ 
550 & e\_bTiO\_alpha\_fit &  --  &  Error in the slope of the gradient of the bTiO index \\ 
551 & D4000\_Re\_fit &  --  &  Value of the D4000 stellar index at 1Re \\ 
552 & e\_D4000\_Re\_fit &  --  &  Error of the D4000 stellar index at 1Re   \\ 
553 & D4000\_alpha\_fit &  --  &  Slope of the gradient of the D4000 index \\ 
554 & e\_D4000\_alpha\_fit &  --  &  Error of the slope of the gradient of the D4000 index   \\ 
\hline 
\end{tabular}\label{tab:cat}
\end{center}
\end{table*}

\end{document}